\documentclass[aps, 12pt, preprint, preprintnumbers,amsmath,amssymb,nofootinbib,superscriptaddress,hyperref]{revtex4-1} 
\usepackage{amssymb}
\usepackage{slashed}
\usepackage{amsmath}
\usepackage{braket}
\usepackage{float}
\usepackage[utf8]{inputenc}
\usepackage{graphicx}
\usepackage{subfigure}
\usepackage{latexsym}
\usepackage{epic}
\usepackage{eepic}
\usepackage{epsfig}
\usepackage{color}
\usepackage{natbib}
\usepackage[colorlinks=true,citecolor=darkred,urlcolor=darkred, pdfborder={0 0 0}]{hyperref}
\usepackage[normalem]{ulem}

\newcommand{\bea}{\begin{eqnarray}}
\newcommand{\eea}{\end{eqnarray}}

\definecolor{darkred}{rgb}{0.6,0,0}

\newcommand{\AddrAHEP}{
  AHEP Group, Institut de F\'{i}sica Corpuscular,
  CSIC/Universitat de Val\`{e}ncia, Parc Cient\'ific de Paterna.\\
 C/ Catedr\'atico Jos\'e Beltr\'an, 2 E-46980 Paterna (Valencia), Spain}

\definecolor{linkcolor}{rgb}{0,0,0.5}

\begin{document}
\title{Bounds on the triplet fermions in type-III seesaw and implications for collider searches}
\author{Arindam Das}\email{arindam.das@het.phys.sci.osaka-u.ac.jp}
\affiliation{Department of Physics, Osaka University, Toyonaka, Osaka 560-0043, Japan}
\author{Sanjoy Mandal}\email{smandal@ific.uv.es}
\affiliation{\AddrAHEP}
\preprint{\textbf{OU-HET-1062}}
\bibliographystyle{unsrt} 
\begin{abstract}
Type-III seesaw is a simple extension of the Standard Model~(SM) with the SU$(2)_\text{L}$ triplet fermion with zero hypercharge. It can explain the origin of the tiny neutrino mass and flavor mixing. After the electroweak symmetry breaking the light neutrino mass is generated by the seesaw mechanism which further ensures the mixings between the light neutrino and heavy neutral lepton mass eigenstates. If the triplet fermions are around the electroweak scale having sizable mixings with the SM sector allowed by the correct gauge symmetry, they can be produced at the high energy colliders leaving a variety of characteristic signatures. Based on a simple and concrete realizations of the model we employ a general parametrization for the neutrino Dirac mass matrix and perform a parameter scan to identify the allowed regions satisfying the experimental constraints from the neutrino oscillation data, the electroweak precision measurements and the lepton-flavor violating processes, respectively considering the normal and inverted neutrino mass hierarchies. These parameter regions can be probed at the different collider experiments.
\end{abstract}
\vspace{-3cm}
\maketitle
\section{Introduction}
\label{introduction}
The neutrino masses and the flavor mixings are some of the missing pieces in the SM which have been observed in different experiments \cite{Anselmann:1995ag,Fukuda:1996sz,Athanassopoulos:1997pv, Fukuda:1998tw,Abdurashitov:1994bc,Ahmad:2002jz,Ahmad:2002ka,Hosaka:2006zd,Eguchi:2002dm,Ahn:2002up,Catanesi:2013fxa,Adamson:2011qu,Abe:2011fz,An:2012eh,Ahn:2012nd,Patrignani:2016xqp} consistently. Such experimental results are allowing us to think about the Beyond the Standard Model (BSM) scenarios which can explain the neutrino oscillation phenomena. A simple realization of the neutrino mass generation scenario was inspired by the introduction of the dimension-$5$ Weinberg operator \cite{Weinberg:1979sa} within the SM which led to extend the SM with an SM-singlet Majorana right handed neutrinos \cite{Minkowski:1977sc,Mohapatra:1979ia,Schechter:1980gr,Yanagida:1979as,GellMann:1980vs,Glashow:1979nm} which can explain the neutrino oscillation data, however, there is no experimental observation of the seesaw mechanism or no definite answer of the question of the origin of the neutrino masses. As a result, variety of models have been proposed to address this open question on the origin of the neutrino masses and the nature of the neutrinos.   

Type-III seesaw is amongst such proposals where the SM is extended by an SU$(2)_L$ triplet fermion with zero hypercharge to generate small neutrino mass \cite{Foot:1988aq} through the seesaw mechanism. The triplet fermion consists of a charge neutral multiplet and a singly charged multiplet where the neutral multiplet participates in the seesaw mechanism to generate the tiny neutrino mass and flavor mixing after the electroweak symmetry breaking. As a result the neutral multiplets can mix with the SM neutrinos and through the mixing they can interact with the SM gauge bosons. Like the neutral multiplet, the charged multiplets can also interact with the SM gauge bosons through the mixing at the time of associated with the SM leptons. Therefore high energy colliders can study the productions of such particles when interacting with the SM gauge bosons. The charged multiplets can be also produced directly~(i. e., not suppressed by the light-heavy
mixing angle) in pair at various colliders from SM gauge bosons mediated process. A variety of phenomenological aspects for studying the triplet fermions at the colliders have been discussed in \cite{delAguila:2008hw,Franceschini:2008pz,Biggio:2011ja,Bandyopadhyay:2011aa,Aguilar-Saavedra:2013twa,Bandyopadhyay:2009xa,Bandyopadhyay:2010wp,Ruiz:2015zca,Goswami:2017jqs,Jana:2019tdm,Das:2020gnt} followed by the experimental searches at the Large Hadron Collider (LHC) \cite{CMS:2012dza,CMS:2012ra,CMS:2015mza,CMS:2016hmk,Sirunyan:2017qkz,CMS:2017wua,CMS:2019xud,Aad:2015cxa,ATLAS:2018ghc}.

The rich phenomenology of the type-III seesaw model has been studied in the past addressing the effective neutrino mass including the threshold effect in \cite{Chakrabortty:2008zh}.
The stability of the scalar potential under the perturbativity bounds for a set of degenerate triplet fermions had been studied in \cite{Gogoladze:2008ak} using the evolutions of the renormalization group equations. The electroweak vacuum stability for the nonzero neutrino mass, naturalness and lepton flavor violation have been studied in \cite{Goswami:2018jar}
for the two generations of the triples which can successfully reproduce the neutrino oscillation data for the normal and inverted orderings of the light neutrino mass spectra.
Type-III seesaw has been motivated under an U$(1)$ extension of the SM where a heavy resonantly produced pair of the triplet fermions can be successfully studied and followed by that a BSM neutral gauge boson can be probed. Type-III seesaw scenario has been realized in the grand unified theories where a triplet and a singlet fermions were proposed to be added in \cite{Ma:1998dn,Dimopoulos:1981zb,Sakai:1981gr} where the triplet can reproduce the neutrino oscillation data being in the intermediate scale. Additionally a development of the type-III seesaw scenario was proposed in the SU$(5)$ theory through the inclusion of the adjoint fermionic multiplet in \cite{Bajc:2006ia} and further phenomenological analyses were performed in \cite{Bajc:2006ia,Adhikari:2008uc,Arhrib:2009mz}. The supersymmetric version of this theory had been proposed in \cite{Perez:2007iw} followed by the nonsupersymmetric counterpart in \cite{Perez:2007rm} to find a renormalizable framework to investigate the origin of the small neutrino mass under the grand unification inspired SU$(5)$ theory. Alternatively an inverse seesaw mechanism has been proposed in the type-III framework \cite{Ibanez:2009du} adding a U$(1)_\text{Y}$ hyperchargeless singlet fermion and an SU$(2)_\text{L}$ triplet fermion in \cite{Ma:2009kh} using an additional U$(1)$ gauge group with the anomaly free scenario \cite{Barr:1986hj,Ma:2001kg,Barr:2005je,Biswas:2019ygr} to the SM. There are a verity of indirect search strategies 
prescribed for the type-III seesaw scenario including Lepton Flavor Violation (LFV) \cite{Abada:2008ea,Lindner:2016bgg,Endo:2020tkb,Capdevila:2020rrl} and nonunitarity effects to  \cite{Abada:2007ux,Biggio:2019eeo}. In this context we also mention that such studies have been made in the context of the type-I seesaw in  \cite{Antusch:2007zza,Forero:2011pc,Basso:2013jka,Antusch:2014woa,Fernandez-Martinez:2015hxa,Antusch:2016brq,Fong:2017gke,Hernandez-Garcia:2017pwx,Escrihuela:2019mot,Blennow:2016jkn,Fernandez-Martinez:2016lgt,Antusch:2006vwa,Antusch:2014woa} where only a Majorana type, heavy, and SM singlet right handed neutrino was introduced in the SM. Limits on the light heavy neutrino mixing from the Eletroweak Precision Data (EWPD) were studied in \cite{Raidal:2008jk,delAguila:2008pw}.

In this paper we study the type-III model generalizing the Dirac Yukawa coupling following the Casas-Ibarra conjecture \cite{Casas:2001sr} under the constraints obtained from the nonunitary effects, LFV and EWPD applying the neutrino oscillation data. In our study we consider three degenerate generations of the $SU(2)_L$ triplet fermions which are involved in the neutrino mass generations mechanism form the seesaw mechanism considering the normal and inverted hierarchies of the light neutrino masses. In the type-III seesaw the mixings between the light and heavy mass eigenstates play important roles to study the triplets at different high energy colliders, for example, proton-proton $(pp)$, electron-positron $(e^-e^+)$ and electron-proton $(e^-p)$. There are some production processes where the production cross section of the triplet might not be affected by mixings, however, their branching ratios will depend upon the mixings. As an example we may consider the pair production triplets (charged multiplets in pair and charged and neutral multiplets productions) where the productions processes do not depend upon the mixing directly, however, the dependence of the the mixing comes at the time of the decay of the triplets. The generation of the neutrino mass mechanism in the type-III seesaw is a type of seesaw mechanism where the Dirac Yukawa coupling is always non-diagonal which gives rise to the Flavor Non-diagonal (FND) scenario to correctly reproduce the neutrino oscillation data which will be considered in this article. Depending upon the constraints we will show the allowed parameter space which can be probed by the collider based experiments in the near future.

The paper is organized in the following way. In Sec.~\ref{model}, we discuss the model and the interactions of the triplet fermions with the SM particles. 
In the Sec.~\ref{CI} we discuss general parametrization of the Yukawa coupling and its effect on the different production modes and decay of the triplets. 
In the Sec.~\ref{BR} we discuss about the branching ratios of the triplet fermions under the general parameters.
We study the possibility of the displaced vertices from the type-III seesaw in Sec.~\ref{DV}.
We compare the upper and lower bounds on the mixings in Sec.~\ref{bds} with the current limits and discuss about their implications in the collider study. 
Finally conclude the article in Sec.~\ref{conclusion}.
\section{Model}
\label{model}
\begin{figure}[]
\centering
\includegraphics[width=1.0\textwidth]{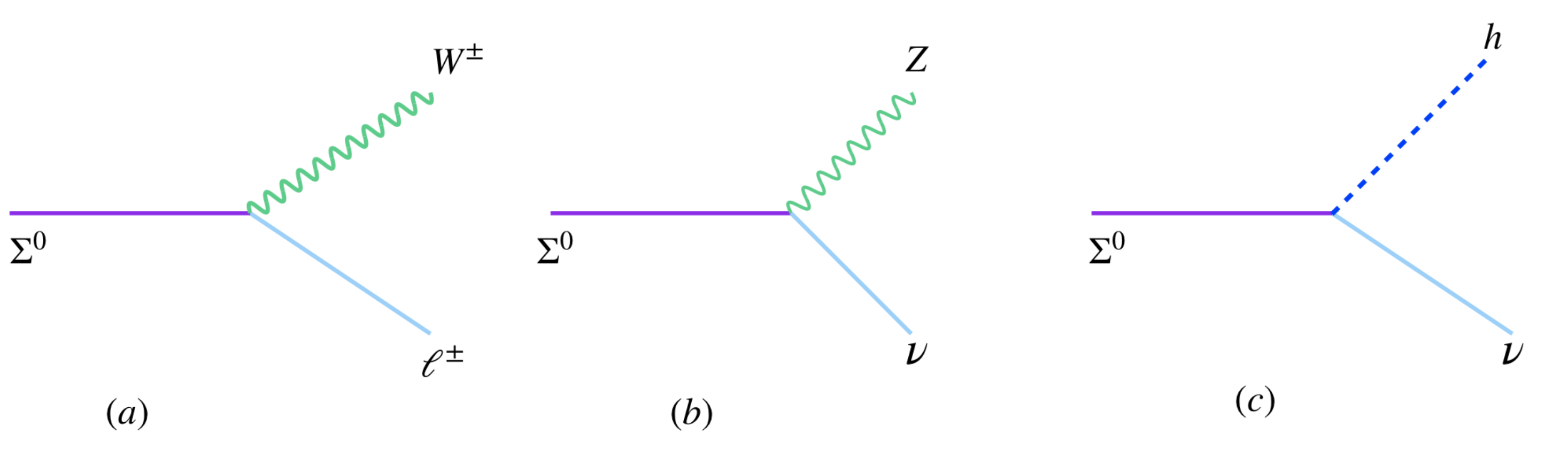}
\caption{Decay modes of $\Sigma^0$.}
\label{Decay1}
\end{figure}

\begin{figure}[]
\centering
\includegraphics[width=1.0\textwidth]{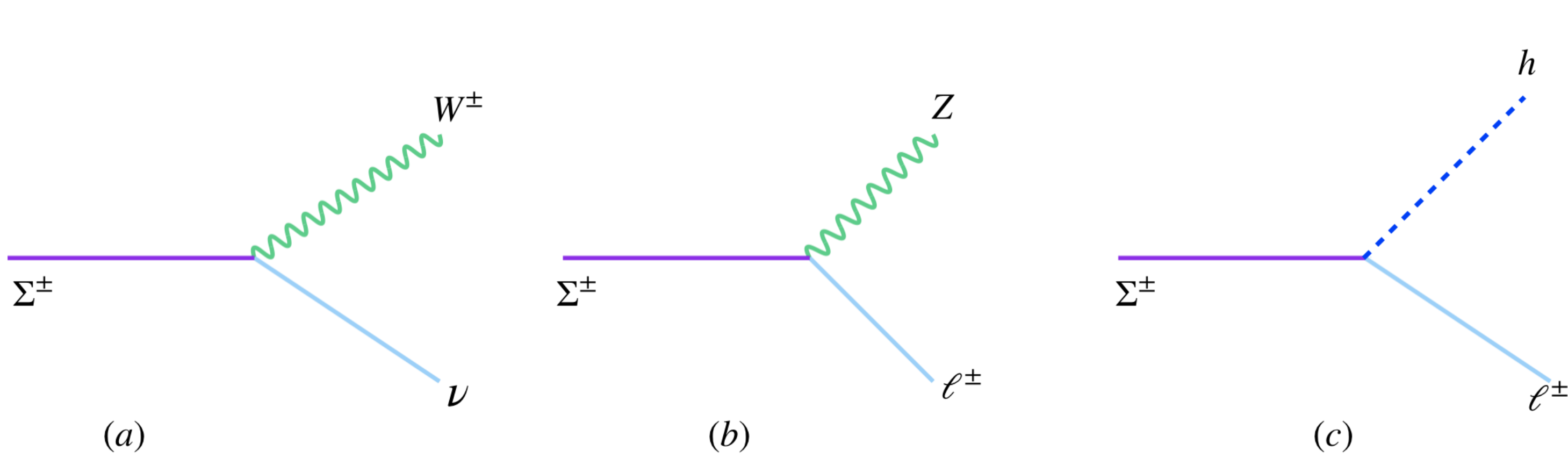}
\caption{Decay modes of $\Sigma^\pm$.}
\label{Decay2}
\end{figure}

\begin{figure}[]
\centering
\includegraphics[width=1.0\textwidth]{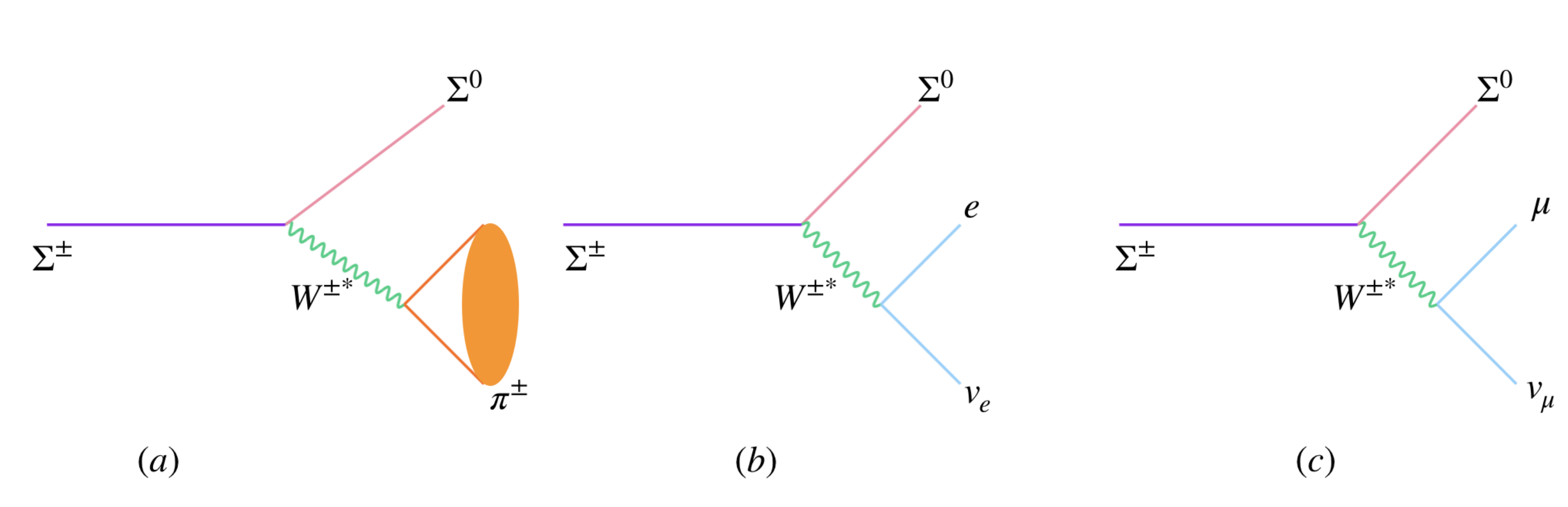}
\caption{Decay modes of $\Sigma^\pm$ evolved from the mass splitting.}
\label{Decay3}
\end{figure}

In the type-III seesaw model SM is extended by three generations of an $SU(2)_L$ triplet fermion $(\Psi)$ with zero hypercharge.
Inclusion of such triplets helps the generation of nonzero but tiny neutrino mass through the seesaw mechanism. 
The Lagrangian can be written as
\begin{align}
\mathcal{L}=\mathcal{L}_{\text{SM}}+ \text{Tr}(\overline{\Psi}i \gamma^\mu D_\mu \Psi)-\frac{1}{2}M \text{Tr}(\overline{\Psi}\Psi^c+\overline{\Psi^c}\Psi)-\sqrt{2}(\overline{\ell_L}Y_D^\dagger \Psi H + H^\dagger \overline{\Psi} Y_D \ell_L)
\label{L}
\end{align}
where $D_\mu$ represents the covariant derivative, $M$ is the Majorana mass term. $\mathcal{L}_{\text{SM}}$ is the relevant part of the SM Lagrangian. We consider three degenerate generation of the triplets. Therefore $M$ is proportional to $\bf{1}_{3\times3}$. $Y_D$ is the Dirac Yukawa coupling between the SM lepton doublet $(\ell_L)$, SM Higgs doublet $(H)$ and the triplet fermion $(\Psi)$. For brevity, we have suppressed the generation indices. In this analysis we represent the relevant SM candidates, the triplet fermion and its charged conjugate $(\Psi^c = C\overline{\Psi}^T)$ as  in the following way 
\bea
\ell_L
 =
 \begin{pmatrix}
  \nu_{L}\\
  e_{L} \\
 \end{pmatrix}\,\,\,
H = 
 \begin{pmatrix}
 \phi^0\\
  \phi^-\\
 \end{pmatrix}\,\,\, 
\Psi=
 \begin{pmatrix}
  \Sigma^0/\sqrt{2}  &  \Sigma^+ \\
 \Sigma^-           &   -\Sigma^0/\sqrt{2}  \\
 \end{pmatrix}\,\,\text{and}\,\,
  \Psi^c=
 \begin{pmatrix}
  \Sigma^{0c}/\sqrt{2}  &  \Sigma^{-c} \\
 \Sigma^{+c}           &   -\Sigma^{0c}/\sqrt{2}  \\
 \end{pmatrix}
 \label{L2}
\eea
After the breaking of the electroweak symmetry $\phi^0$ acquires a vacuum expectation value and we can express it as $\phi^0=\frac{v+h}{\sqrt{2}}$ with $v=246$~GeV.
To study the mixing between the SM charged leptons and $\Sigma^\pm$ we write the four degrees of freedom of each $\Sigma^\pm$ in terms of a Dirac spinor such as 
$\Sigma=\Sigma_R^-+\Sigma_R^{+c}$ where as $\Sigma^0$ are two component fermions with two degrees of freedom. The corresponding Lagrangian after the electroweak symmetry breaking can be written as 
\begin{align}
 -\mathcal{L}_{\text{mass}}=
  \begin{pmatrix}
  \overline{e}_L & \overline{\Sigma}_L \\
 \end{pmatrix}
 \begin{pmatrix}
  m_\ell & Y_D^\dagger v\\
  0 & M \\
 \end{pmatrix}
 \begin{pmatrix}
  e_R \\
  \Sigma_R \\
 \end{pmatrix} 
+
 \frac{1}{2}\begin{pmatrix}
  \overline{\nu_L^c} & \overline{\Sigma_R^0} \\
 \end{pmatrix}
 \begin{pmatrix}
  0 & Y_D^T \frac{v}{\sqrt{2}} \\
  Y_D\frac{v}{\sqrt{2}} & M \\
 \end{pmatrix}
 \begin{pmatrix}
  \nu_L \\
  \Sigma_R^{0c} \\
 \end{pmatrix}
 +h. c.
 \label{n1}
\end{align}
where $m_\ell$ is the Dirac type SM charged lepton mass. The $3\times3$ Dirac mass of the triplets can be written as 
\bea
M_D=\frac{Y_D^T v}{\sqrt{2}}.
\label{mDI}
\eea
Diagonalizing the neutrino mass matrix in Eq.~\ref{n1} we can write the light neutrino mass eigenvalue as 
\bea
m_\nu \simeq -\frac{v^2}{2} Y_D^T M^{-1} Y_D = M_D M^{-1} M_D^{T}
\label{n3}
\eea
hence the mixing between light and heavy mass eigenstates can be obtained as $\mathcal{O}(M_D M^{-1})$. Hence the light neutrino flavor eigenstate can be expressed in terms of the light and heavy mass eigenstates in the following way
\bea
\nu= \mathcal{A} \nu_m + V \Sigma_m
\label{n44}
\eea
where $\nu_m$ and $\Sigma_m$ represent the light and heavy mass eigenstates respectively where $V= M_D M^{-1}$  and $\mathcal{A}= \Big(1-\frac{1}{2} \tilde{\epsilon}\Big)V_{\text{PMNS}}$ with $\tilde{\epsilon} = V^\ast V^T$ and $V_{\text{PMNS}}$ is the $3\times 3$ neutrino mixing matrix which diagonalizes
the light neutrino mass matrix as 
\bea
V_{\text{PMNS}}^T  m_\nu V_{\text{PMNS}}= \text{diag}(m_1, m_2, m_3).
\label{n4}
\eea
Due to the presence of $\tilde{\epsilon}$ the mixing matrix $(\mathcal{A})$ becomes non-unitary, $\mathcal{A}^\dagger\mathcal{A}\neq1$. The charged current (CC) interactions can be expressed in terms of the mass eigenstates including the light heavy mixings as
\bea
 -\mathcal{L}_{\text{CC}}&=&\frac{g}{\sqrt{2}}
 \begin{pmatrix}
  \overline{e} & \overline{\Sigma} \\
 \end{pmatrix}
 \gamma^\mu W_\mu^- P_L  
 \begin{pmatrix}
 (1+\frac{\epsilon}{2}) V_{\text{PMNS}} &-\frac{Y_D^\dagger M^{-1} v}{\sqrt{2}}\\
 0&\sqrt{2}(1-\frac{\epsilon^\prime}{2})\\
 \end{pmatrix}
 \begin{pmatrix}
  \nu \\
  \Sigma^0 \\
 \end{pmatrix} \nonumber \\
 &+&\frac{g}{\sqrt{2}}
 \begin{pmatrix}
  \overline{e} & \overline{\Sigma} \\
 \end{pmatrix}
 \gamma^\mu W_\mu^- P_R  
  \begin{pmatrix}
 0&-\sqrt{2}m_{\ell} Y_D^\dagger M^{-2}v \\
 -\sqrt{2}M^{-1}Y_D(1-\frac{\epsilon^\ast}{2})V_{\text{PMNS}}^\ast&\sqrt{2}(1-\frac{\epsilon^{\prime^\ast}}{2})
 \end{pmatrix} 
 \begin{pmatrix}
  \nu \\
  \Sigma^0 \\
 \end{pmatrix}
  \label{CC} 
   \eea
and the modified neutral current (NC) interaction for the charged sector can be written as 
 \bea
- \mathcal{L}_{\text{NC}}&=&\frac{g}{\cos\theta_W}
 \begin{pmatrix}
  \overline{e} & \overline{\Sigma} \\
 \end{pmatrix}
 \gamma^\mu Z_\mu P_L  
 \begin{pmatrix}
  \frac{1}{2}-\cos^2\theta_W-\epsilon&\frac{Y_D^\dagger M^{-1} v}{2}\\
 \frac{M^{-1} Y_D v}{2}& \epsilon^\prime-\cos^2\theta_W
 \end{pmatrix}
 \begin{pmatrix}
  e \\
  \Sigma \\
 \end{pmatrix} \nonumber \\
 &+&\frac{g}{\cos\theta_W}
 \begin{pmatrix}
  \overline{e} & \overline{\Sigma} \\
 \end{pmatrix}
 \gamma^\mu Z_\mu P_R  
  \begin{pmatrix}
 1-\cos^2\theta_W&m_\ell Y_D^\dagger M^{-2} v \\
 M^{-2} Y_D m_{\ell} v&-\cos^2\theta_W
 \end{pmatrix} 
 \begin{pmatrix}
  e \\
  \Sigma \\
 \end{pmatrix} \nonumber \\
&+& \begin{pmatrix}
  \overline{\nu} & \overline{\Sigma^{0}} \\
 \end{pmatrix}
 \gamma^\mu Z_\mu P_L  
 \begin{pmatrix}
 1-V_{\text{PMNS}}^\dagger \epsilon V_{\text{PMNS}} & \frac{V_{\text{PMNS}}^\dagger Y_D^\dagger M^{-1} v}{\sqrt{2}}\\
 \frac{ M^{-1} Y_D V_{\text{PMNS}} v}{\sqrt{2}}& \epsilon^\prime
 \end{pmatrix}
 \begin{pmatrix}
  \nu \\
  \Sigma^{0} \\
 \end{pmatrix}
 \label{NC} 
 \eea
where $\theta_W$ is the Weinberg angle or weak mixing angle. 
Finally we write the interaction Lagrangian of the SM leptons, triplet fermions with the SM Higgs $(h)$ boson.
The interaction Lagrangian can be written as 
\bea
 -\mathcal{L}_{H}&=&\frac{g}{2M_W}
 \begin{pmatrix}
  \overline{e} & \overline{\Sigma} \\
 \end{pmatrix}
 h P_L
 \begin{pmatrix}
  -\frac{m_\ell}{v}(1-3\epsilon)& m_{\ell} Y_D^\dagger M^{-1}\\
Y_D(1-\epsilon)+M^{-2} Y_D m_\ell^2 & Y_D Y_D^\dagger M^{-1} v
 \end{pmatrix}
  \begin{pmatrix}
  e \\
  \Sigma \\
 \end{pmatrix}  \nonumber \\ 
 &+&
 \frac{g}{2M_W}
 \begin{pmatrix}
  \overline{e} & \overline{\Sigma} \\
 \end{pmatrix}
 P_R 
  \begin{pmatrix}
  -\frac{m_\ell}{v}(1-3\epsilon^\ast)& M^{-1}Y_D^\dagger m_\ell\\
(1-\epsilon^\ast) Y_D^\dagger+ m_\ell^2 Y_D^\dagger M^{-2} &  M^{-1}Y_D Y_D^\dagger v
 \end{pmatrix} 
   \begin{pmatrix}
  e\\
  \Sigma \\
 \end{pmatrix} \nonumber \\
&+& \begin{pmatrix}
  \overline{\nu} & \overline{\Sigma^0} \\
 \end{pmatrix}
 h P_L
 \begin{pmatrix}
\frac{\sqrt{2} m_\nu}{v}& V_{\text{PMNS}}^T m_\nu Y_D^\dagger M^{-1} \\
(Y_D-\frac{Y_D \epsilon}{2} -\frac{\epsilon^{\prime T}Y_D}{2})V_{\text{PMNS}}&\frac{Y_D Y_D^\dagger M^{-1} v}{\sqrt{2}}
 \end{pmatrix}
  \begin{pmatrix}
  \nu \\
  \Sigma^0 \\
 \end{pmatrix}  \nonumber \\ 
 &+&
 \begin{pmatrix}
  \overline{e} & \overline{\Sigma^0} \\
 \end{pmatrix}
 P_R 
  \begin{pmatrix}
\frac{\sqrt{2} m_\nu}{v}&  M^{-1} Y_D m_\nu V_{\text{PMNS}}^\ast \\
V_{\text{PMNS}}^\ast(Y_D^\dagger-\frac{\epsilon^\ast Y_D^\dagger}{2} -\frac{Y_D^\dagger \epsilon^{\prime \ast}Y_D}{2})&\frac{M^{-1} Y_D Y_D^\dagger v}{\sqrt{2}}
 \end{pmatrix}
 \begin{pmatrix}
  \nu\\
  \Sigma^0 \\
 \end{pmatrix}
 \label{H1}
\eea
The charged multiplets of the triplet fermions can interact with photons $(A_\mu)$. The corresponding Lagrangian derived from Eq.~\ref{L} can be written as  
\bea
- \mathcal{L}_{\gamma \Sigma\Sigma}&=&g\sin\theta_W
 \begin{pmatrix}
  \overline{e} & \overline{\Sigma} \\
 \end{pmatrix}
 \gamma^\mu A_\mu P_L  
 \begin{pmatrix}
 1&0\\
 0&1
 \end{pmatrix}
 \begin{pmatrix}
  e \\
  \Sigma \\
 \end{pmatrix} \nonumber \\
 &+&g\sin\theta_W
 \begin{pmatrix}
  \overline{e} & \overline{\Sigma} \\
 \end{pmatrix}
 \gamma^\mu A_\mu P_R  
  \begin{pmatrix}
  1&0\\
  0&1
 \end{pmatrix} 
 \begin{pmatrix}
  e \\
  \Sigma \\
 \end{pmatrix}.
\label{Ph}
\eea
In the Eqs.~\ref{CC}-\ref{H1} the parameters $\epsilon= \frac{v^2}{2} Y_D^\dagger M^{-2} Y_D$, $\epsilon^\prime= \frac{v^2}{2} M^{-1} Y_DY_D^\dagger M^{-1}$ are the small quantities according to \cite{Abada:2007ux,Abada:2008ea,Biggio:2011ja}. We neglect the effects of the higher powers (above $1$) of $\epsilon$ and $\epsilon^\prime$ in the calculations. 
Using the Eq.~\ref{CC} to Eq.~\ref{H1} and the expression for the mixing $(V_{\ell \Sigma})$ we calculate the partial decay widths of $(\Sigma^0)$ as
\bea
\Gamma(\Sigma^0 \to \ell^+ W)&=&\Gamma(\Sigma^0 \to \ell^- W)=\frac{g^2 |V_{\ell \Sigma}|^2}{64 \pi} \Big(\frac{M^3}{M_W^2}\Big) \Big(1-\frac{M_W^2}{M^2}\Big)^2 \Big(1+2\frac{M_W^2}{M^2}\Big) \nonumber \\
\Gamma(\Sigma^0 \to \nu Z)&=&\Gamma(\Sigma^0 \to \overline{\nu} Z)=\frac{g^2 |V_{\ell \Sigma}|^2}{128 \pi \cos^2\theta_W} \Big(\frac{M^3}{M_Z^2}\Big) \Big(1-\frac{M_Z^2}{M^2}\Big)^2 \Big(1+2\frac{M_Z^2}{M^2}\Big) \nonumber \\
\Gamma(\Sigma^0 \to \nu h)&=&\Gamma(\Sigma^0 \to \overline{\nu} h)=\frac{g^2 |V_{\ell \Sigma}|^2}{128 \pi} \Big(\frac{M^3}{M_W^2}\Big) \Big(1-\frac{M_h^2}{M^2}\Big)^2, 
\label{decay1}
\eea
respectively for the Majorana neutrinos. The corresponding Feynman Diagrams have been shown in Fig.~\ref{Decay1}. 
Similarly the partial decay widths of $(\Sigma^\pm)$ are calculated as
\bea
\Gamma(\Sigma^\pm \to \nu W)&=&\frac{g^2 |V_{\ell \Sigma}|^2}{32 \pi} \Big(\frac{M^3}{M_W^2}\Big) \Big(1-\frac{M_W^2}{M^2}\Big)^2 \Big(1+2\frac{M_W^2}{M^2}\Big) \nonumber \\
\Gamma(\Sigma^\pm \to \ell Z)&=&\frac{g^2 |V_{\ell \Sigma}|^2}{64 \pi \cos^2\theta_W} \Big(\frac{M^3}{M_Z^2}\Big) \Big(1-\frac{M_Z^2}{M^2}\Big)^2 \Big(1+2\frac{M_Z^2}{M^2}\Big) \nonumber \\
\Gamma(\Sigma^\pm \to \ell h)&=&\frac{g^2|V_{\ell \Sigma}|^2}{64 \pi} \Big(\frac{M^3}{M_W^2}\Big) \Big(1-\frac{M_h^2}{M^2}\Big)^2,
\label{decay2}
\eea
respectively. $M_W$, $M_Z$ and $M_h$ in the above expressions are the SM $W$, $Z$ and Higgs boson masses respectively.
The corresponding Feynman Diagrams have been shown in Fig.~\ref{Decay2}. 
The charged multiplet $\Sigma^{\pm}$ and neutral multiplet $\Sigma^0$ are degenerate in mass at the tree-level. This degeneracy is lifted up due to the radiative corrections induced by the SM gauge boson in the loop. The estimation of this mass difference $\Delta M$ is found in Ref.~\cite{Cirelli:2005uq} and is given by:
\begin{align}
\Delta M=\frac{\alpha_2 M}{4\pi}\Big(f\big(\frac{M_W}{M}\big)-\cos^2\theta_W f\big(\frac{M_Z}{M}\big)\Big)
\end{align}
where the function $f$ is defined as $f(r)=\frac{r}{2}\big(2r^3\text{ln}\,r-2r+\sqrt{r^2-4}(r^2+2)\text{ln}\,A\big)$ and $A=\big(r^2-2-r\sqrt{r^2-4}\big)/2$. This mass splitting saturates at the value 
$\Delta M\approx 170$ MeV for mass $M>500$ GeV. If this mass splitting $\Delta M$ is larger than pion mass, then $\Sigma^\pm$ will have the following additional decay modes \cite{Cirelli:2005uq} 
\bea
\Gamma(\Sigma^\pm \to \Sigma^0 \pi^\pm)&=& \frac{2 G_F^2 V_{ud}^2 \Delta M^3 f_\pi^2}{\pi} \sqrt{1-\frac{m_\pi^2}{\Delta M^2}} \nonumber \\
\Gamma(\Sigma^\pm \to \Sigma^0 e \nu_e) &=& \frac{2 G_F^2 \Delta M^5}{15 \pi} \nonumber \\
\Gamma(\Sigma^\pm \to \Sigma^0 \mu \nu_\mu) &=&0.12 \Gamma(\Sigma^\pm \to \Sigma^0 e \nu_e)
\label{decay3}
\eea
which are independent of the free parameters. The corresponding Feynman Diagrams have been shown in Fig.~\ref{Decay3}.
The value of the Fermi Constant, $G_F$, is $1.1663787 \times10^{-5}$ GeV$^{-2}$, the value of the CKM parameter $(V_{ud})$ is $0.97420 \pm 0.00021$ and the decay constant of the $\pi$ meson, $f_\pi$, is $0.13$ GeV from \cite{Tanabashi:2018oca}. Notice that for vanishing mixing angles $V_{\ell\Sigma}$, the $\Sigma^{\pm}$ dominantly decay into $\Sigma^0$, hence the decay width or the decay length is determined by $\Delta M$ and is constant. On the contrary, for very large mixing angles, $\Sigma^0$ decay width~(decay length) is very large~(very small). \\
The elements of the matrices $\mathcal{A}$ and $V$ in Eq.~\ref{n44} can be constrained by the experimental data. 
In this analysis we take the global fit results at $3\sigma$ level~\cite{Esteban:2018azc} for the neutrino oscillation parameters: 
\bea
   \Delta m_{12}^2 = m_2^2-m_1^2 = \Big[6.79 \times 10^{-5} \text{eV}^2,\, 8.01\times 10^{-5}\text{eV}^2\Big]   \nonumber  \ \\
   \Delta m_{23}^2= |m_3^2-m_2^2|=\Big[2.432 \times 10^{-3} \text{eV}^2,\,2.618\times 10^{-3}\text{eV}^2\Big]  \nonumber \ \\
   \sin^2 \theta_{12}=\Big[0.275,\,0.350\Big] \nonumber \\
   \sin^2 \theta_{23}=\Big[0.427,\,0.609\Big] \nonumber \ \\
   \sin^{2}{\theta_{13}}=\Big[0.02046,\,0.02440\Big].
   \label{data}
\eea   
The $3\times 3$ neutrino mixing matrix $V_{\rm PMNS}$ is given by 
\bea
V_{\rm{PMNS}}= \begin{pmatrix} c_{12} c_{13}&s_{12}c_{13}&s_{13}e^{i\delta_{\text{CP}}}\\-s_{12}c_{23}-c_{12}s_{23}s_{13}e^{i\delta_{\text{CP}}}&c_{12}c_{23}-s_{12}s_{23}s_{13}e^{i\delta_{\text{CP}}}&s_{23} c_{13}\\ s_{12}c_{23}-c_{12}c_{23}s_{13}e^{i\delta_{\text{CP}}}&-c_{12}s_{23}-s_{12}c_{23}s_{13}e^{i\delta_{\text{CP}}}&c_{23}c_{13} \end{pmatrix} \begin{pmatrix} 1&0&0\\0&e^{i\rho_1}&0\\0&0&e^{i\rho_2}\end{pmatrix} 
\label{pmns}
\eea
where $c_{ij}=\cos\theta_{ij}$ and $s_{ij}=\sin\theta_{ij}$. In our analysis the Dirac CP-phase $(\delta_{\rm CP})$ is a free parameter running between the limit $[-\pi, \pi]$.
However, in the recent experiments by NO$\nu$A~\cite{Adamson:2016tbq} and T2K~\cite{Abe:2017uxa} indicate that $\delta_{\rm CP}$ can be $-\frac{\pi}{2}\pm \frac{\pi}{2}$.
Due to non-unitarity \cite{Abada:2007ux} the elements of $\mathcal{A}$ are severely constrained at $90\%$ C. L.:
\bea
|\mathcal{A}\mathcal{A}^\dagger| =
\begin{pmatrix} 
 1.001\pm0.002& <1.1 \times 10^{-6} &  < 1.2\times 10^{-3}\\
 <1.1 \times 10^{-6} & 1.002\pm 0.002 & <1.2\times 10^{-3}\\
 < 1.2 \times 10^{-3}& < 1.2 \times 10^{-3} & 1.002\pm 0.002
\end{pmatrix}.
\label{non-unit}
\eea
 The diagonal elements of Eq.~\ref{non-unit} are obtained from the precision studies of the SM weak boson where as the SM prediction is $1$. 
The off-diagonal entries of Eq.~\ref{non-unit}  are the upper bounds obtained from the cLFV studies, 
for example, the constraints on the $12$ and $21$ elements of Eq.~\ref{non-unit} are coming from the the $\mu \to 3e $ process \cite{Bellgardt:1987du}, 
the constraints on the $23$ and $32$ elements are coming from the $\tau \to 3\mu$ process
and finally the constraints on the $13$ and $31$ elements are originated from the $\tau \to 3e $ process respectively.
These bounds are taken from \cite{Abada:2007ux}. The diagonal elements are obtained from LEP \cite{ALEPH:2005ab,Tanabashi:2018oca}.
As a result we have $\mathcal{A}\mathcal{A}^\dagger \simeq 1 - \tilde{\epsilon}$ and we can calculate the constraints on $\tilde{\epsilon}$ from Eq.~\ref{non-unit}
as
\bea
|\tilde{\epsilon}| =
\begin{pmatrix} 
 0.001\pm0.002& < 1.1 \times 10^{-6} & < 1.2 \times 10^{-3}\\
 < 1.1\times 10^{-6} & 0.002\pm0.002 & < 1.2 \times 10^{-3}\\
 < 1.2\times 10^{-3}& < 1.2 \times 10^{-3} & 0.002\pm0.002
\end{pmatrix} .
\label{eps}
\eea
where we have used the central values for the diagonal elements. 
Note that the stringent bound is given by the $12$-element which is originated from the $\mu \to 3e$ cLFV process.\\

\section{Bounds on the mixing angles under the general parametrization and its effect on the decay of the triplet fermions}
\label{CI}
In this analysis we generalize of the Dirac Yukawa mass matrix of  Eq.~\ref{mDI} using the Casas-Ibarra \cite{Casas:2001sr} conjecture as follows 
\bea 
M_D^{\text{NH/IH}} =V_{\rm{PMNS}}^\ast \sqrt{D_{\rm{NH/IH}}} \; O \sqrt{M},  
\label{mDg}
\eea
where $O$ is a general orthogonal matrix and it can be written as 
\bea
O \ = \ 
\begin{pmatrix}
1&0&0\\
0&\cos[x]& \sin[x]\\
0&-\sin[x]& \cos[x]
\end{pmatrix}
\begin{pmatrix}
\cos[y]&0&\sin[y]\\
0&1& 0\\
-\sin[y]& 0&\cos[y]
\end{pmatrix}
\begin{pmatrix}
\cos[z]&\sin[z]&0\\
-\sin[z]&\cos[z]&0\\
0&0&1
\end{pmatrix}
\label{Omatrix}
\eea
where the angles $x, y, z$ are the complex numbers. Now using $\tilde{\epsilon}=(V^\ast V^T)_{\rm{NH/IH}}$, $(V_{\alpha i})_{\rm{NH/IH}} = M_{D_{\rm{NH/IH}}} M^{-1}$ and Eqs.~\ref{n3} and \ref{n4} for the two different hierarchies we can write
\bea
\tilde{\epsilon}^{\rm{NH/IH}}
    = V_{\rm{PMNS}} \sqrt{D_{\rm{NH/IH}}} O^{\ast} M^{-1} O^{T} \sqrt{D_{\rm{NH/IH}}} V_{\rm{PMNS}}^{\dagger}. 
\label{RR}    
\eea 
where NH is the normal hierarchy $(m_3 > m_2 > m_1)$ and IH is the inverted hierarchy $(m_2 > m_1 > m_3)$.
The light neutrino mass eigenvalue matrices $(\sqrt{D_{\rm NH/IH}})$ for the NH and IH cases are written as 
\bea 
\sqrt{D^{\rm{NH}}}=
\begin{pmatrix}
\sqrt{m_1}&0&0\\
0&\sqrt{m_2^{\rm{NH}}}&0\\
0&0&\sqrt{m_3^{\rm{NH}}}
\end{pmatrix}, 
\sqrt{D^{\rm{NH}}}=
\begin{pmatrix}
\sqrt{m_1^{\rm IH}}&0&0\\
0&\sqrt{m_2^{\rm{IH}}}&0\\
0&0&\sqrt{m_3}
\end{pmatrix}  
\label{DNHIH}
\eea 
where $m_2^{\rm{NH}}=\sqrt{\Delta m_{12}^2+m_1^2}$, $m_3^{\rm{NH}}=\sqrt{\Delta m_{23}^2 + (m_2^{\rm{NH}})^2}$, $m_2^{\rm{IH}}=\sqrt{ \Delta m_{23}^2 + m_3^2}$ and $m_1^{\rm{IH}}=\sqrt{(m_2^{\rm{IH}})^2- \Delta m_{12}^2}$ for the NH and IH respectively. In both cases, the triplet mass matrix is defined as  $M= M (\bf{1}_{3\times3})$ which is proportional to a $3\times3$ unit matrix for the three degenerate triplets. In Eq.~\ref{DNHIH} the lightest mass eigenvalue is a free parameter and bounded from the PLANCK data \cite{Aghanim:2018eyx} and $m_1 (m_3)$ is the lightest light neutrino mass eigenvalue for the NH (IH) case. In this analysis $\delta_{CP}$ and $\rho_{1,2}$ vary between $[-\pi, \pi]$. In this context we mention that seesaw mechanism has been extensively studied utilizing the general parametrization under the Casas-Ibarra conjecture in \cite{Perez:2009mu, Ibarra:2010xw,Ibarra:2011wi,Ibarra:2011xn,Dinh:2012bp,Cely:2012bz,Das:2017nvm,Das:2012ze,Bhardwaj:2018lma,Das:2019fee,Das:2018tbd} and following that to study the vacuum stability in type-III seesaw with two generations of the triplet fermions using the Casas-Ibarra conjecture has been studied in \cite{Goswami:2018jar}, however, in our analysis we study three degenerate triplets under the constraints obtained from the indirect searches.
\begin{figure}[]
\centering
\includegraphics[width=0.49\textwidth]{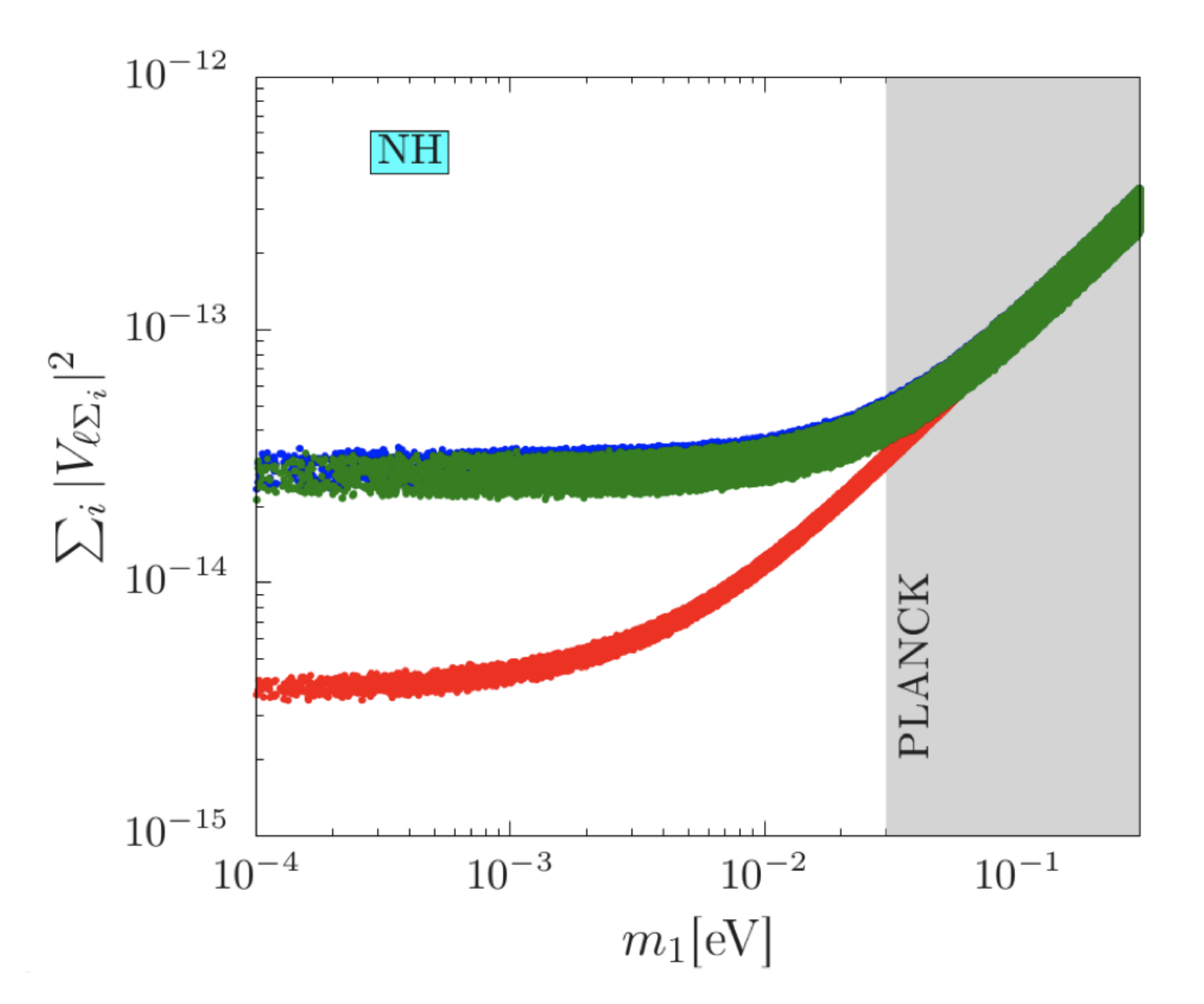}
\includegraphics[width=0.49\textwidth]{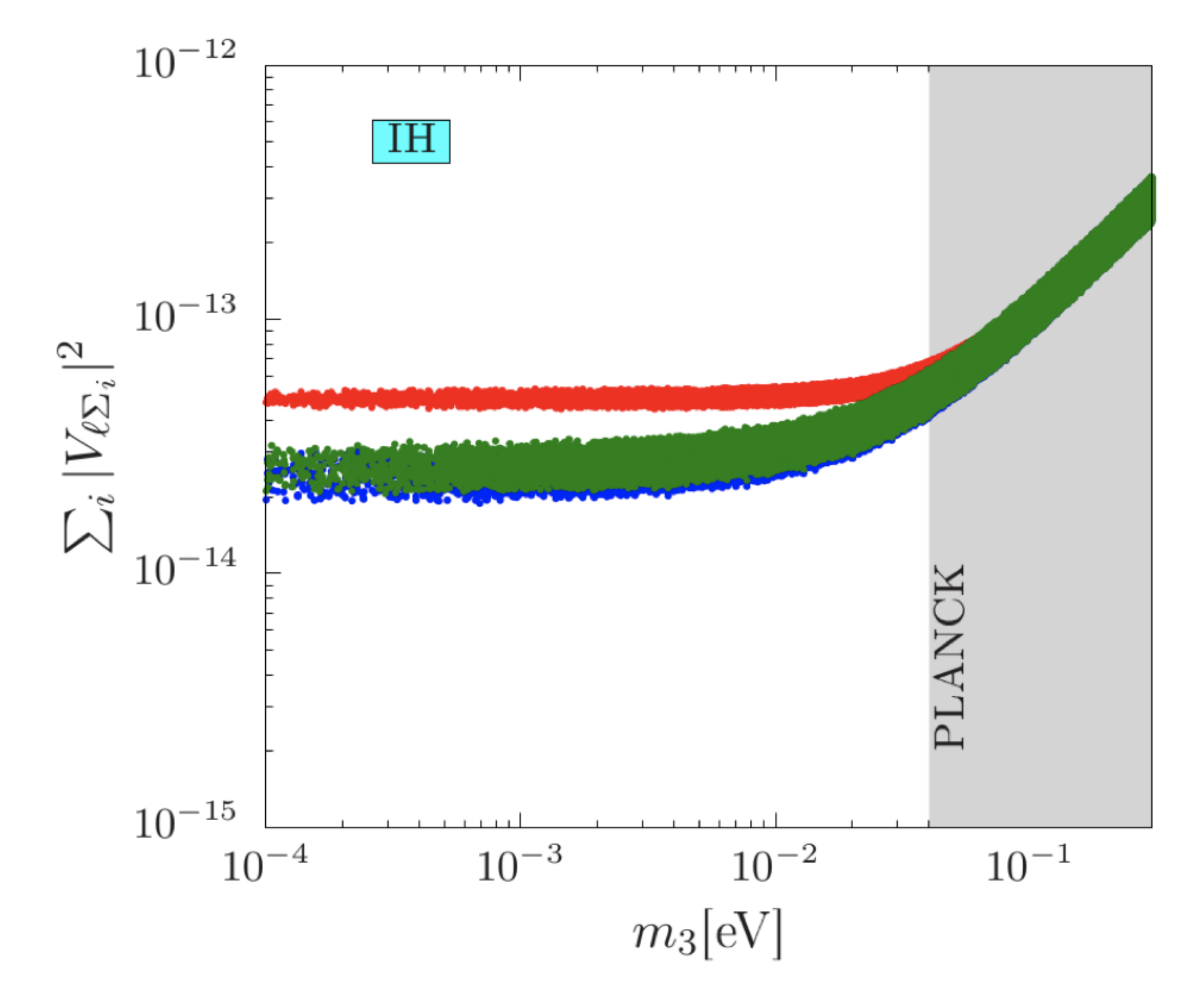}
\caption{Bounds on $\Sigma_i |V_{\ell \Sigma_i}|^2$ as a function of the $m_1 (m_3)$ NH (IH) case in the left (right) panel for fixed SM lepton flavors. The red band represents electron $(e)$, the blue band represents the muon $(\mu)$ and the green band represents the tau $(\tau)$. In this case we consider $O=\bf 1_{3\times 3}$ as a identity matrix. The same nature will be obtained from case when $O$ is a real orthogonal matrix. We fix the triplet mass $M=1$ TeV. The shaded region in gray is ruled out by the PLANCK data.}
\label{Mix1}
\end{figure}
We have three different choices for the orthogonal matrix in Eq.~\ref{Omatrix} as follows:
\begin{itemize}
\item[(i)]$O$ is a identity matrix, $O= \bf 1_{3\times3}$. In this case Eq.~\ref{mDg} will be 
\bea
M_D^{\text{NH/IH}}=V_{\text{PMNS}}^\ast \sqrt{D_{\text{NH/IH}}} \sqrt{M}.
\eea
This will further affect the light-heavy mixing. In this case there is no dependence on $x,~y,~z$.
\item[(ii)]$O$ is a real orthogonal matrix with diagonal and off-diagonal entries, $(x,~y,~z)$ are real and vary between $[-\pi, \pi ]$
\item[(iii)] $O$ is a complex orthogonal matrix where $x,~y,~z$ are the complex numbers, i. e., $x_i+ i y_i$ and $-\pi \leq x_i, y_i \leq \pi$.
Needless to say, the application of the non-unitarity effects will restrict the unboundedness of the complex quantities in the trigonometric functions.
\end{itemize}
For the cases (i) and (ii) using the two hierarchies of the neutrino masses (NH and IH) we calculate the modulus square of the mixing between a triplet and the corresponding lepton flavors. Then fixing the lepton flavor, we sum over the triplets as
\bea
\Sigma_i |V_{\ell \Sigma_i}|^2= |V_{\ell \Sigma_1}|^2 + |V_{\ell \Sigma_2}|^2+ |V_{\ell \Sigma_3}|^2  
\eea
where $\ell=e,\mu$ and $\tau$. Note that $\sum_{i}|V_{\ell\Sigma_i}|^2$ is same if $O$ is identity or real orthogonal matrix.
For both of these cases, $\Sigma_i |V_{\ell \Sigma_i}|^2$ have been plotted as a function of the lightest light neutrino mass eigenvalue in Fig.~\ref{Mix1}.
The NH (IH) case is shown in the left (right) panel as a function of $m_1~(m_3)$ where the electron flavor is presented by the red band and 
the muon and tau flavors are represented by the blue and green bands. In the NH (IH) case the bounds on the electron flavor (muon and tau flavors) are stronger
for the decreasing $m_1~(m_3)$. In this analysis we fix the triplet mass $M=1$ TeV. 

\begin{figure}[]
\centering
\includegraphics[width=0.31\textwidth]{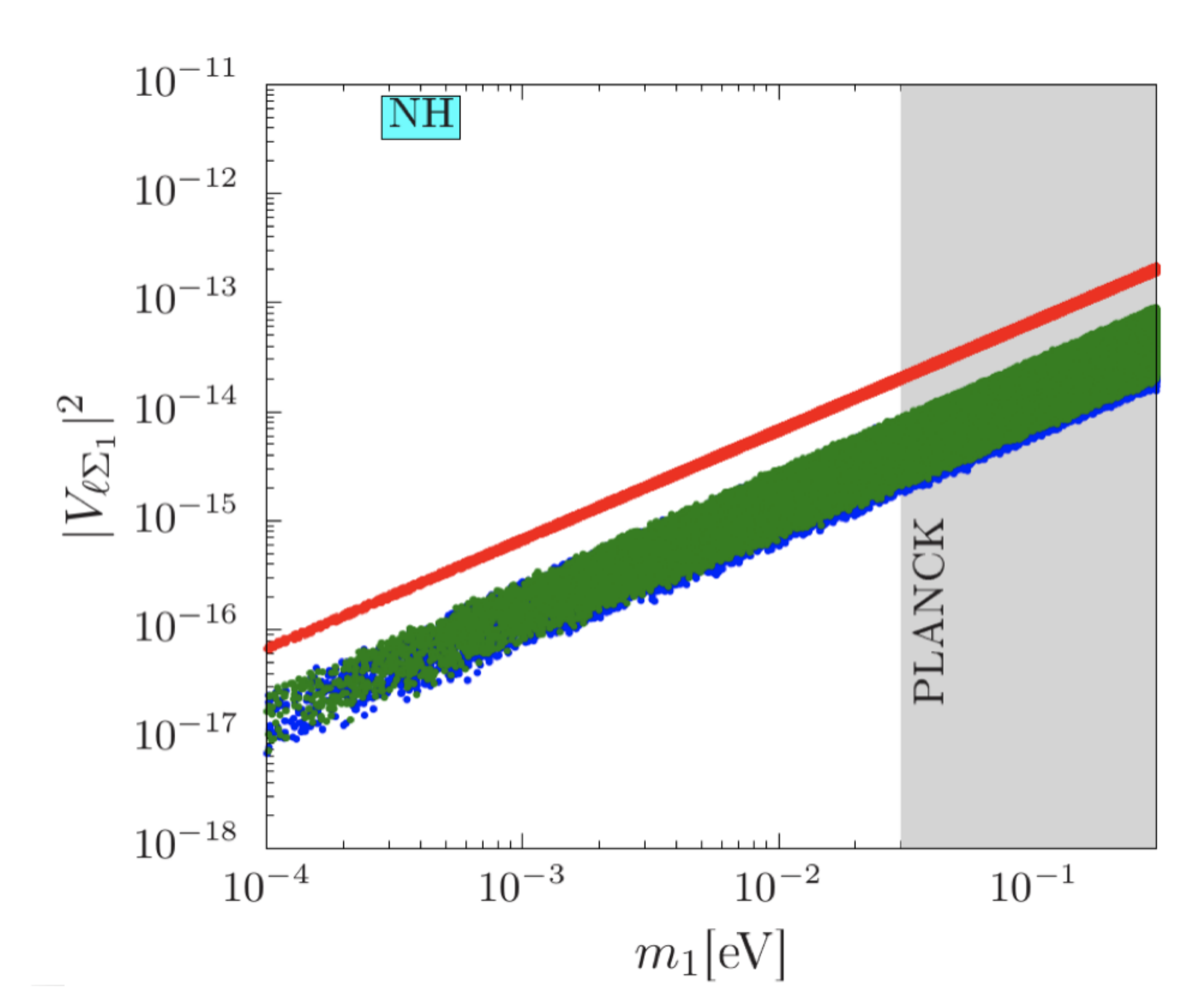}
\includegraphics[width=0.31\textwidth]{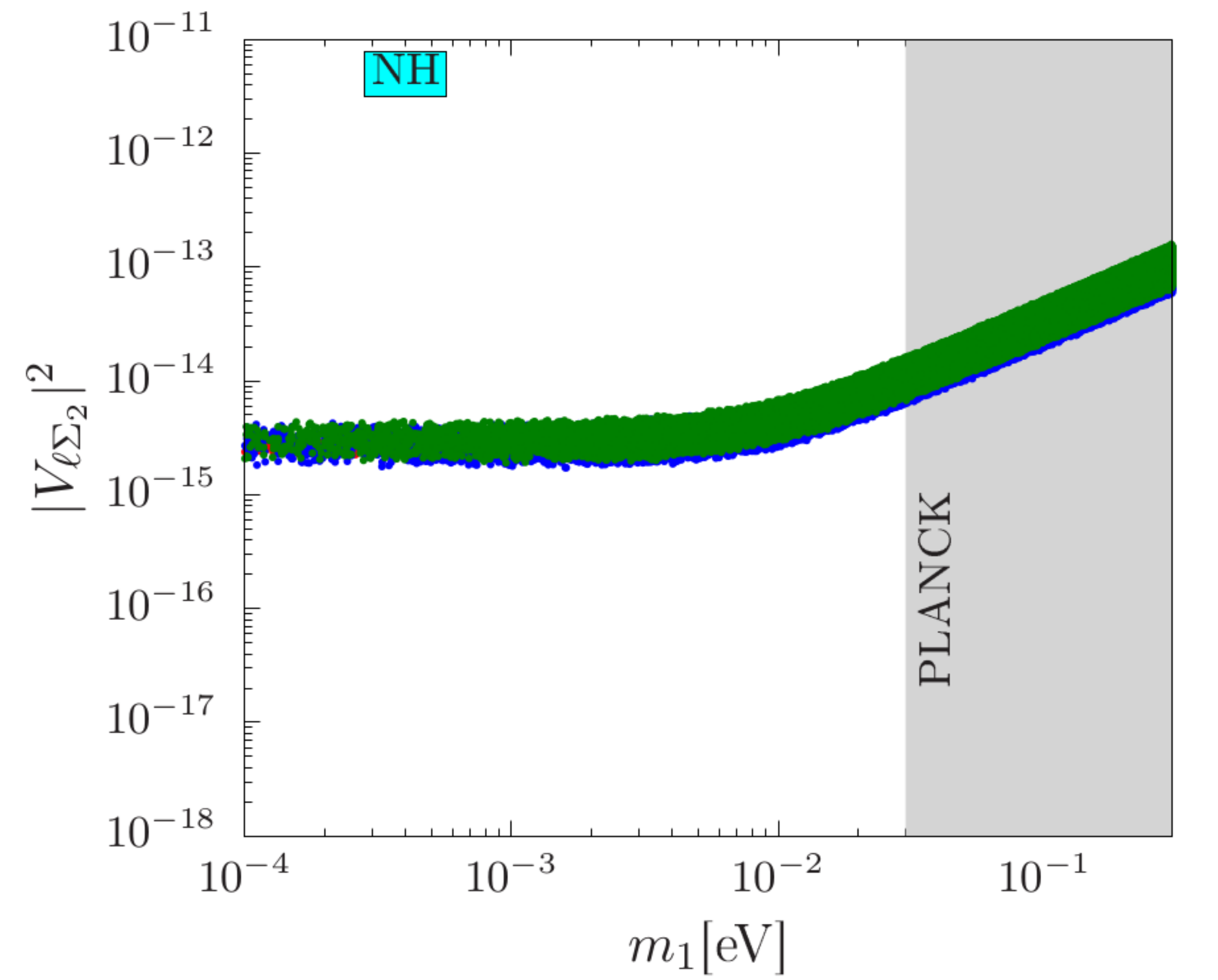}
\includegraphics[width=0.31\textwidth]{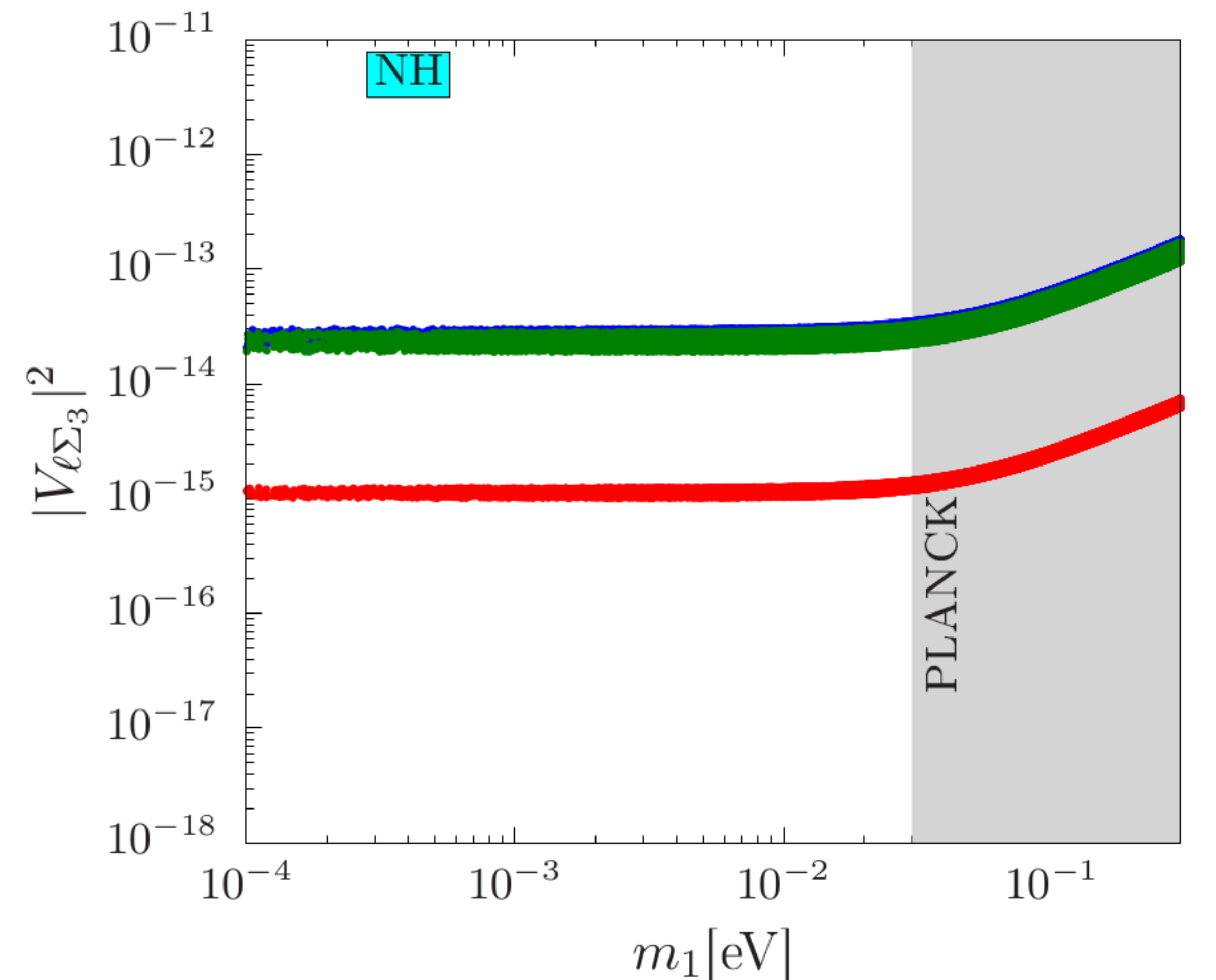}\\
\includegraphics[width=0.31\textwidth]{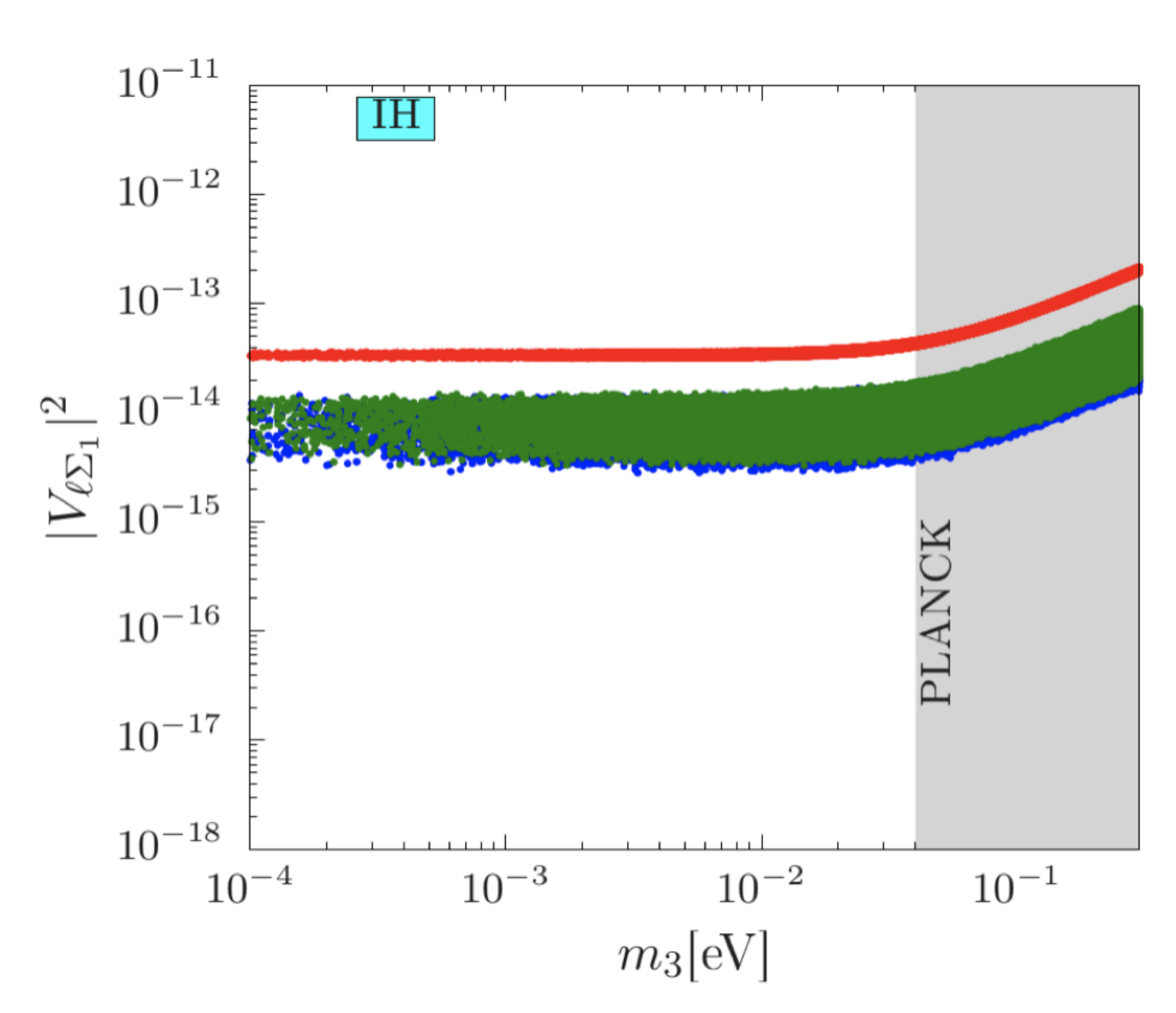}
\includegraphics[width=0.31\textwidth]{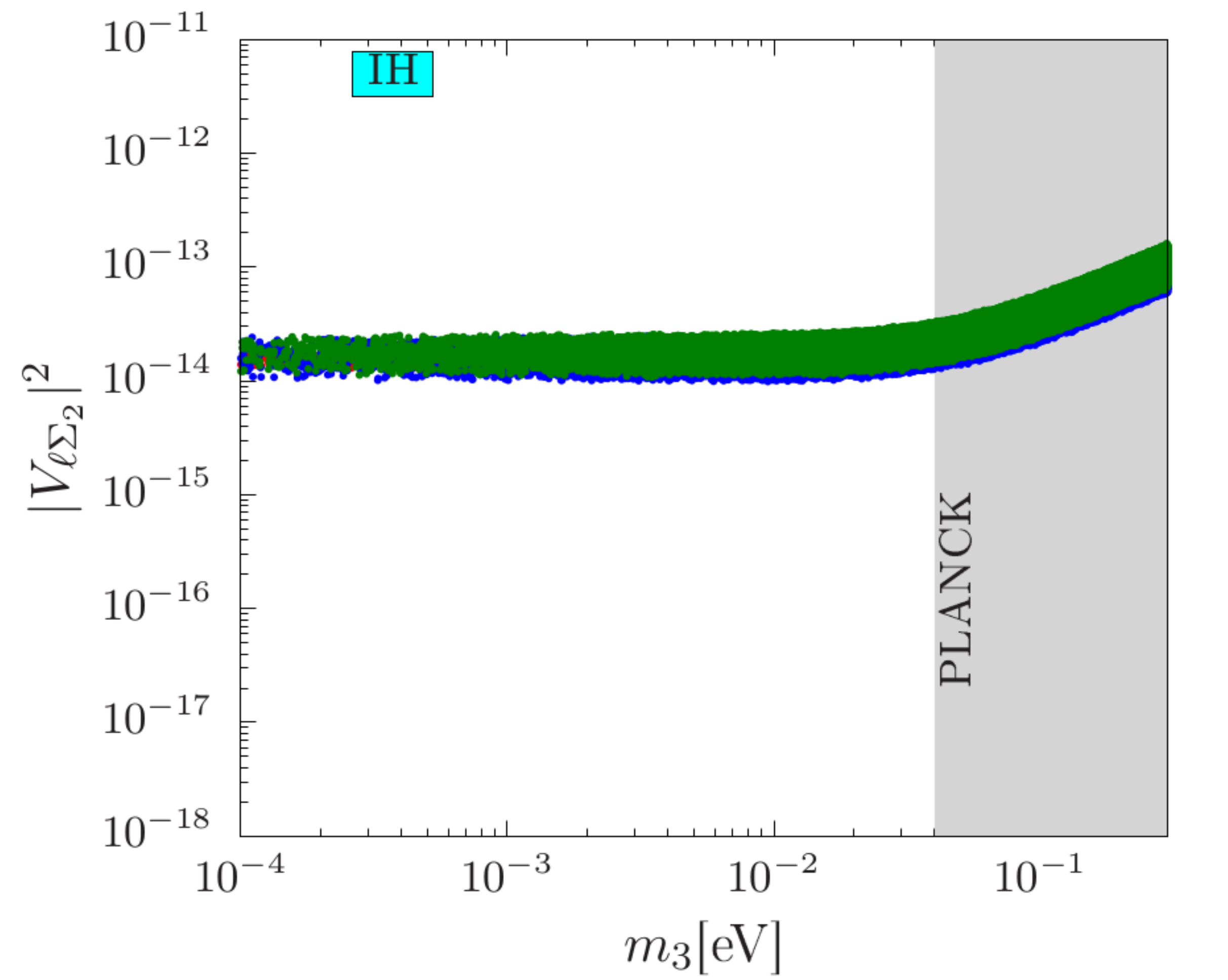}
\includegraphics[width=0.31\textwidth]{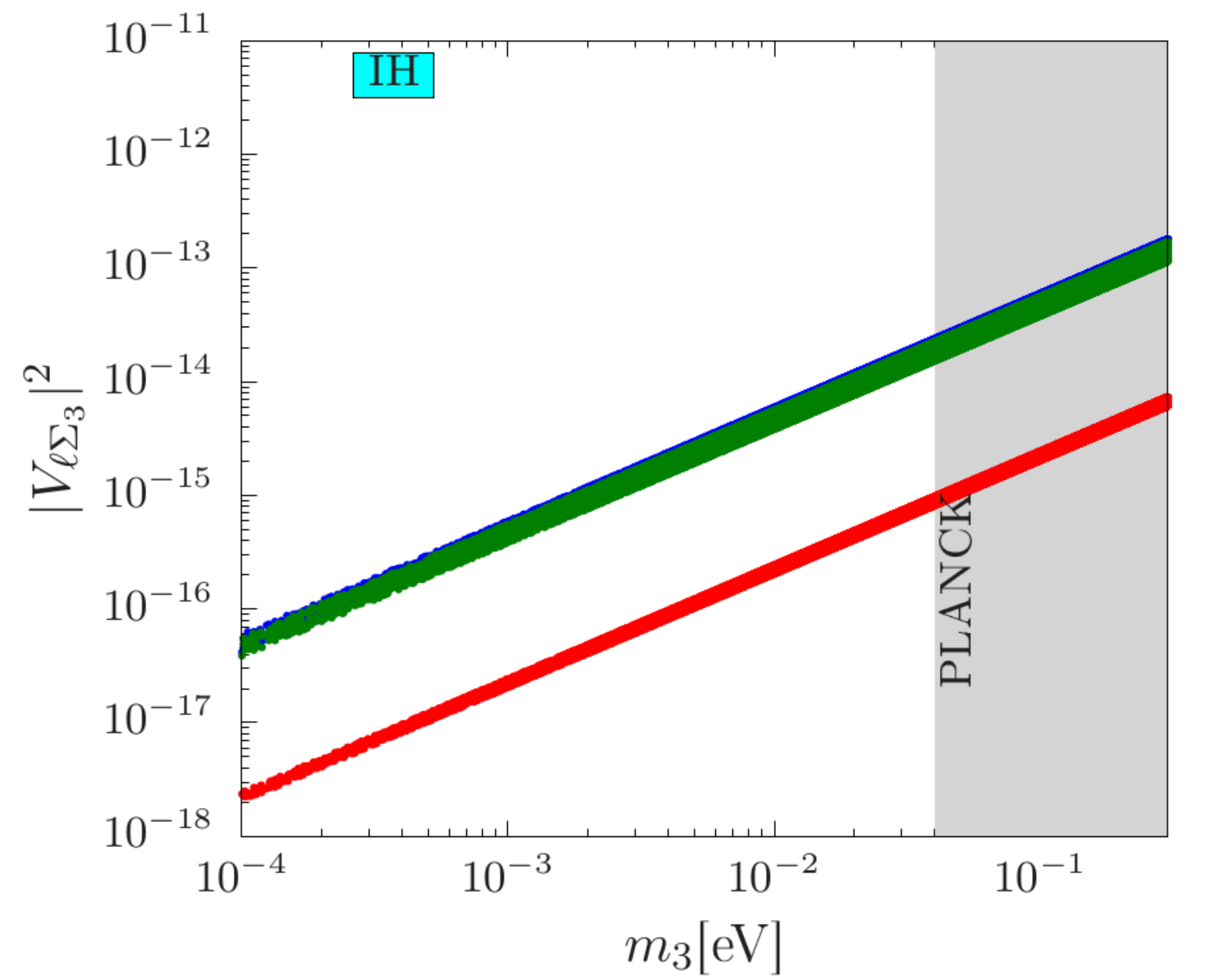}
\caption{Bounds on the individual mixing $|V_{\ell \Sigma_i}|^2$ for $O=\bf 1_{3\times3}$ as a function of $m_1(m_3)$ in the NH (IH) case. 
In this case we fix the triplet flavor $(\Sigma_{i})$ and find the bounds on its mixing with electron (red), muon (blue) and tau (green). We have considered $M=1$ TeV. 
The shaded region in gray is ruled out by the PLANCK data.}
\label{Mix2}
\end{figure}

We also plot the individual mixing as a function of the $m_1(m_3)$ for the NH (IH) case in the top (bottom) panel of the Fig.~\ref{Mix2} for case (i).
We find that $|V_{\ell\Sigma_1}|^2$ for electron (red), muon (blue) and tau (green) in the NH case are related to $m_1$, lower the value of $m_1$ lowers the individual mixing in the NH case 
whereas in the IH case the mixings are parallel to the horizontal axis below the PLANCK limit. In both of the cases the $|V_{e\Sigma_1}|^2$ is less stronger than the other mixings. 
The mixings for other two flavors overlap with each other. 
The nature of the $|V_{\ell\Sigma_2}|^2$ is same for the three flavors of the leptons in both of the NH and IH cases, where all flavors overlap with each other. 
On the other hand for $|V_{\ell \Sigma_3}|^2$ the mixing with the electron flavor is stronger than those with the other two flavors whereas $|V_{\mu\Sigma_3}|^2$ and $|V_{\tau \Sigma_3}|^2$ 
overlap with each other in both of the NH and IH cases, however, in the NH case all three mixings are parallel to the horizontal axis below the PLANCK limit. 
On the other hand in the IH case mixing decreases with the decreasing $m_3$. 

In the following we write down the individual mixings between the $\Sigma_1$ and the three generations of the leptons for the case of $O=\bf 1_{3\times3}$:
\bea
|V_{e\Sigma_1}|^2 &=& m_1 \frac{c_{12}^2 c_{13}^2}{ M} \nonumber \\
|V_{\mu \Sigma_1}|^2 &=& m_1\frac{|c_{12} s_{12}+ c_{12} e^{i\delta_{\text{CP}}} s_{13} s_{23}|^2}{ M} \nonumber \\
|V_{\tau \Sigma_1}|^2 &=& m_1\frac{|c_{12} c_{23} e^{i\delta_{\text{CP}}} s_{13}-s_{12} s_{23}|^2}{ M}.
\label{m1}
\eea
We write down the individual mixings between the $\Sigma_2$ and the three generations of the leptons for the case of $O=\bf 1_{3\times3}$:
\bea
|V_{e\Sigma_2}|^2 &=& m_2\frac{c_{13}^2 s_{12}^2 }{ M} \nonumber \\
|V_{\mu \Sigma_2}|^2 &=& m_2\frac{|c_{23} e^{i\delta_{\text{CP}}} s_{12} s_{13}+ c_{12} s_{23}|^2}{ M} \nonumber \\
|V_{\tau \Sigma_2}|^2 &=& m_2\frac{|c_{12} c_{23} -e^{i\delta_{\text{CP}}} s_{12}s_{13} s_{23}|^2}{ M}
\label{m2}
\eea
and we write down the individual mixings between the $\Sigma_3$ and the three generations of the leptons for the case of $O=\bf 1_{3\times3}$:
\bea
|V_{e\Sigma_3}|^2 &=& m_3\frac{s_{13}^2 }{ M} \nonumber \\
|V_{\mu \Sigma_3}|^2 &=& m_3\frac{c_{13}^2 s_{23}^2}{ M} \nonumber \\
|V_{\tau \Sigma_3}|^2 &=& m_3\frac{c_{13}^2 c_{23}^2}{ M} 
\label{m3}
\eea
Hence we can calculate $\Sigma_i |V_{\ell \Sigma_i}|^2$ from the Eqs.~\ref{m1}-\ref{m3} for $i=1,~2,~3$ and $\ell=e,~\mu,~\tau$. 

We notice that $|V_{\ell \Sigma_1}|^2$ is proportional to $m_1$ and $|V_{\ell \Sigma_3}|^2$ is proportional to $m_3$. Hence in the NH and IH cases the corresponding individual mixings in Eqs.~\ref{m1} and \ref{m3} will tend to zero as $m_1 \to 0$ for the NH and $m_3 \to 0$ for the IH cases, which is clearly visible in Fig.~\ref{Mix2}. The behavior for the other mixings in the NH $(|V_{\ell\Sigma_2}|^2, |V_{\ell\Sigma_3}|^2)$ and IH $(|V_{\ell\Sigma_1}|^2, |V_{\ell\Sigma_2}|^2)$ cases do not have this behavior because they depend on $(m_2, m_3)$ for the NH and on $(m_1, m_3)$ for the IH cases respectively. They almost independent of lightest light neutrino mass eigenvalue for the respective NH $(m_1)$ and IH $(m_3)$ cases slightly below the PLANCK limit. 
\begin{figure}
\centering
\includegraphics[width=0.31\textwidth]{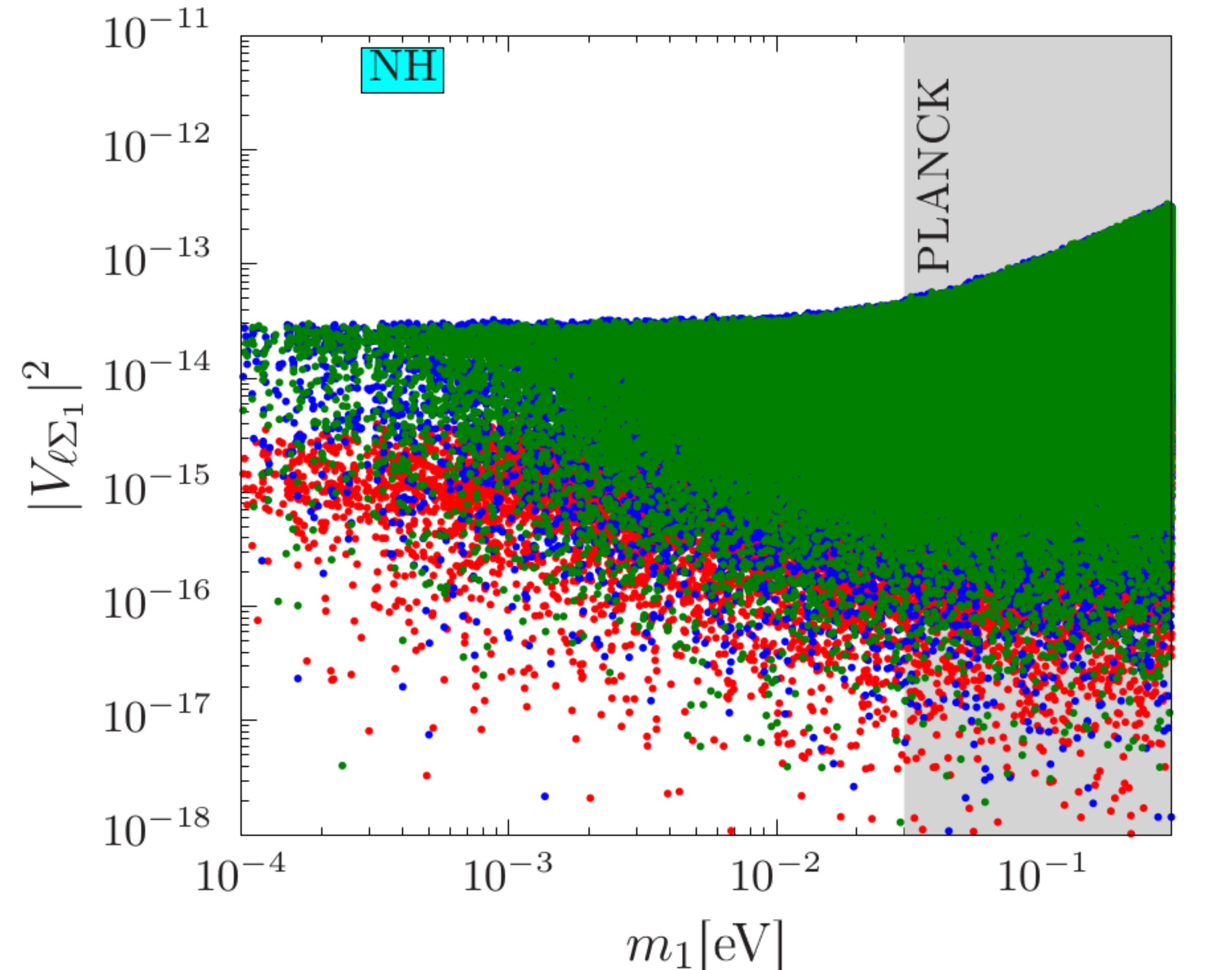}
\includegraphics[width=0.31\textwidth]{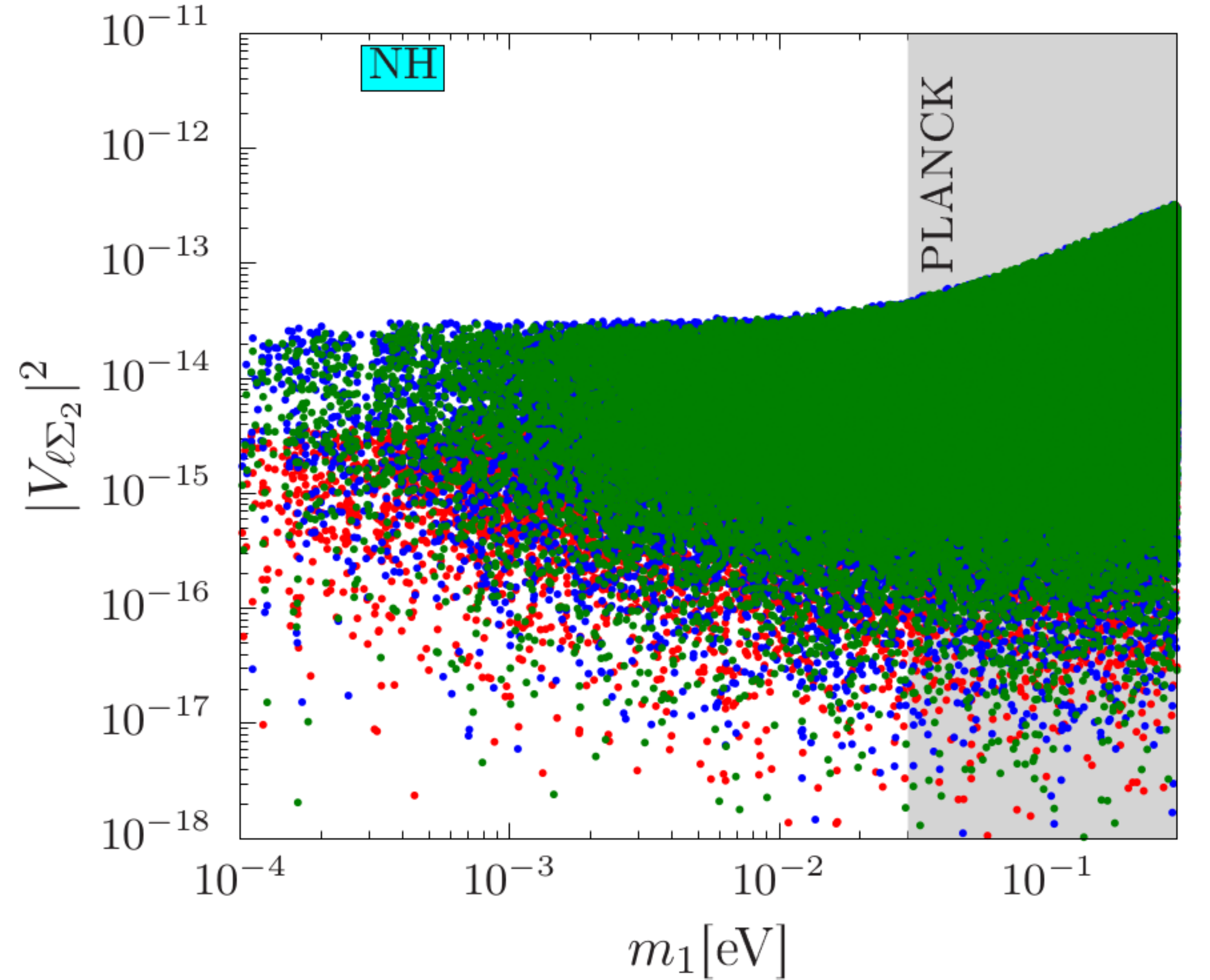}
\includegraphics[width=0.31\textwidth]{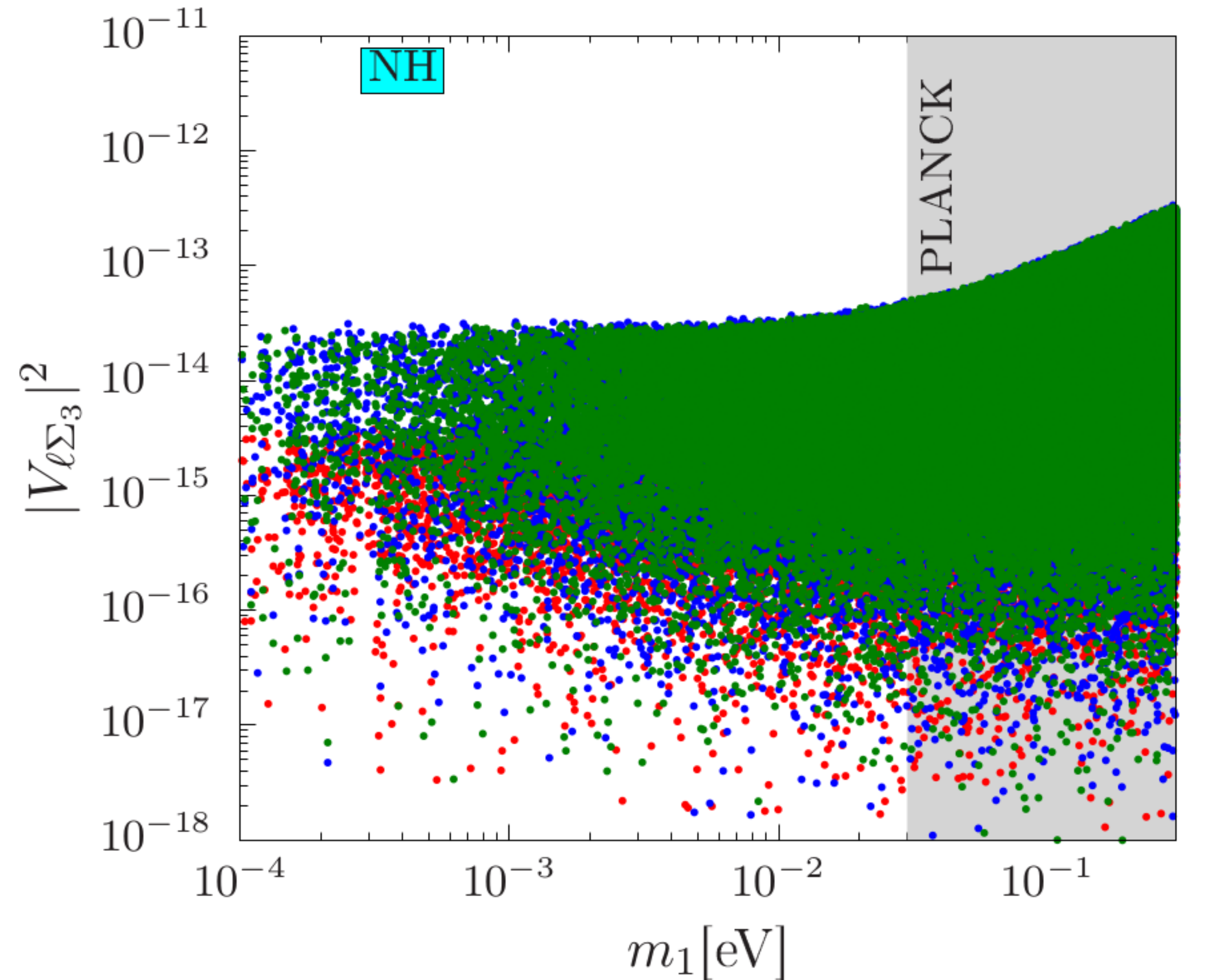}\\
\includegraphics[width=0.31\textwidth]{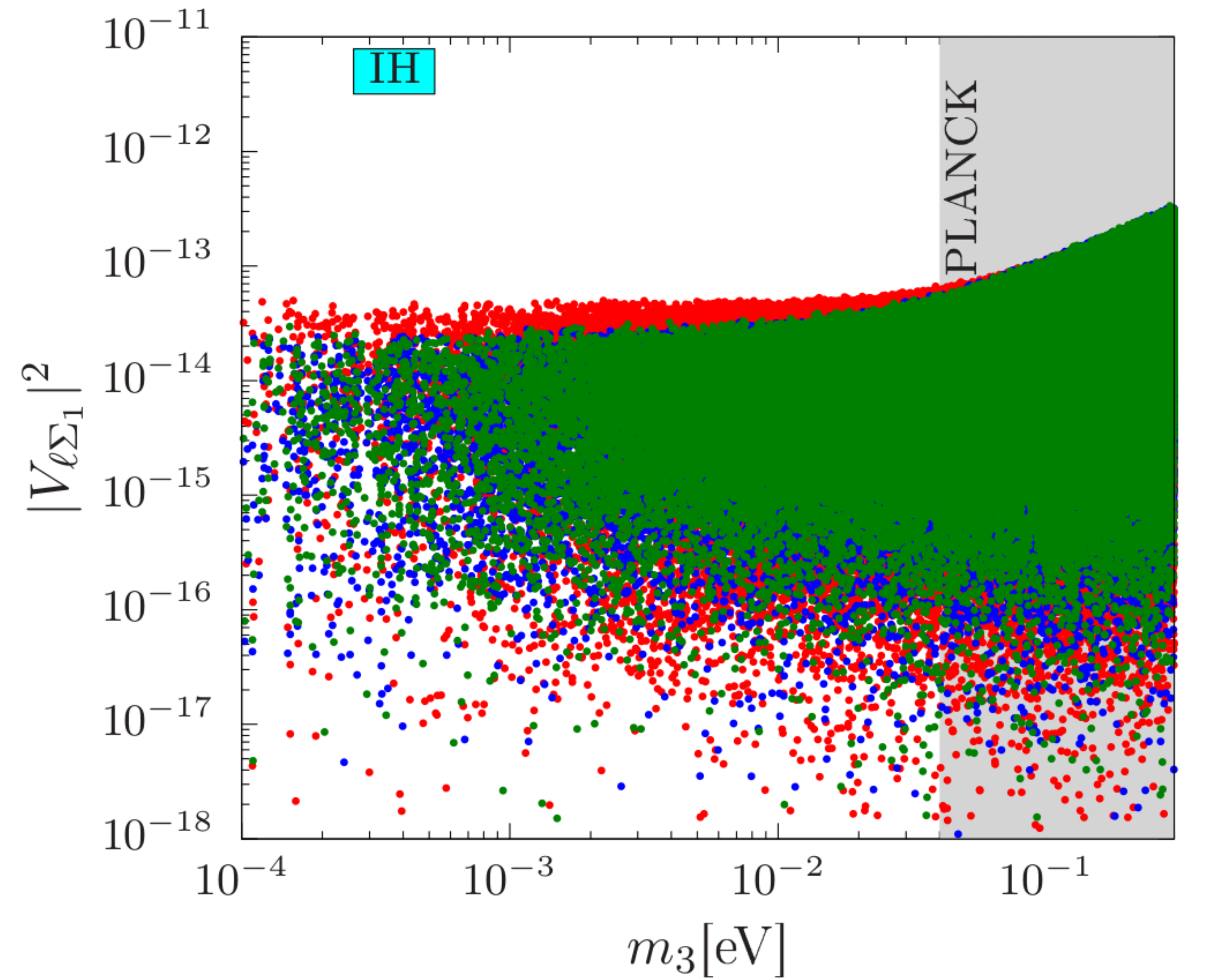}
\includegraphics[width=0.31\textwidth]{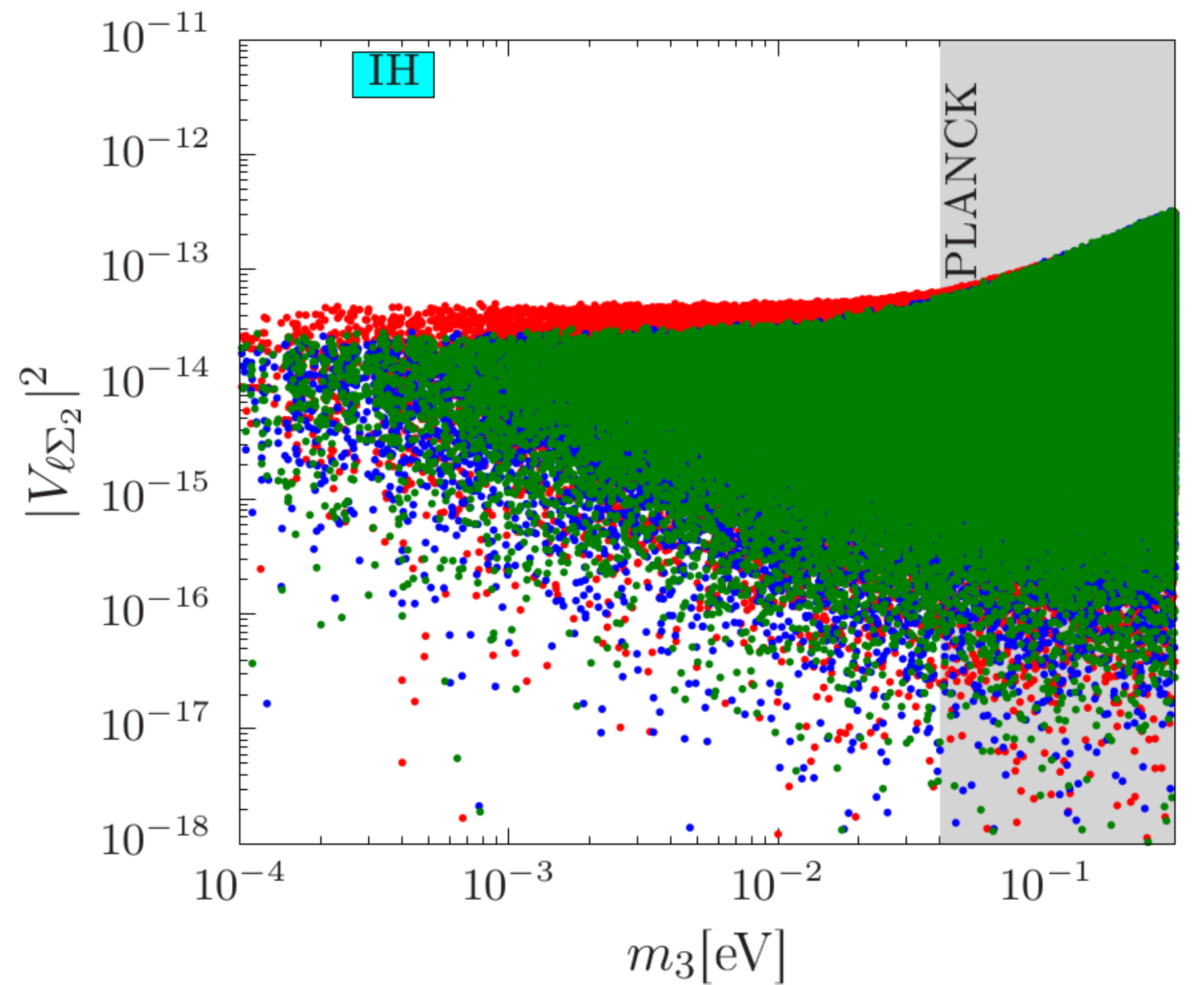}
\includegraphics[width=0.31\textwidth]{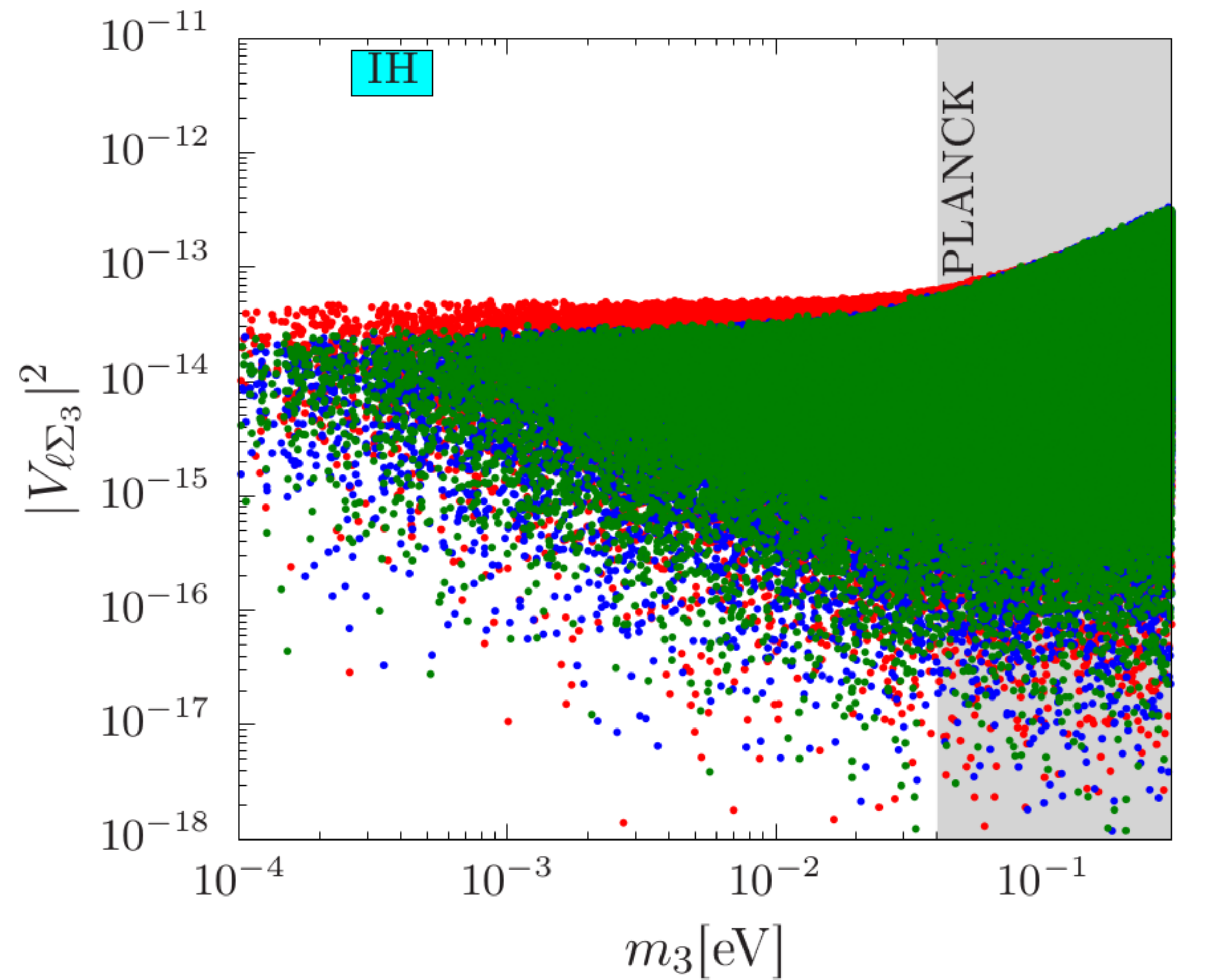}
\caption{Bounds on the individual mixing $|V_{\ell \Sigma_i}|^2$ for a real orthogonal matrix $O$ as a function of $m_1(m_3)$ in the NH (IH) case. 
In this case we fix the triplet flavor $(\Sigma_{i})$ and find the bounds on its mixing with electron (red), muon (blue) and tau (green). We have considered $M=1$ TeV. 
The shaded region in gray is ruled out by the PLANCK data. }
\label{Mix3}
\end{figure}

In the similar fashion we study the case (ii) where $O$ is a real orthogonal matrix of the form Eq.~\ref{Omatrix} where the elements are the real parameters.
The corresponding parameter regions for individual mixing angles are shown in Fig.~\ref{Mix3}.
We notice that the mixing $|V_{\ell\Sigma_1}|^2$($|V_{\ell\Sigma_3}|^2$) does not go to zero even with the limit $m_1\to 0$($m_3\to 0$) for the NH (IH) case.
In the NH case the upper limit of the $|V_{e\Sigma_1}|^2$ parameter space stays below the other two mixings for the three generations of the triplets.
This is opposite in the IH case.  

In the following we write down the individual mixings between the $\Sigma_1$ and the three generations of the leptons with $O$ as real orthogonal matrix: 
\bea
|V_{e\Sigma_1}|^2 &=&\frac{1}{M_{\Sigma}} |c_{12} c_{13} \cos[y] \cos[z] \sqrt{m_1}- \nonumber \\
&&e^{i(\delta_{\text{CP}}-\rho_{2})} s_{13} \sin[y]\sqrt{m_3} -c_{13} e^{-i\rho_1} s_{12} \cos[y] \sin[z] \sqrt{m_2}|^2 \nonumber \\
|V_{\mu\Sigma_1}|^2 &=&\frac{1}{M_{\Sigma}} |-e^{i(\delta_{\text{CP}}-\rho_2)} s_{13} \cos[y] \sin[x] \sqrt{m_3}+ c_{12} c_{13} \sqrt{m_2} \nonumber \\
&& (-\cos[z] \sin[x] \sin[y]+ \cos[x] \sin[z])+ e^{-i\rho_1} c_{13} \sqrt{m_2} s_{12}\nonumber \\
&& (\cos[x] \cos[z]+\sin[x]\sin[y]\sin[z])|^2 \nonumber \\
|V_{\tau\Sigma_1}|^2 &=&\frac{1}{M_{\Sigma}} |e^{i(\delta_{\text{CP}}-\rho_2)} \sqrt{m_3} s_{13} \cos[x] \cos[y]+ c_{12} c_{13} \sqrt{m_1}\nonumber \\
&& (\cos[x] \cos[z] \sin[y]+\sin[x]\sin[z])+c_{12} e^{-i\rho_1}\nonumber \\
&&\sqrt{m_2} s_{12} (\cos[z] \sin[x]-\cos[x]\sin[y]\sin[z])|^2.
\label{m4}
\eea
We write down the individual mixings between the $\Sigma_2$ and the three generations of the leptons with $O$ as real orthogonal matrix: 
\bea
|V_{e\Sigma_2}|^2 &=&\frac{1}{M_{\Sigma}} |\sqrt{m_1} (-c_{23} s_{12}-c_{12} e^{i\delta_{\text{CP}}} s_{13} s_{23}) \cos[y] \cos[z]-c_{13} e^{-i\rho_2} \sqrt{m_3} s_{23} \sin[y]\nonumber \\
&&-e^{-i\rho_1} \sqrt{m_2} (c_{12} c_{23}-e^{i\delta_{CP}} s_{12} s_{13} s_{23}) \cos[y] \sin[z]|^2 \nonumber \\
|V_{\mu\Sigma_2}|^2 &=&\frac{1}{M_{\Sigma}} |-c_{13} e^{-i \rho_2} \sqrt{m_3} s_{23} \cos[y] \sin[x]+\sqrt{m_1} (-c_{23} s_{12}-c_{12} e^{i \delta_{\text{CP}}} s_{13} s_{23}) \nonumber \\
&&(-\cos[z] \sin[x] \sin[y]+ \cos[x] \sin[z])+ e^{-i\rho_1} \sqrt{m_2} (c_{12} c_{23} -e^{i\delta_{\text{CP}}} s_{12} s_{13} s_{23}) \nonumber \\
&&(\cos[x] \cos[z]+\sin[x] \sin[y] \sin[z])|^2 \nonumber \\
|V_{\tau \Sigma_{2}}|^2&=& \frac{1}{M_{\Sigma}} |c_{13} e^{-i\rho_2} \sqrt{m_3} s_{23} \cos[x] \cos[y]+ \sqrt{m_1} (-c_{23} s_{12}-c_{12} e^{i\delta_{\text{CP}}} s_{13} s_{23}) \nonumber \\
&& (\cos[x] \cos[z] \sin[y]+\sin[x] \sin[z])+ e^{-i\rho_2} \sqrt{m_2} (c_{12} c_{23} -e^{i\delta_{\text{CP}}} s_{12} s_{13} s_{23})\nonumber \\
&& (\cos[z] \sin[x] -\cos[x] \sin[y] \sin[z])|^2
\label{m5}
\eea
and we write down the individual mixings between the $\Sigma_3$ and the three generations of the leptons with $O$ as real orthogonal matrix: 
\bea
|V_{e\Sigma_3}|^2 &=&\frac{1}{M_{\Sigma}} |\sqrt{m_1} (-c_{12} c_{23} e^{i\delta_{\text{CP}}} s_{13}+ s_{12} s_{23}) \cos[y] \cos[z] -c_{13} c_{23} e^{-i\rho_2} \sqrt{m_3} \sin[y] \nonumber \\
&-& e^{i\rho_1} \sqrt{m_2} (-c_{12} e^{i \delta_{\text{CP}}} s_{12} s_{13}-c_{12} s_{23}) \cos[y] \sin[z])|^2 \nonumber \\
|V_{\mu \Sigma_3}|^2 &=&\frac{1}{M_{\Sigma}} |-c_{13} c_{23} e^{-i\rho_2} \cos[y] \sin[x]+ \sqrt{m_1} (-c_{12} c_{23} e^{i\delta_{\text{CP}}} s_{13}+ s_{12} s_{23})(-\cos[z] \sin[x] \sin[y]\nonumber \\
&&+\cos[x] \sin[z]) + e^{-i\rho_1} \sqrt{m_2} (-c_{23} e^{i\delta_{\text{CP}}} s_{12} s_{13}- c_{12} s_{23}) (\cos[x] \cos[z] +\sin[x] \sin[y] \sin[z])|^2\nonumber \\
|V_{\tau \Sigma_3}|^2 &=&\frac{1}{M_{\Sigma}}|c_{13} c_{23} e^{-i\rho_2} \sqrt{m_3} \cos[x] \cos[y]+ \sqrt{m_1} (-c_{12} c_{23} e^{i\delta_{\text{CP}}} s_{13}+ s_{12} s_{23}) \nonumber \\
&&( \cos[x] \cos[z] \sin[y]+ \sin[x] \sin[z])+ e^{-i\rho_1} \sqrt{m_2} (-c_{23} e^{i\delta_{\text{CP}}} s_{12} s_{13}-c_{12} s_{23}) \nonumber \\
&& (\cos[z] \sin[x]-\cos[x] \sin[y] \sin[z])|^2.
\label{m6}
\eea
We calculate again $\Sigma_i |V_{\ell \Sigma_i}|^2$ from the Eqs.~\ref{m4}-\ref{m6} and find that this is same with the case of $O=\bf 1_{3\times 3}$. 

We notice that unlike the case of $O=\bf 1_{3\times 3}$, now $|V_{\ell \Sigma_1}|^2$($|V_{\ell\Sigma_3}|^2$) is a function of all three light neutrino mass eigenvalues $(m_1$, $m_2$ and $m_3)$. Hence in the NH and IH cases the corresponding individual mixings in Eqs.~\ref{m4} and \ref{m6} will not tend to zero even for $m_1 \to 0$ for the NH and $m_3 \to 0$ for the IH cases respectively when $O$ is a real general orthogonal matrix, which is clearly visible in Fig.~\ref{Mix3}. Same argument will be applicable for the mixings $|V_{\ell \Sigma_2}|^2$ and $|V_{\ell \Sigma_3}|^2$ in the NH case and the mixings $|V_{\ell \Sigma_1}|^2$ and $|V_{\ell \Sigma_2}|^2$ in the IH case. This behavior can be observed in Fig.~\ref{Mix3}.  We consider $M=1$ TeV in this analysis.

\begin{figure}[h]
\centering
\includegraphics[width=0.49\textwidth]{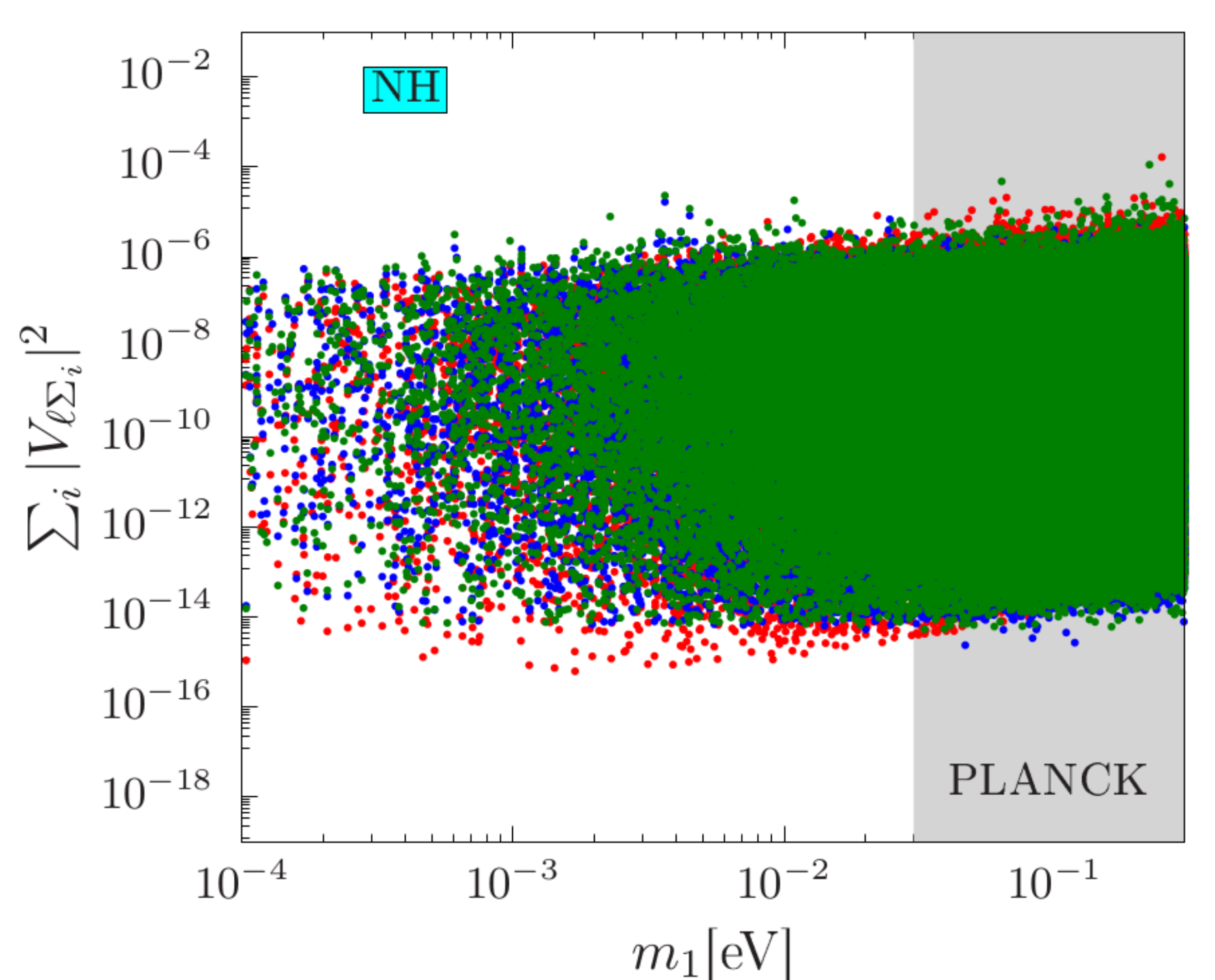}
\includegraphics[width=0.49\textwidth]{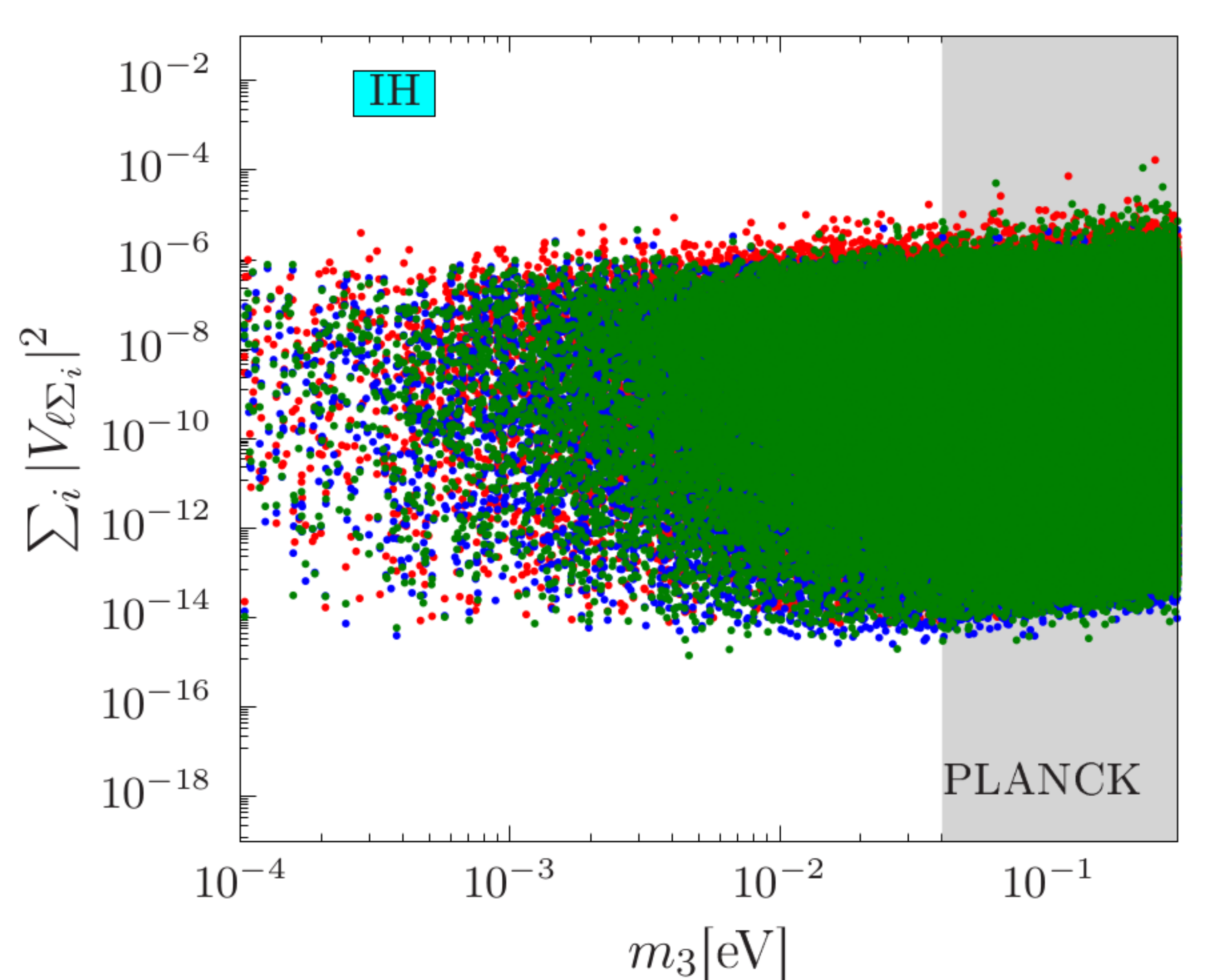}
\caption{Bounds on $\Sigma_i |V_{\ell \Sigma_i}|^2$ as a function of the $m_1 (m_3)$ NH (IH) case in the left (right) panel for fixed SM lepton flavors. The red band represents electron $(e)$, the blue band represents the muon $(\mu)$ and the green band represents the tau $(\tau)$. In this case we consider $O$ as a complex orthogonal matrix. We fix the triplet mass $M=1$ TeV. The shaded region in gray is ruled out by the PLANCK data.}
\label{Mix11a}
\end{figure}

We also study the effect of the general parametrization where $O$ is an orthogonal matrix of the form given in Eq.~\ref{Omatrix}
and using the case (iii). We take the most general form of the entries of the matrix as complex parameters. In this case running over the full set of the parameters 
we find that there is no special correlation between the mixings and $m_1 (m_3)$ for the NH (IH) case. 
The interesting fact is due to Casas-Ibarra conjecture and for the complex orthogonal matrix, the maximum possible mixing is enhanced dramatically. 
Fixing the generation of the SM charged lepton and summing over the triplet generations, we plot the bounds on the mixings satisfying the neutrino oscillation data and the PLANCK limit 
for two hierarchic masses in Fig.~\ref{Mix11a} as a function of the lightest light neutrino mass in each hierarchy.
We notice that the application of the Casas-Ibarra conjecture improves the mixing by several orders of magnitude under the applied constraints.  

\begin{figure}[]
\centering
\includegraphics[width=0.31\textwidth]{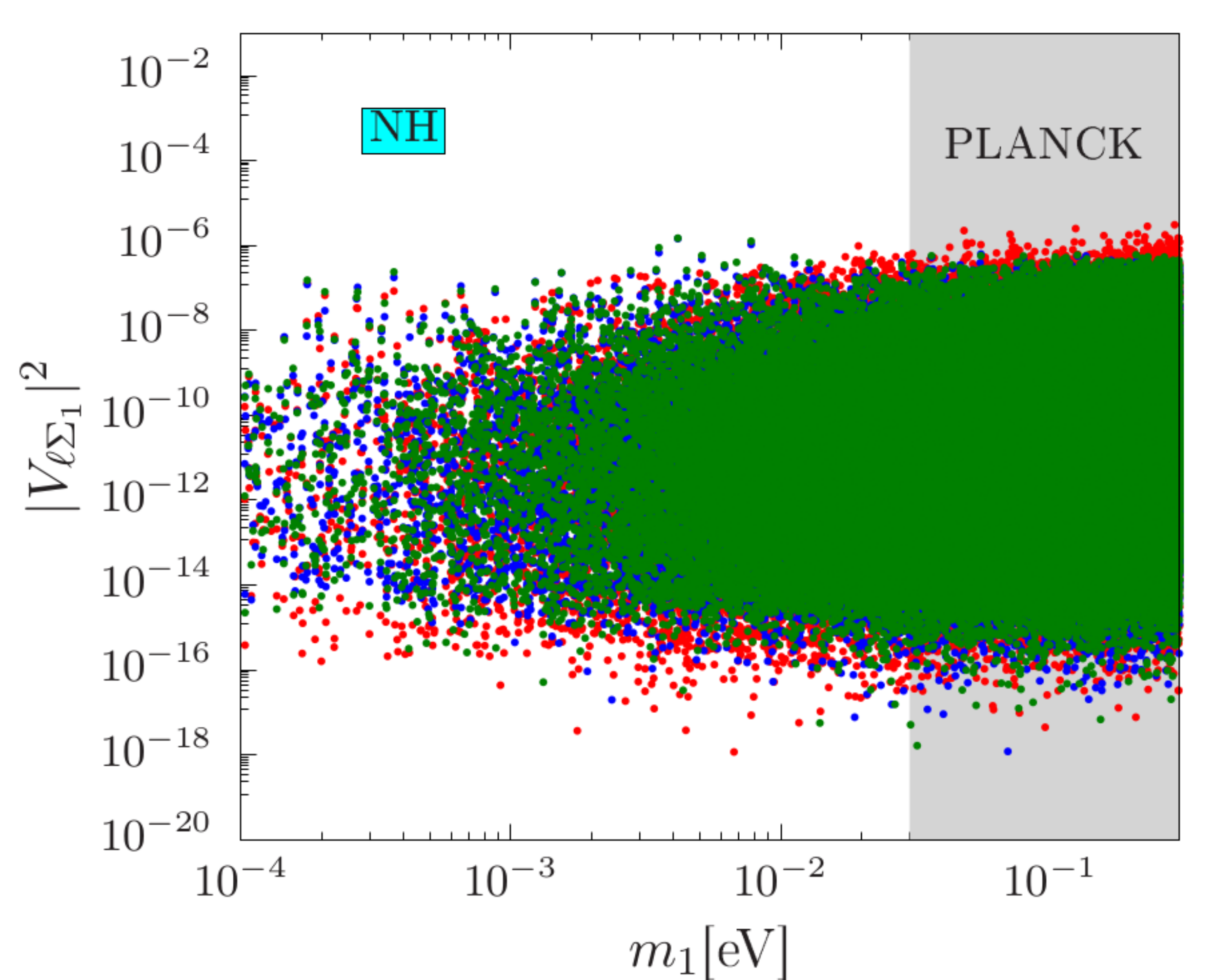}
\includegraphics[width=0.31\textwidth]{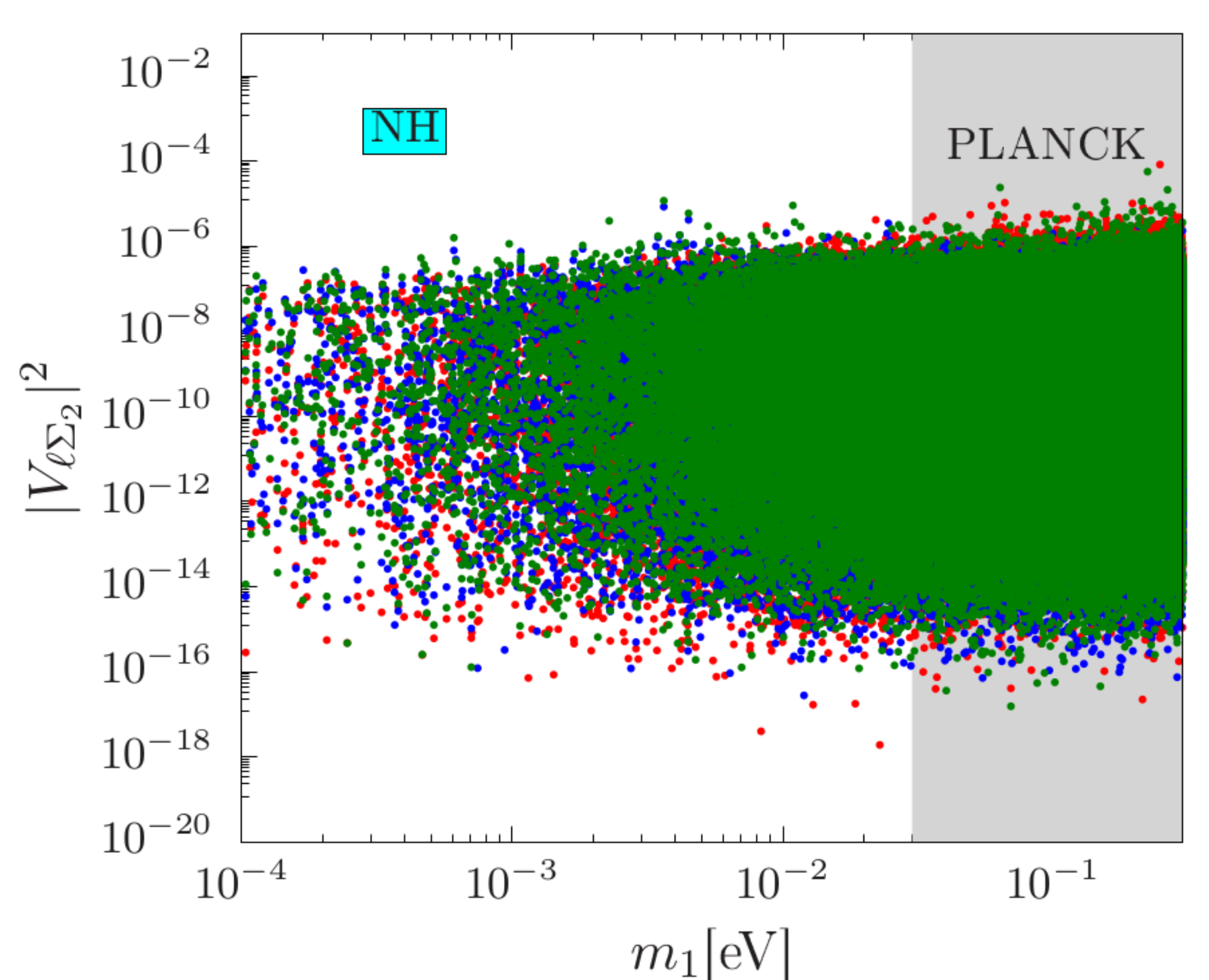}
\includegraphics[width=0.31\textwidth]{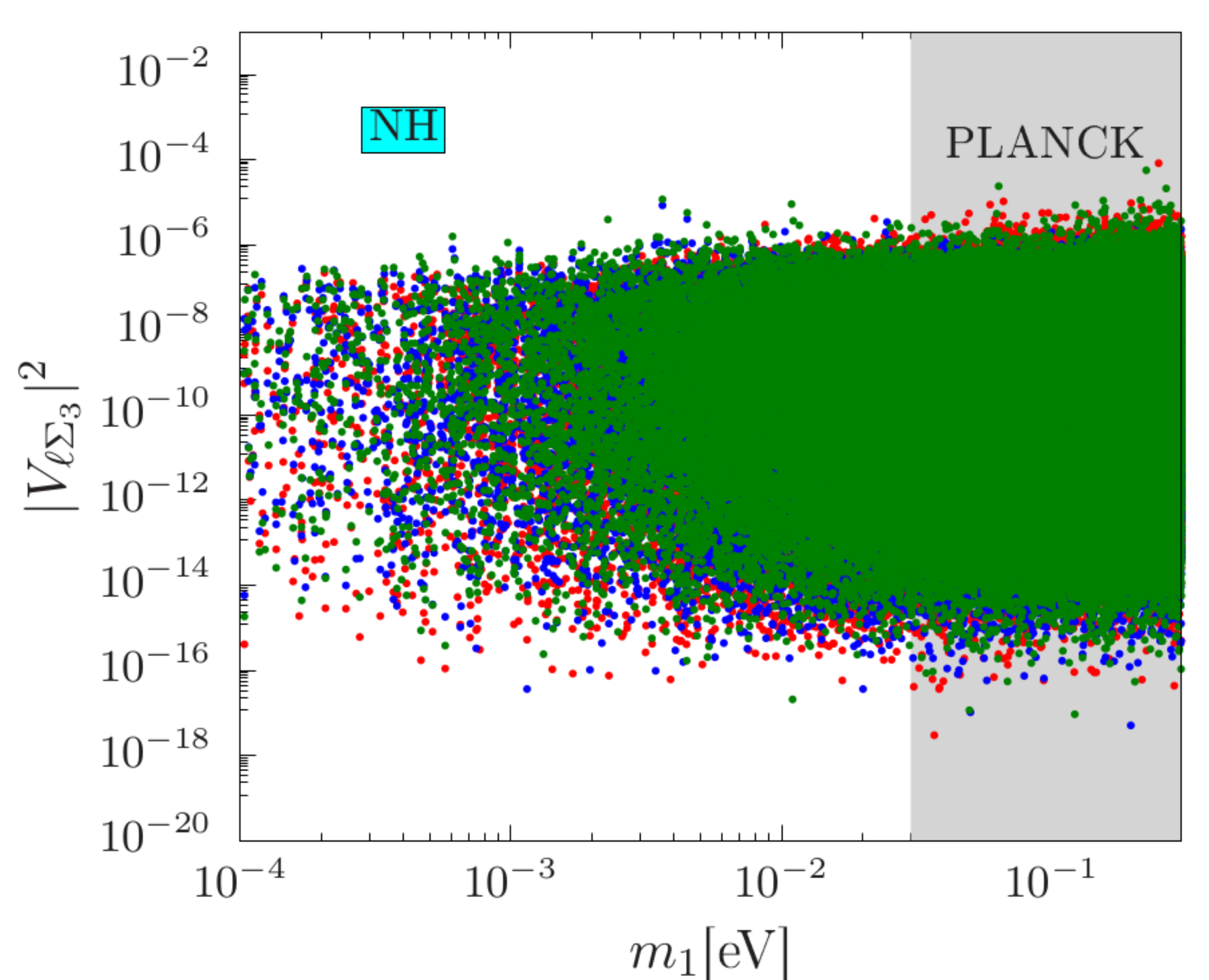}\\
\includegraphics[width=0.31\textwidth]{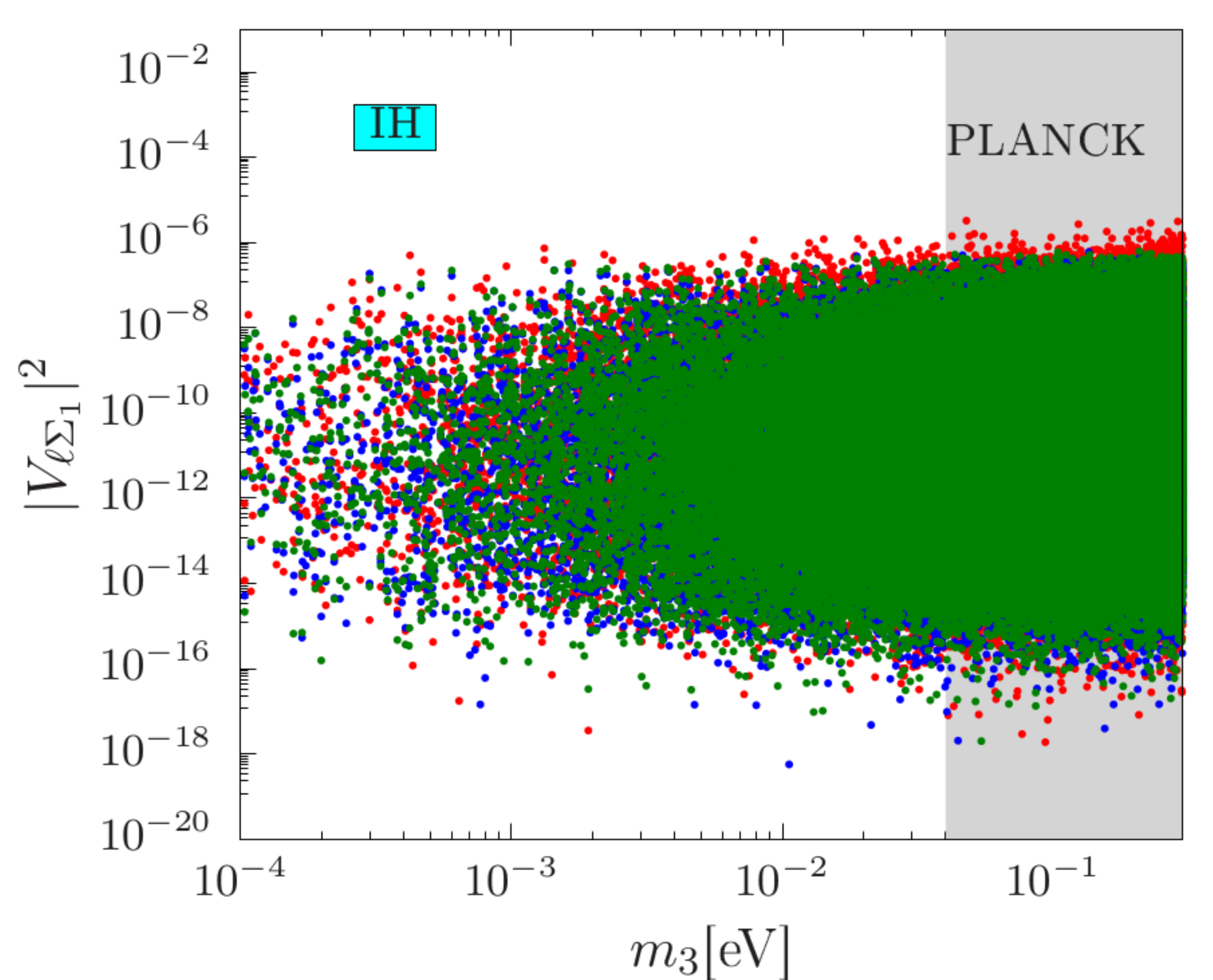}
\includegraphics[width=0.31\textwidth]{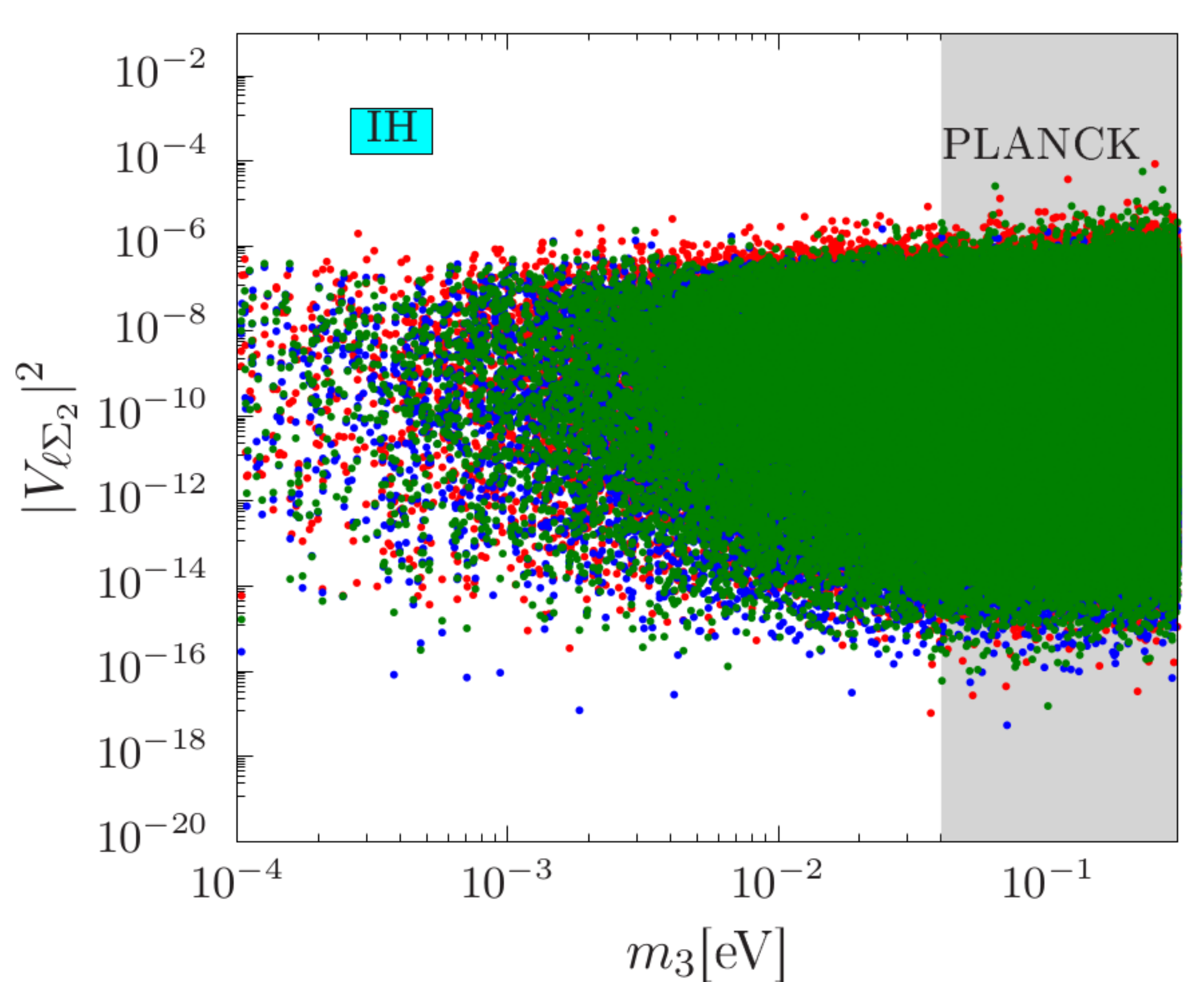}
\includegraphics[width=0.31\textwidth]{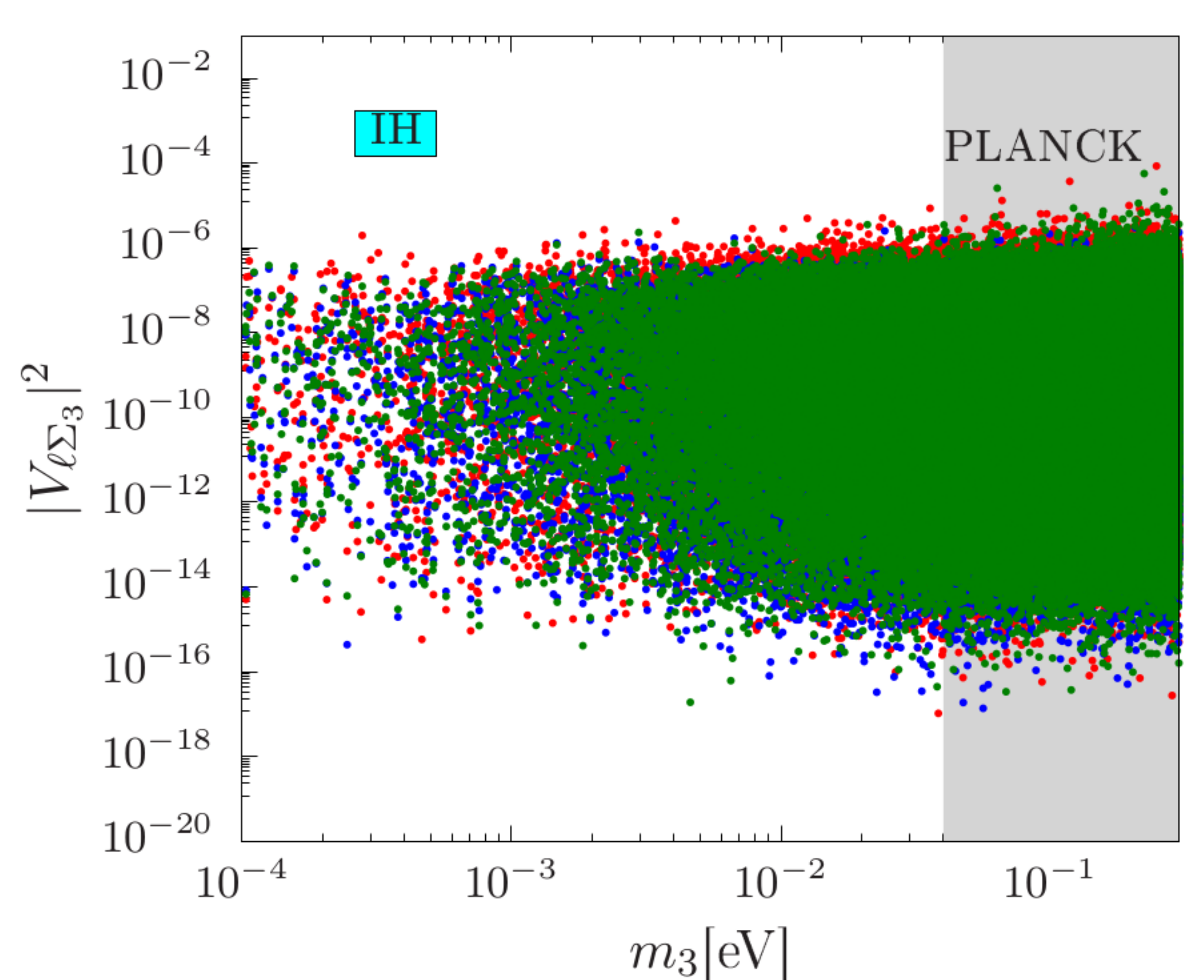}
\caption{Bounds on $|V_{\ell \Sigma_i}|^2$ as a function of the $m_1 (m_3)$ NH (IH) case in the upper (lower) panel for fixed SM lepton flavors. The red band represents electron $(e)$, the blue band represents the muon $(\mu)$ and the green band represents the tau $(\tau)$. In this case we consider $O$ as a complex orthogonal matrix. The same nature will be obtained from case when $O$ is a real orthogonal matrix. We fix the triplet mass $M=1$ TeV. The shaded region in gray is ruled out by the PLANCK data.}
\label{Mix11b}
\end{figure}

We show the individual mixing in Fig.~\ref{Mix11b} for the NH (IH) case as a function of the lightest light neutrino mass $m_1$ $(m_3)$. At this point we mention that using Eq.~\ref{m4}- \ref{m6} we can similarly calculate $\Sigma_{i} |V_{\ell \Sigma_i}|^2$ for the case (iii) using complex values of $x$, $y$ and $z$ and it will not be same as the case of (i) or (ii). $ |V_{\ell \Sigma_i}|^2$ is now a complicated function of the light neutrino mass eigenvalues $m_i$, complex parameters $x$ $y$, $z$ and the CP violating phases $\delta_{\text{cp}}$, $\rho_i$. Therefore the extreme smallness of lightest mass eigenvalues $m_1\to 0$($m_3\to 0$) will not push the mixing to zero because the rest of the two light neutrino mass eigenvalues will not allow to do that. For the individual mixing, in each panel of Fig.~\ref{Mix11b} we show the mixings between the triplet fermion and the charged lepton. 
The important fact of this scenario is the upper bounds of the light heavy mixing squared which can go up to an $\mathcal{O}(10^{-5})$, however, the lower bounds stay around $\mathcal{O}(10^{-18})$. We have showed the individual mixing for the NH (IH) case in the upper (lower) of Fig.~\ref{Mix11b}.

The individual mixings for the orthogonal matrix $O$ with the complex elements can be written as Eqs.~\ref{m4}, \ref{m5} and \ref{m6} respectively taking $x$, $y$ and $z$ as the complex quantities having real and imaginary parts.  
\section{Branching ratios of the triplet fermion for different choices of the orthogonal matrices}
\label{BR}
Using the three typical forms of the orthogonal matrix $O$, we calculate the bounds on the branching ratios of the $\Sigma^0$ and $\Sigma^\pm$ respectively. We consider a degenerate scenario for the three generations of the triplet fermions having mass at $M=1$ TeV. Summing over the three generations of $\Sigma^0_i$ and $\Sigma_i^\pm$ separately, we obtain the total branching ratios of $\Sigma^0_{\text{Tot}}$ and $\Sigma^\pm_{\text{Tot}}$ respectively for the NH and IH cases. 

\begin{figure}
\centering
\includegraphics[width=0.49\textwidth]{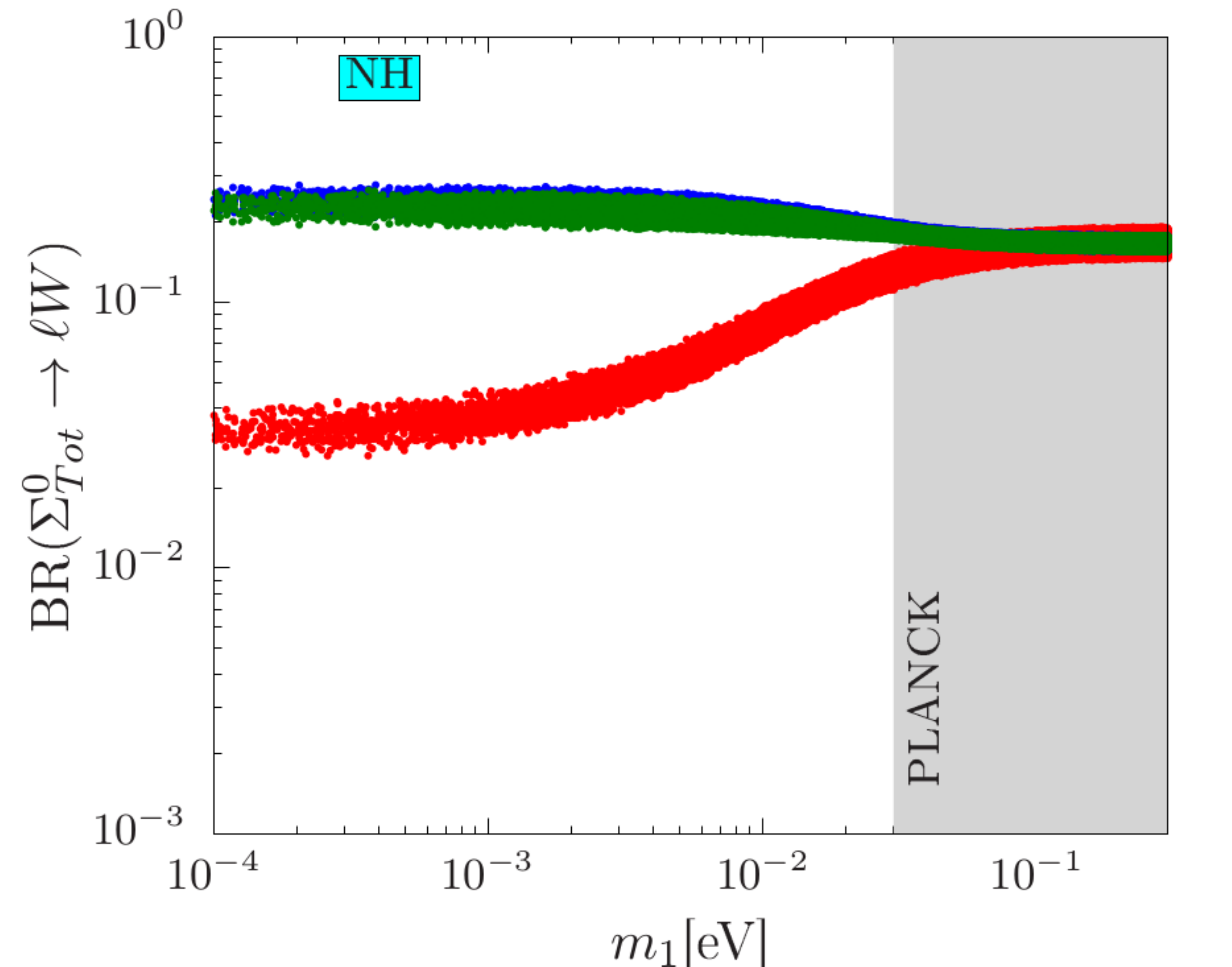}
\includegraphics[width=0.49\textwidth]{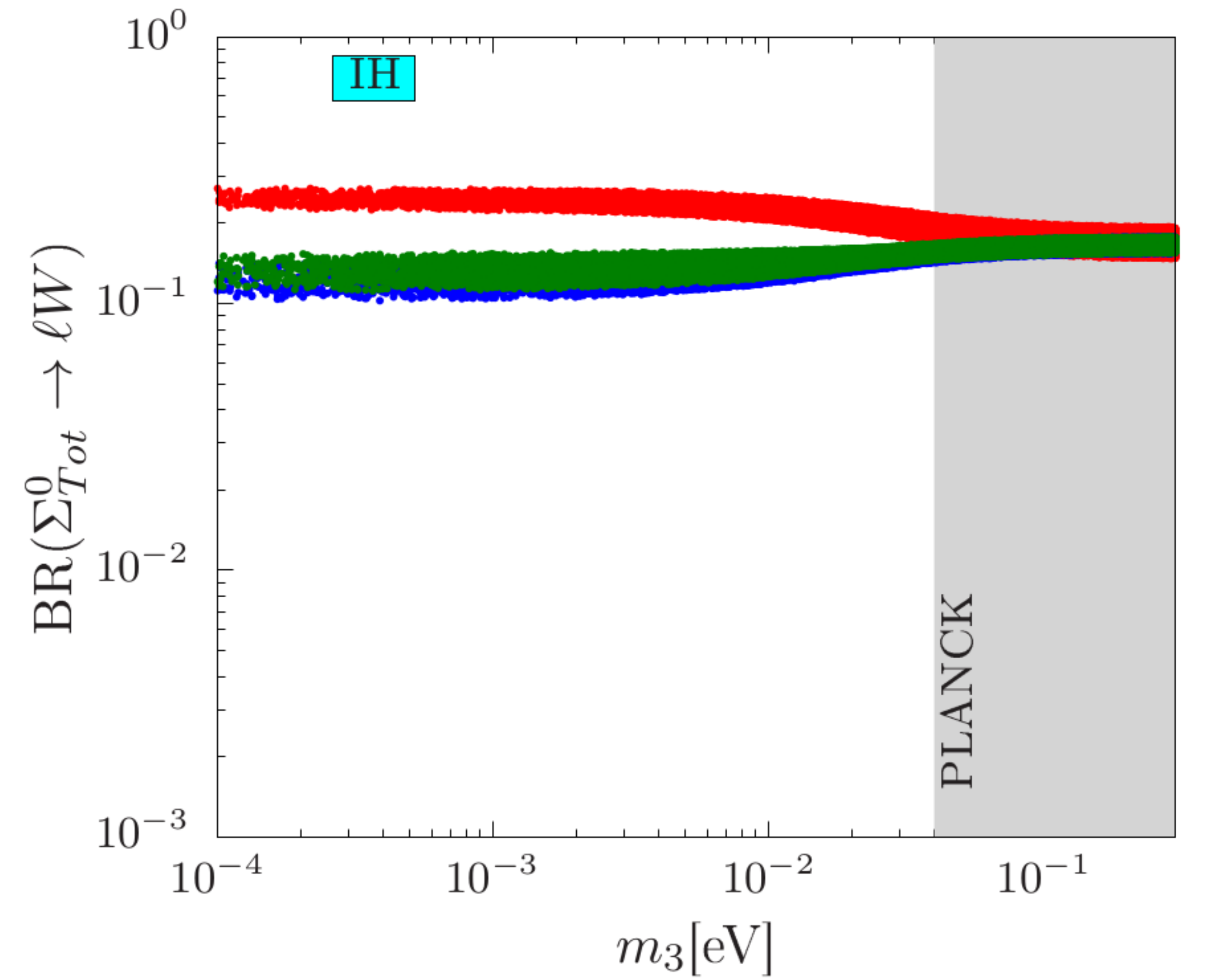}\\
\includegraphics[width=0.49\textwidth]{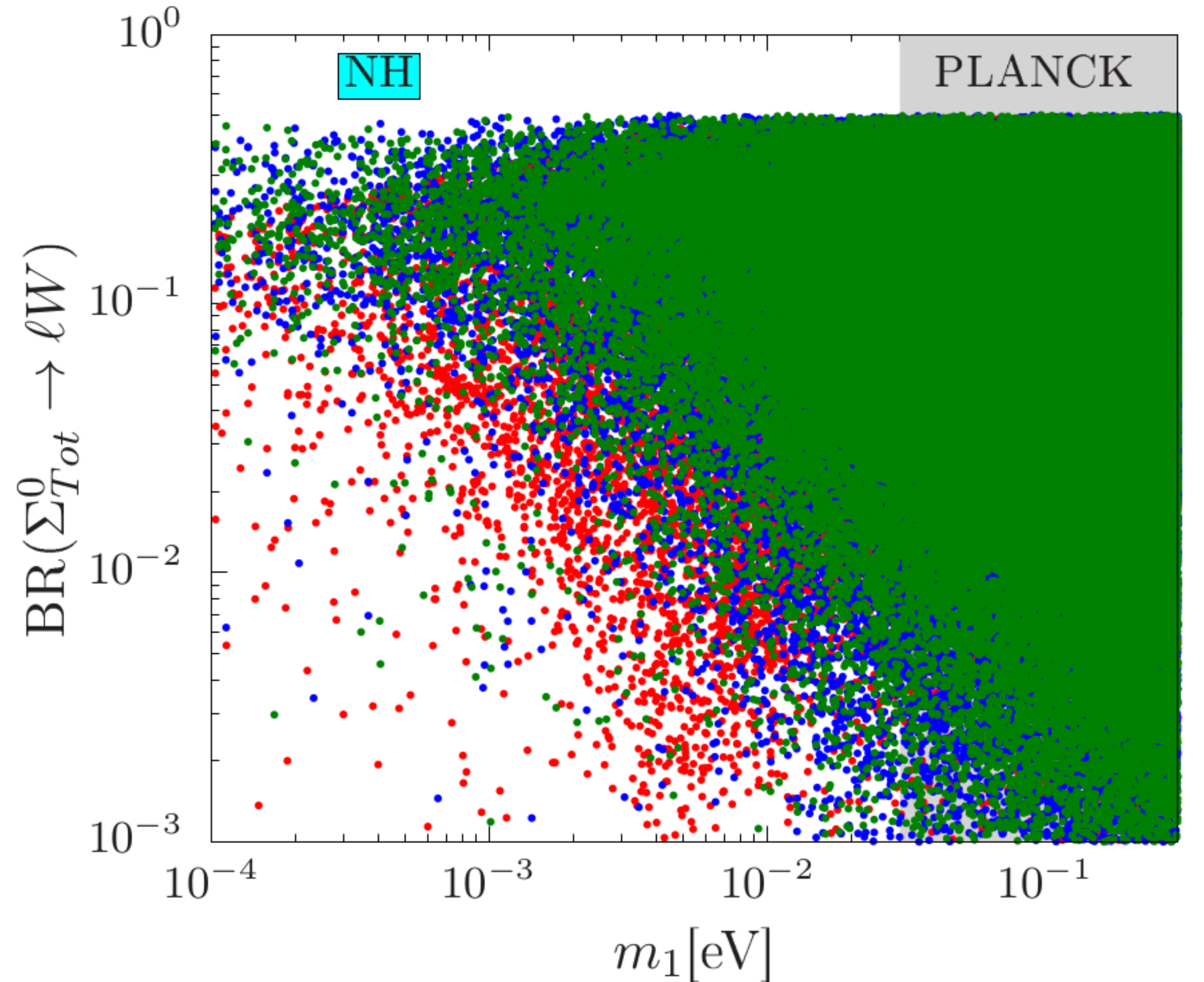}
\includegraphics[width=0.49\textwidth]{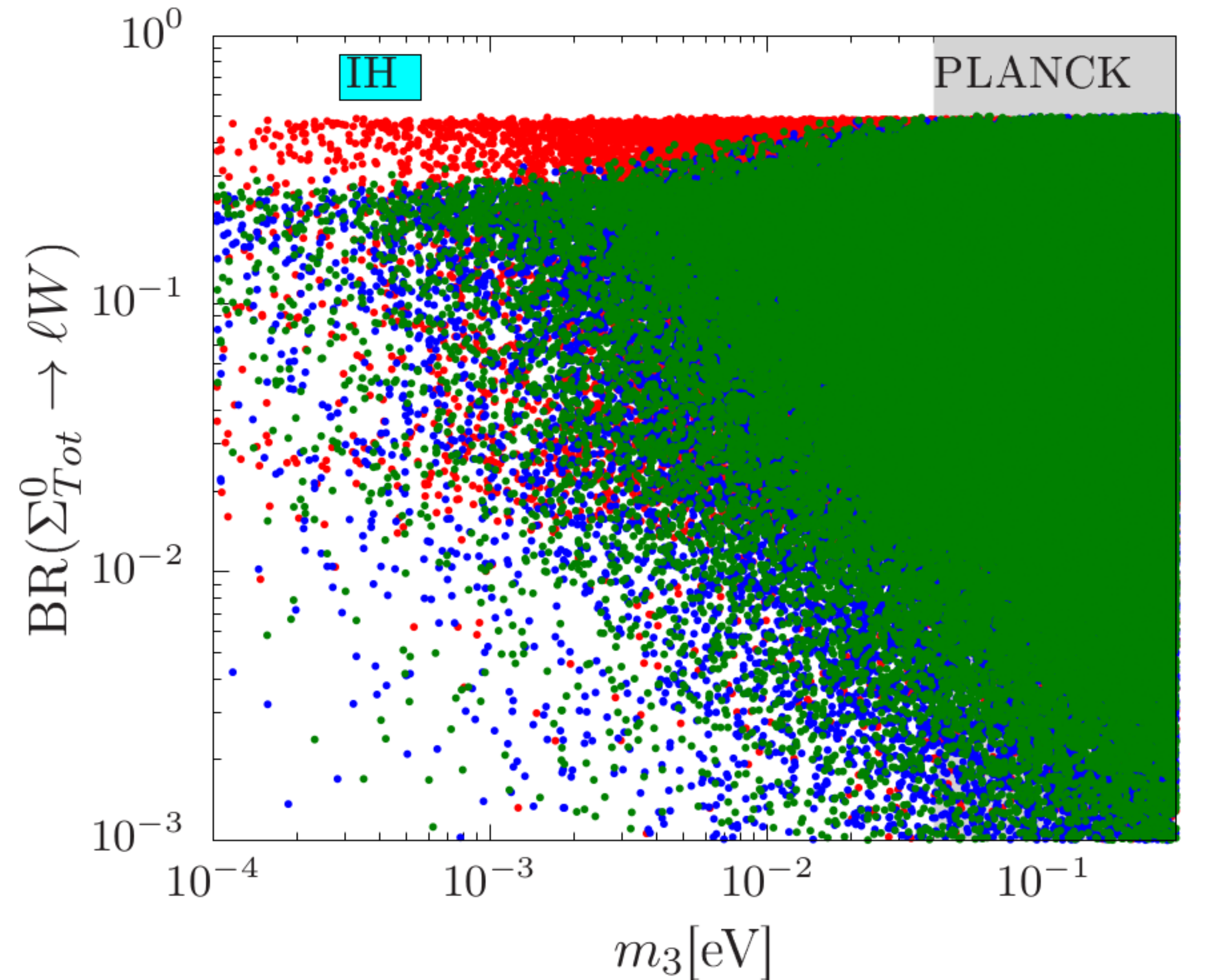}
\caption{Total branching ratio of $\Sigma^0_{\text{Tot}}$~($\sum_i\Sigma^0_i$) into the leading $\ell W$ mode as a function of the lightest neutrino mass for the NH $(m_1)$ and IH $(m_3)$ cases in the left and right panels respectively. We add three generations of $\Sigma_i^0$ to obtain $\Sigma_\text{Tot}$. The mode containing electron is represented by the red dots, that containing muon is shown by blue dots and the tau mode is represented by green dots. This result is same for the orthogonal matrix considered to be a identity matrix and a real matrix. The corresponding result  is shown upper panel. We have also considered the case where $O$ is a general orthogonal matrix. The result is shown in the lower panel. We consider $M=1$ TeV.}
\label{Mix5}
\end{figure}

We consider the leading mode of $\Sigma_i^0$ to $\ell^\pm W$ as this is the visible one. For $\Sigma_i^\pm$ we consider all the decay modes because where $\nu W$ is the leading mode, $\ell^\pm Z$ and $\ell^\pm h$ are the subdominant modes but visible with the charged leptons. BR$(\Sigma^0_{\text{Tot}} \to \ell^\pm W)$ for the NH (IH) case has been plotted in the top-left (top-right) panel of the Fig.~\ref{Mix5}. The muon (blue) and tau (green) modes are dominant over the electron (red) mode of the lepton flavors in the NH case. The IH case is opposite to the NH case. We find that the results are same for the orthogonal matrix as a identity and as a real matrix. 

\begin{figure}[]
\centering
\includegraphics[width=0.31\textwidth]{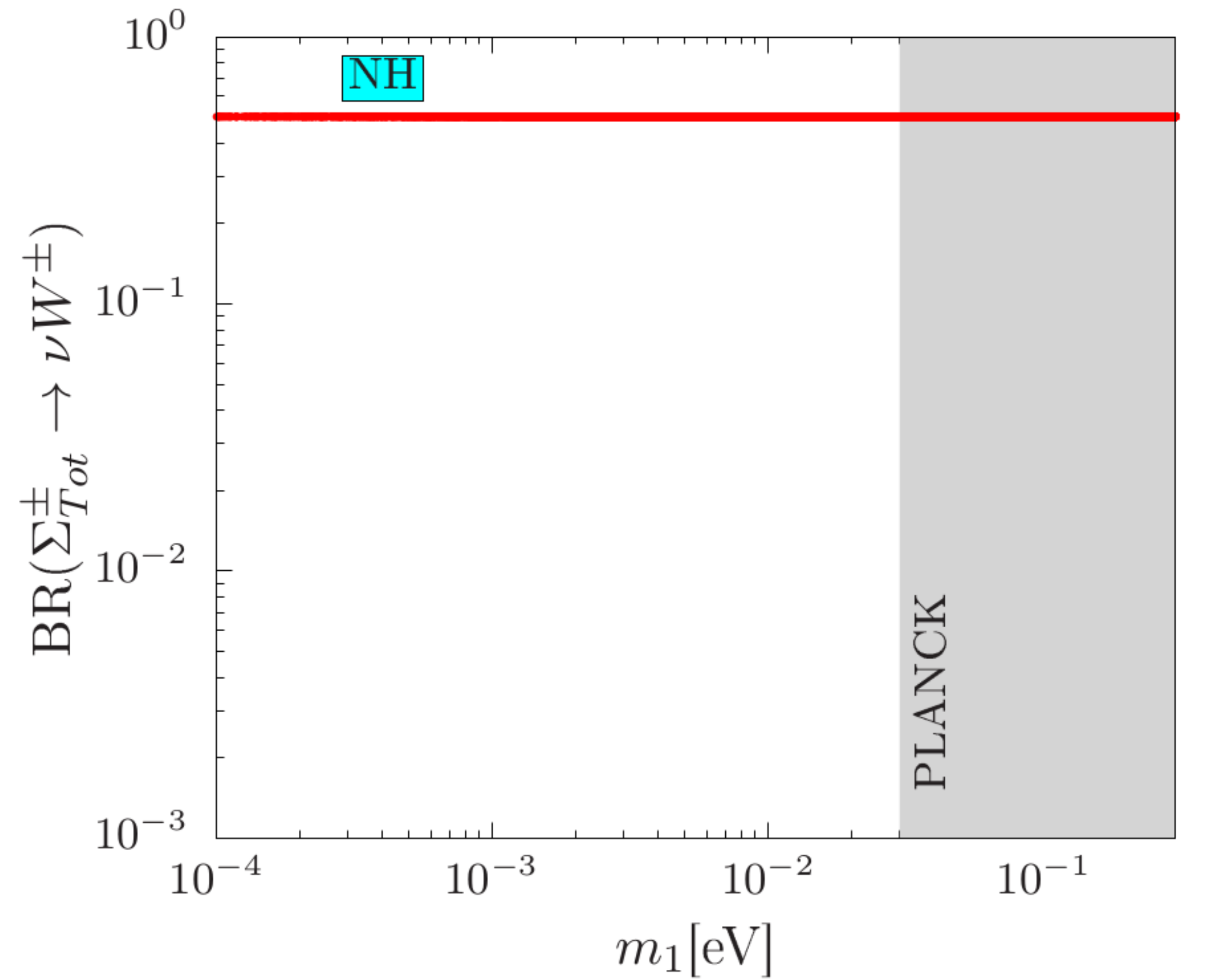}
\includegraphics[width=0.31\textwidth]{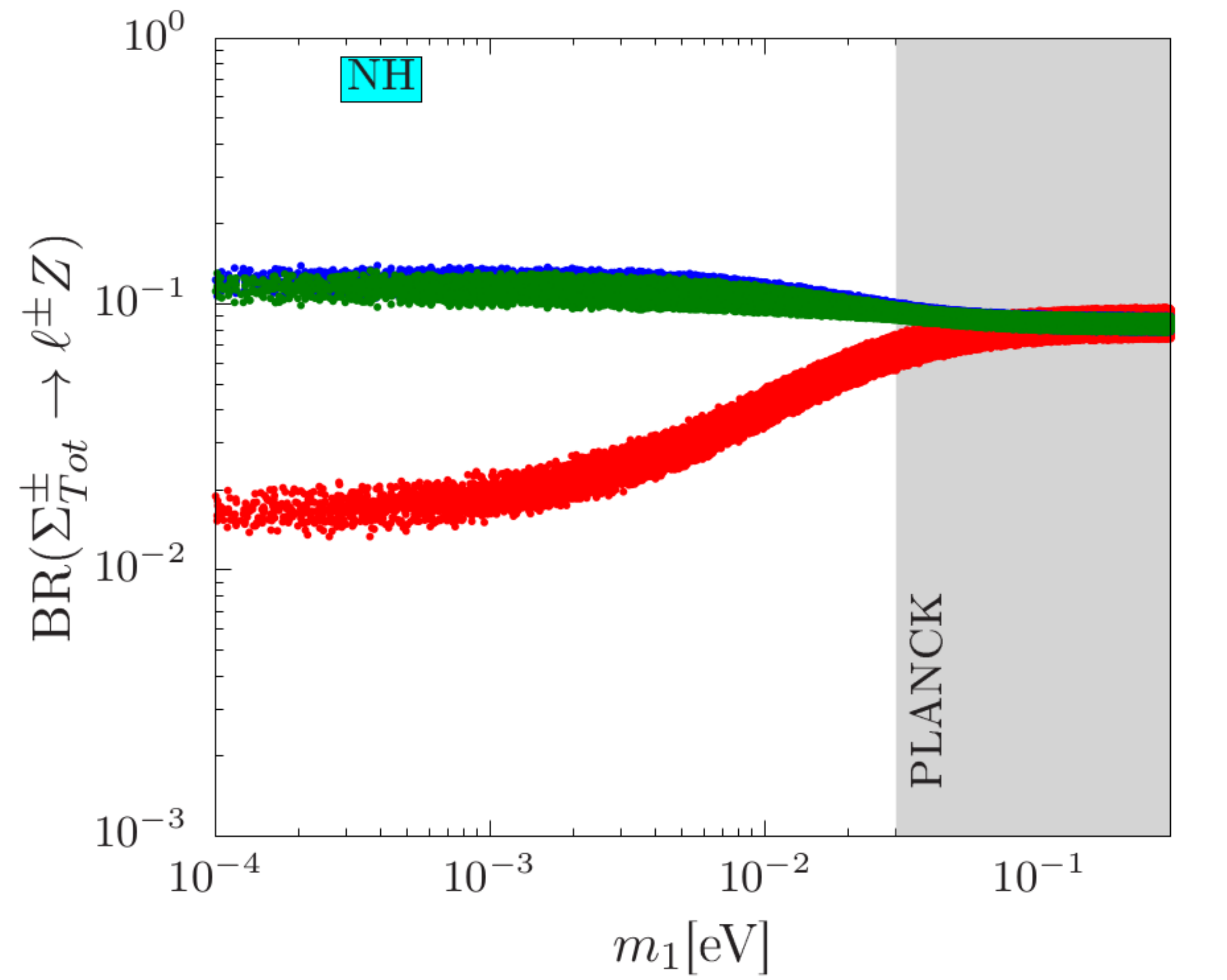}
\includegraphics[width=0.31\textwidth]{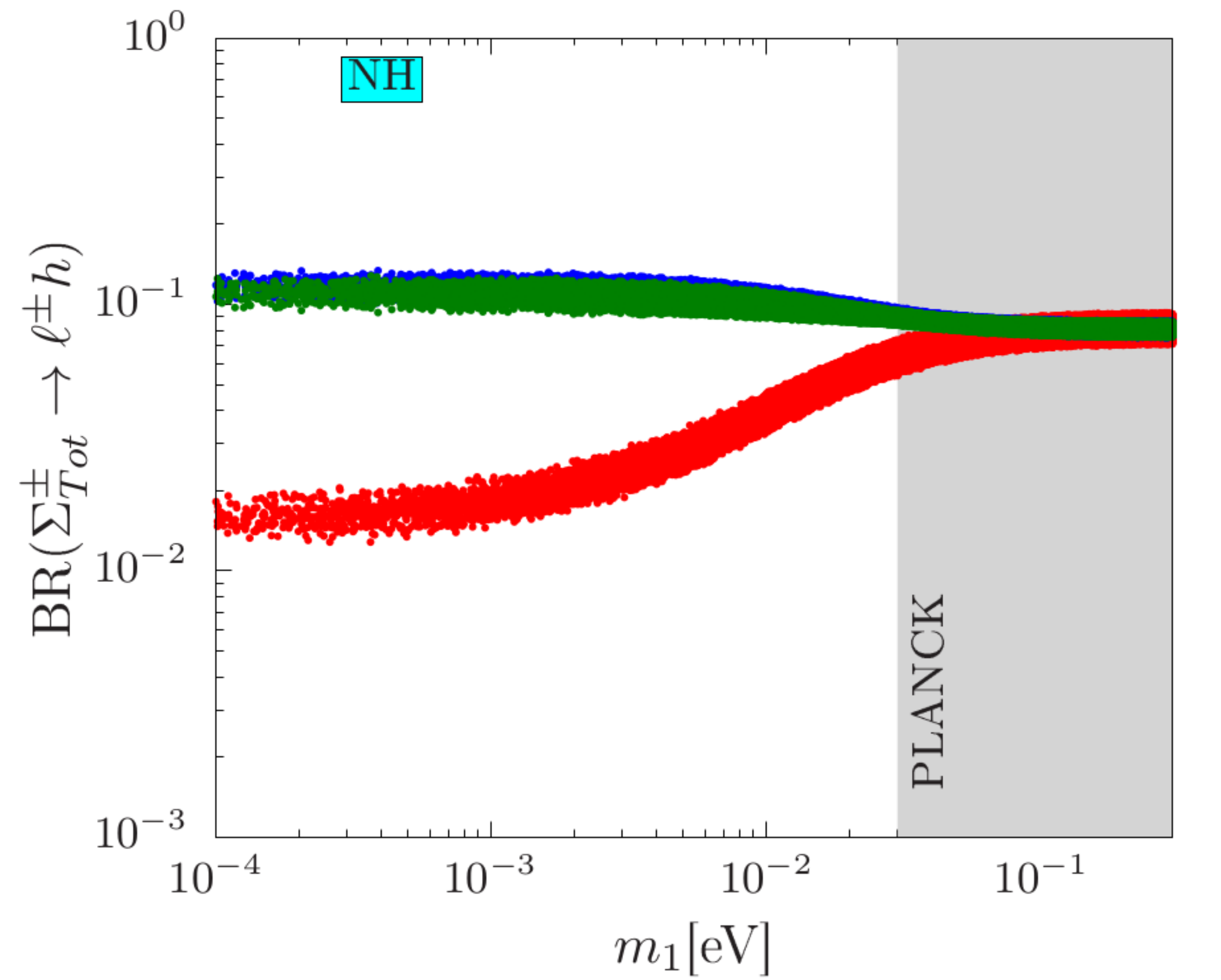}\\
\includegraphics[width=0.31\textwidth]{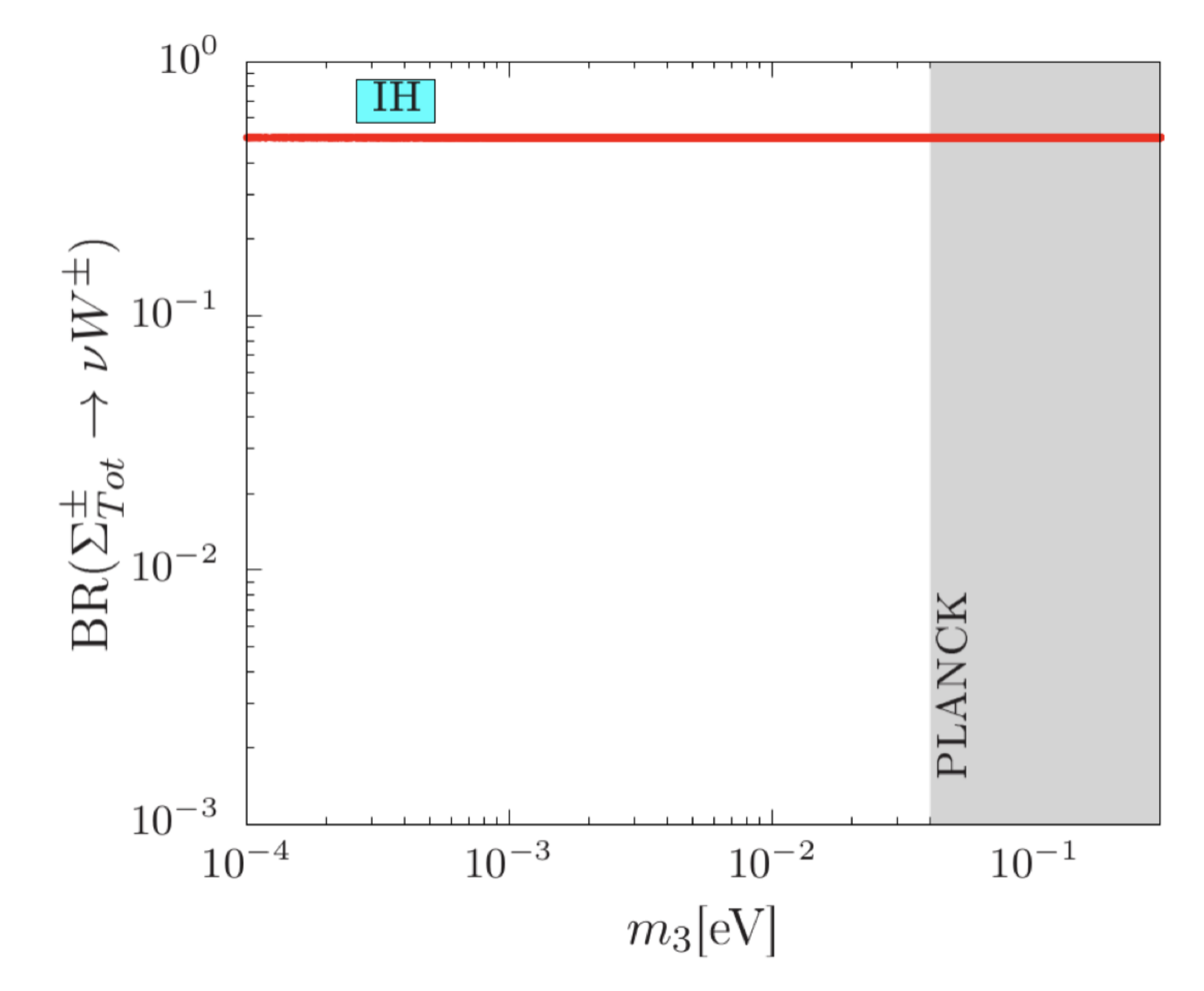}
\includegraphics[width=0.31\textwidth]{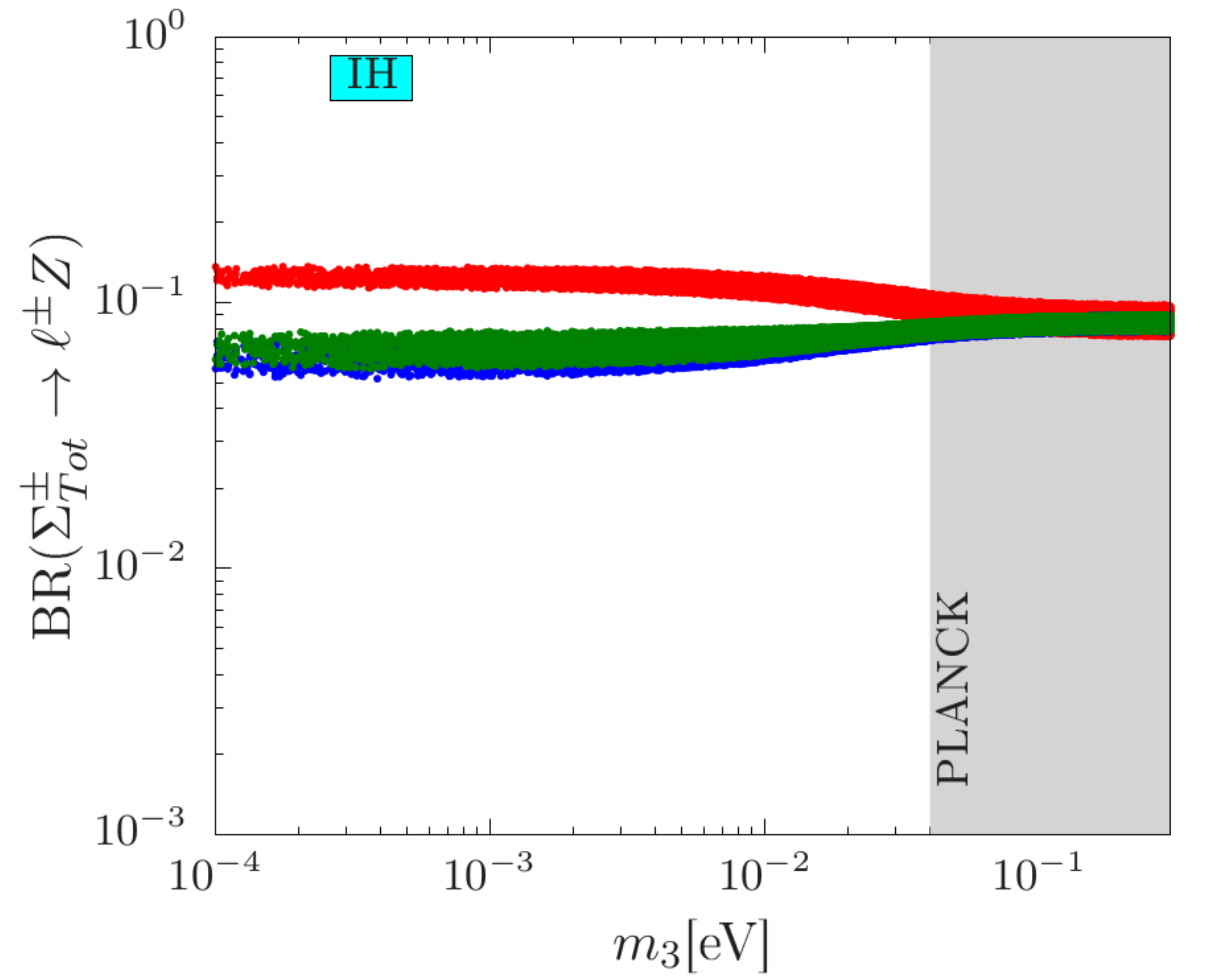}
\includegraphics[width=0.31\textwidth]{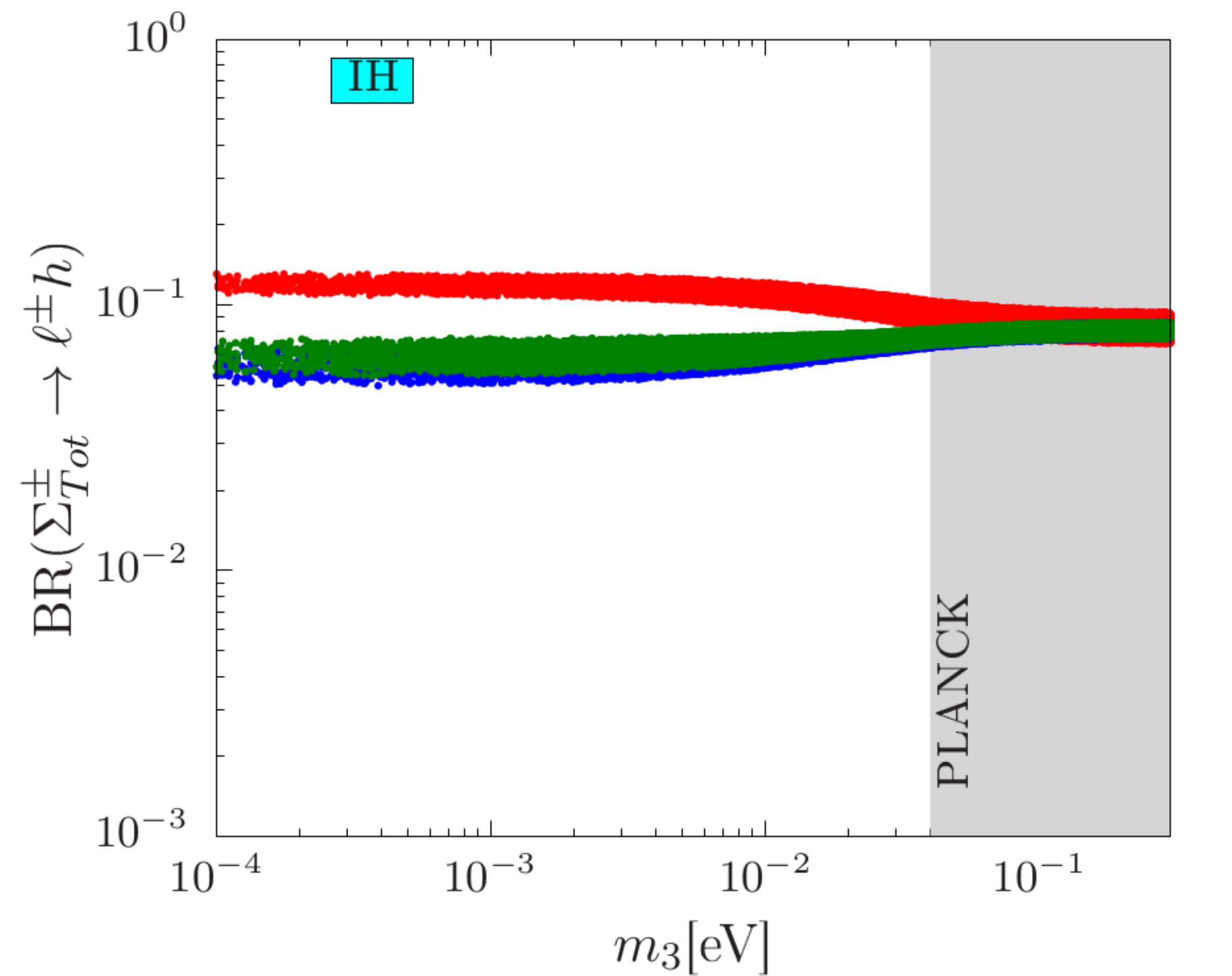}
\caption{Total branching ratio of $\Sigma^\pm$$(\text{BR}(\Sigma_{\text{Tot}}^\pm))$ into the leading $\nu W$ (first column), subleading $\ell Z$ (second column) and $\ell h$ (third column) modes with respect to the lightest neutrino mass for the NH $(m_1)$ and IH $(m_3)$ cases in the top and bottom panels respectively. We add three generations of $\Sigma_i^\pm$ to obtain $\Sigma^{\pm}_\text{Tot}$. The mode containing electron is represented by the red dots, that containing muon is shown by blue dots and the tau mode is represented by green dots. The $\nu W$ mode is indistinguishable from the point of view of the neutrinos. This result is same for the orthogonal matrix considered to be a identity matrix and a real matrix. The shaded region in gray is excluded by the PLANCK data. We consider $M=1$ TeV.}
\label{Mix6}
\end{figure}

Corresponding total branching ratios $\text{BR}(\Sigma_{\text{Tot}}^\pm)$ for the $\nu W$, $\ell^\pm Z$ and $\ell^\pm h$ modes are shown for the NH (IH) case in the top (bottom) panel of Fig.~\ref{Mix6} for the $O$ as a $3\times 3$ identity matrix. This is exactly same for the case when $O$ is a real orthogonal matrix. The $\nu W$ mode is shown in the left column. In this case we do not distinguish between the light neutrinos as they will be obtained as the missing energy. The $\ell^\pm Z $ $(\ell^\pm h)$ mode has been shown in the second (third) column of the Fig.~\ref{Mix6}. For the $\ell^\pm Z$ and $\ell^\pm h$ modes we show the electron (red), muon (blue) and tau (green) leptons separately for the NH (top row) and IH (bottom row) cases. In the NH case muon and tau regions coincide and dominate over the electron mode. In the IH case the electron mode dominates over the muon and tau modes.

\begin{figure}[]
\centering
\includegraphics[width=0.31\textwidth]{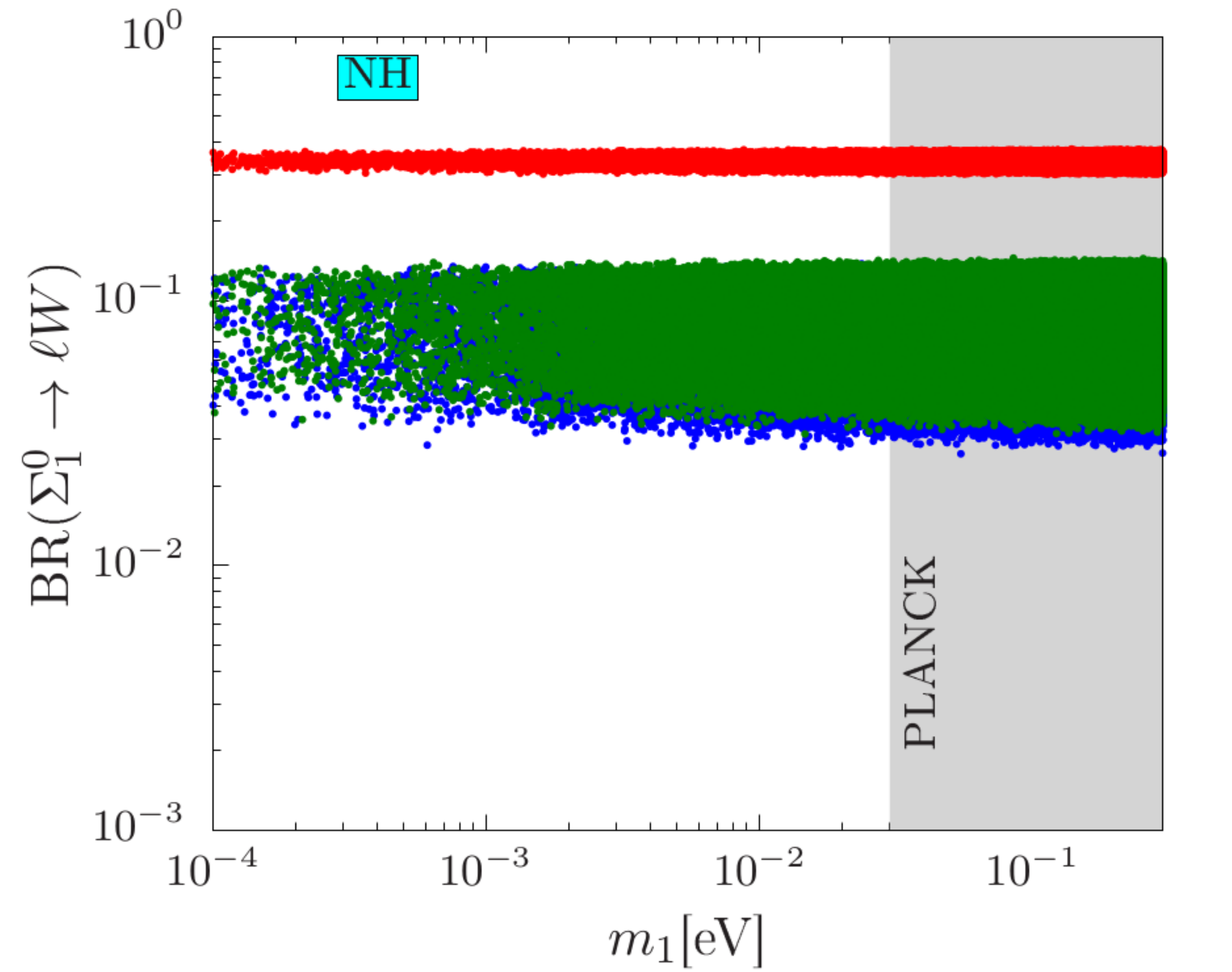}
\includegraphics[width=0.31\textwidth]{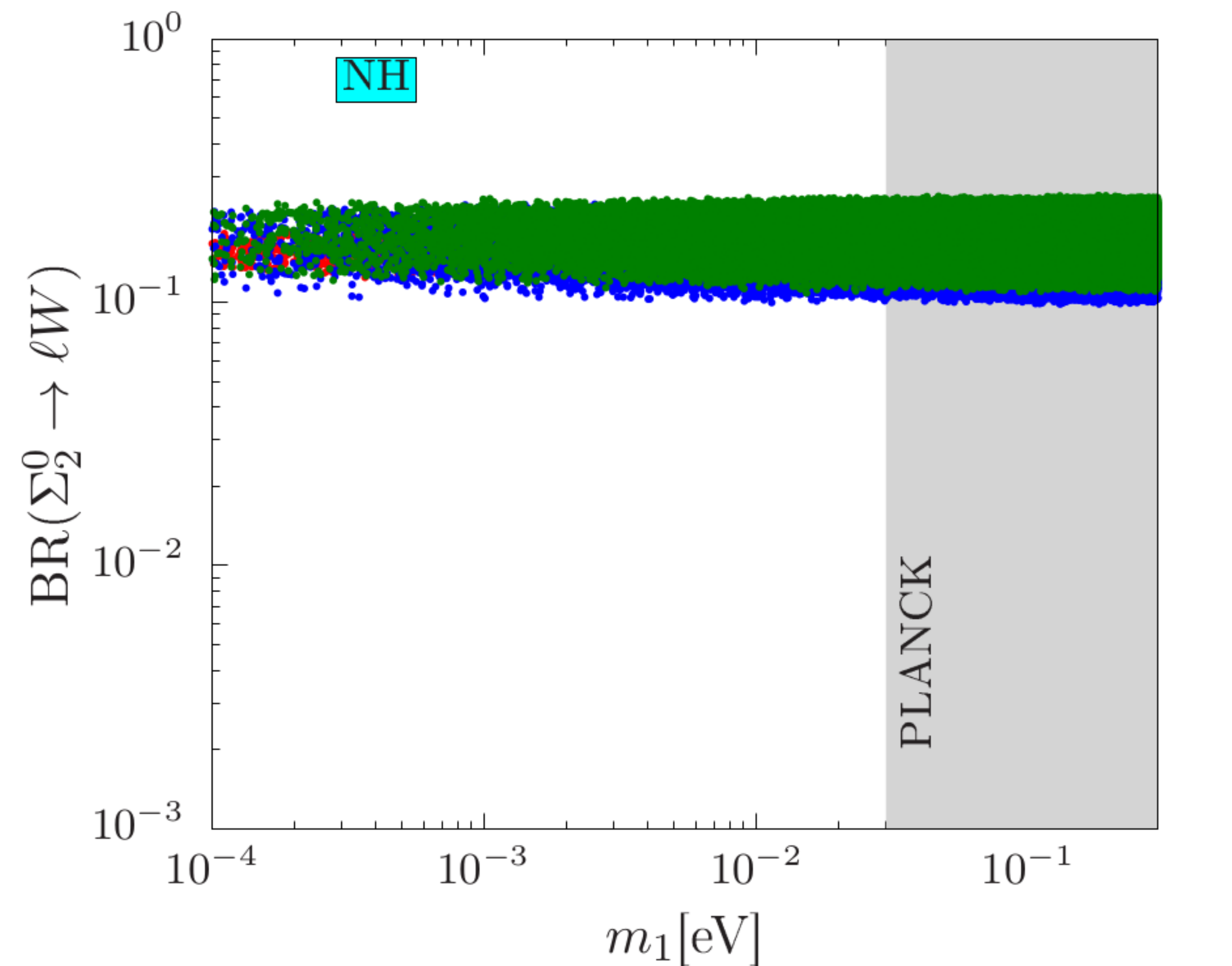}
\includegraphics[width=0.31\textwidth]{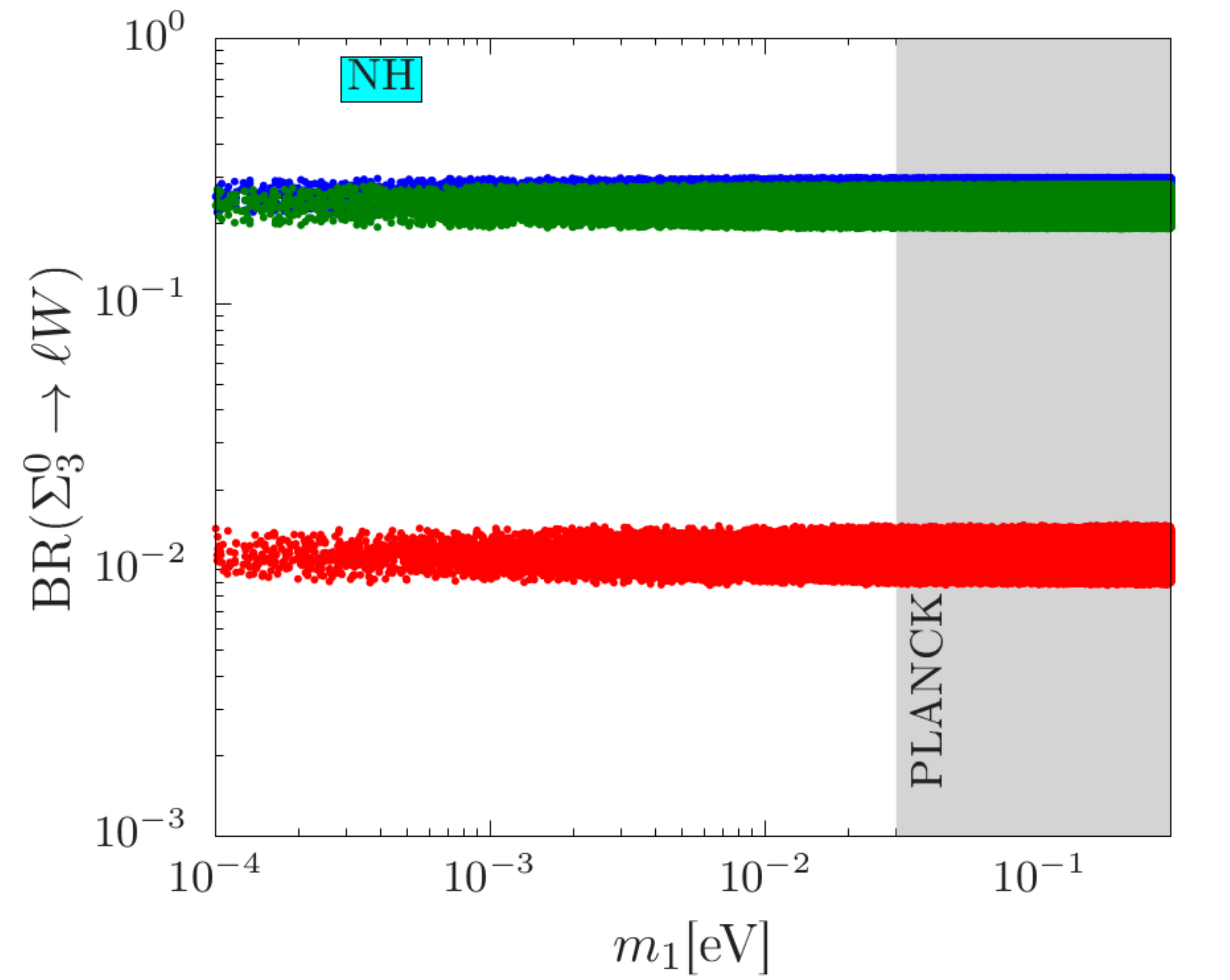}\\
\includegraphics[width=0.31\textwidth]{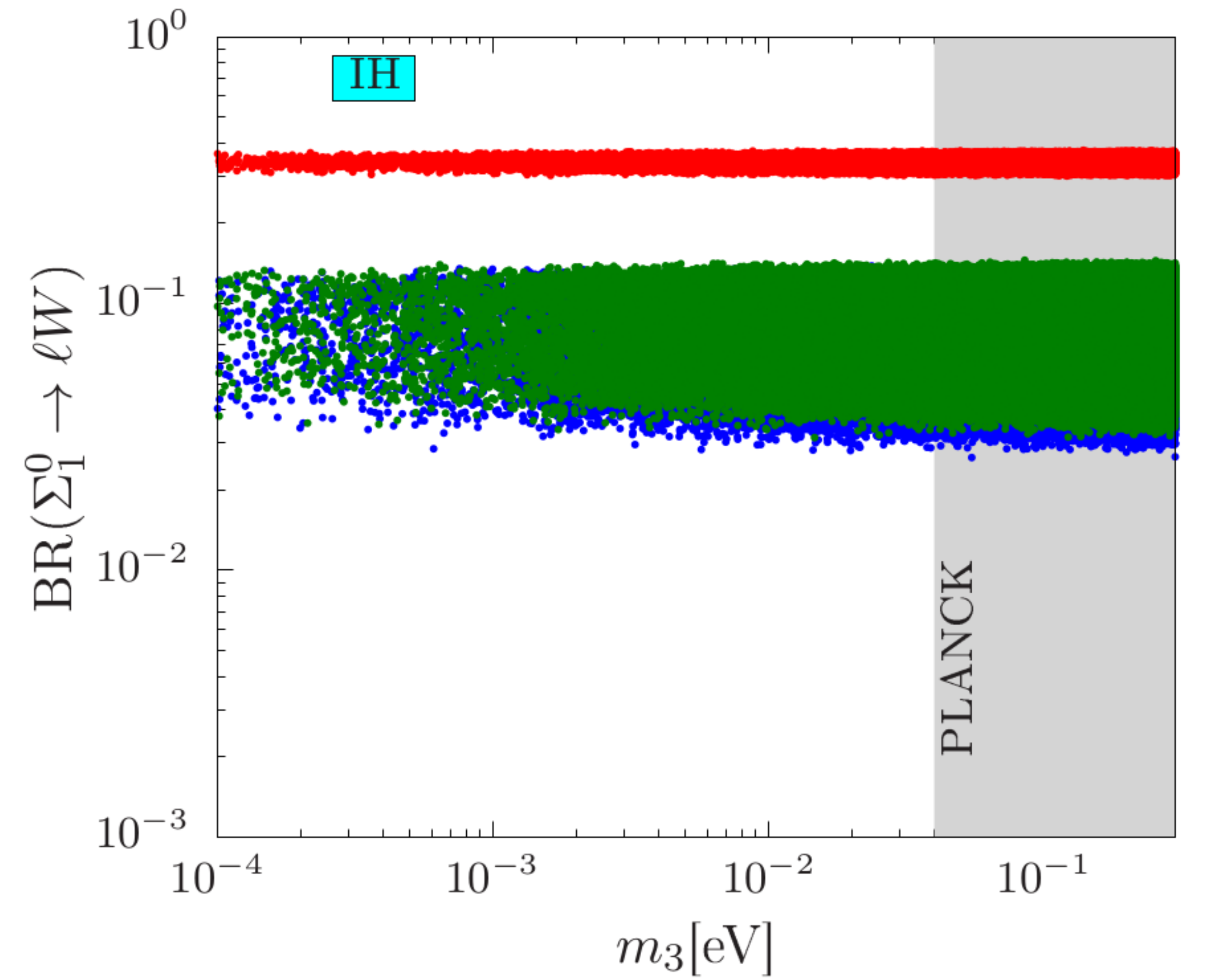}
\includegraphics[width=0.31\textwidth]{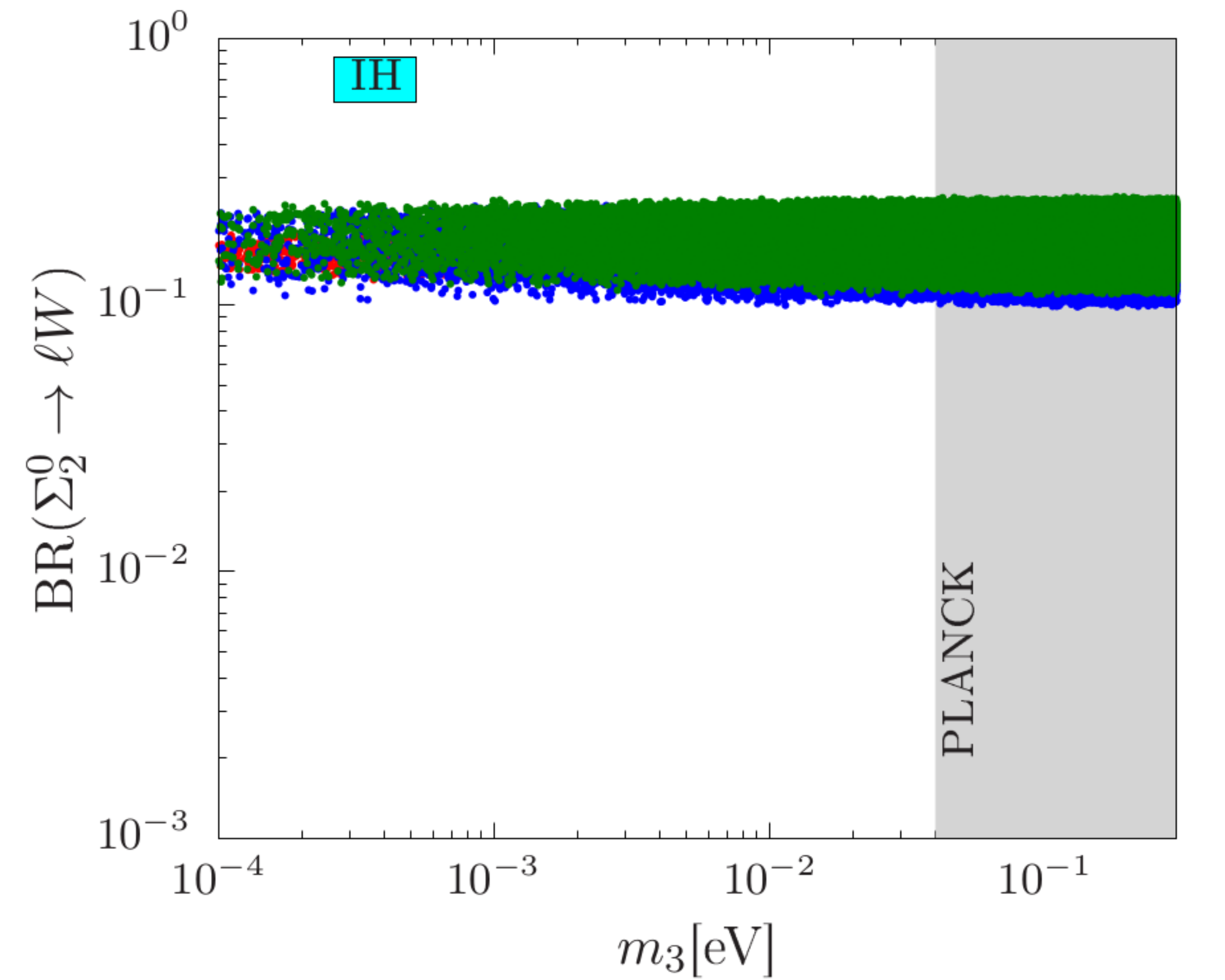}
\includegraphics[width=0.31\textwidth]{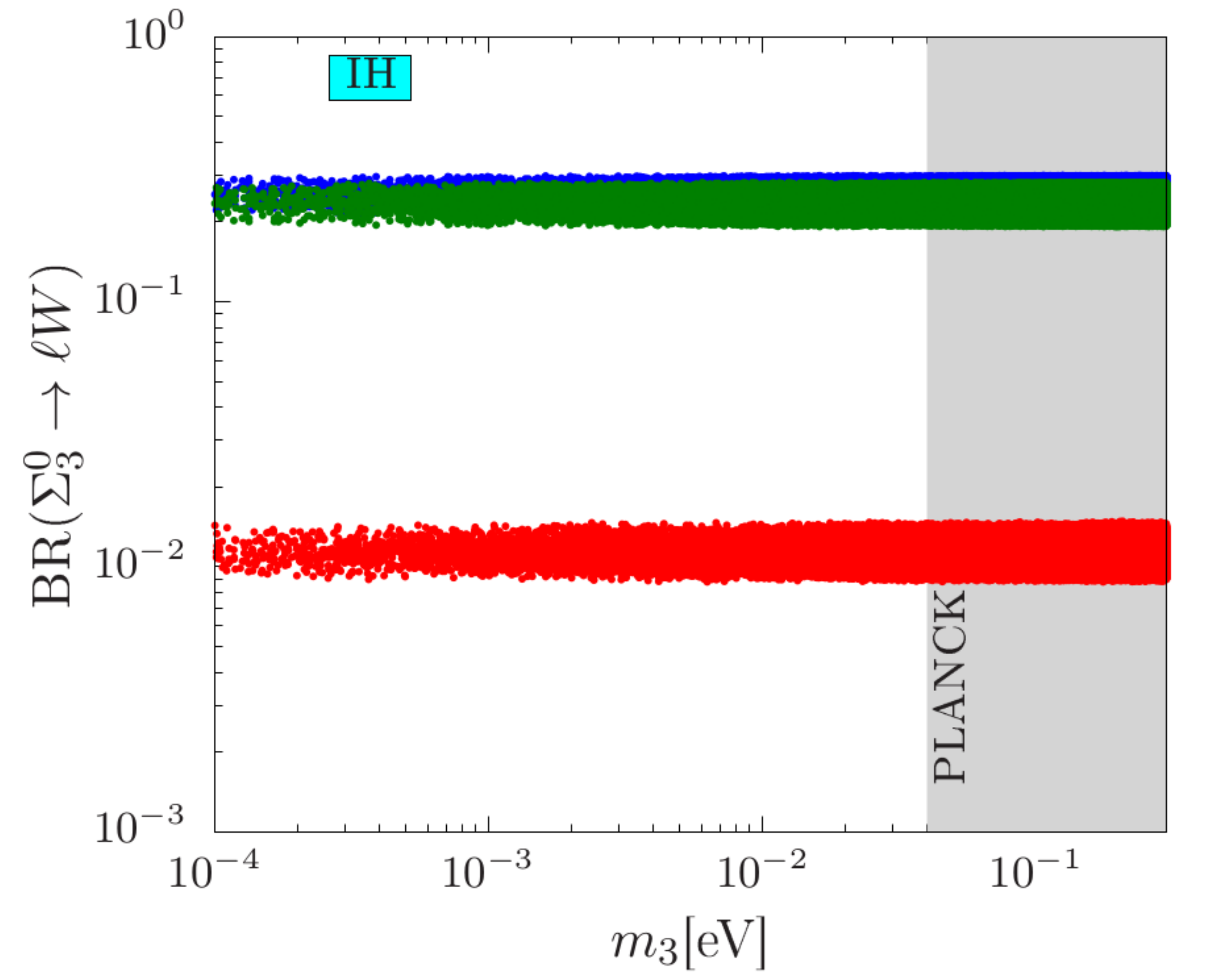}
\caption{Individual branching ratio of $\Sigma^0_i$ into the leading $\ell W$ mode as a function of the lightest neutrino mass for the NH $(m_1)$ and IH $(m_3)$ cases in the top and bottom panels respectively for the orthogonal matrix considered to be a identity matrix. The decay modes contain electron (red), muon (blue) and tau (green) for the $\Sigma_1$ (left column), $\Sigma_2$ (middle column) and $\Sigma_3$ (right column). The shaded region in gray is excluded by the PLANCK data. We consider $M=1$ TeV.}
\label{Mix7}
\end{figure}

We show the individual leading branching ratio of $\Sigma_i^0$ in the Fig.~\ref{Mix7} for the NH (IH) case in the top (bottom) panel for the 
orthogonal matrix as a identity matrix. For the first generation of the neutral multiplet of the triplet $(\Sigma_1^0)$ we show that the branching ratio into the electron (red) mode
dominates over the muon (blue) and tau (green) flavor for the NH and IH cases. For the second generation $(\Sigma_2^0)$ all the modes
coincide with each other for both of the neutrino mass hierarchy. For the third generation $(\Sigma_3^0)$ the muon and tau modes coincide 
and they dominate over the electron mode for the NH and IH cases. 

\begin{figure}[]
\centering
\includegraphics[width=0.33\textwidth]{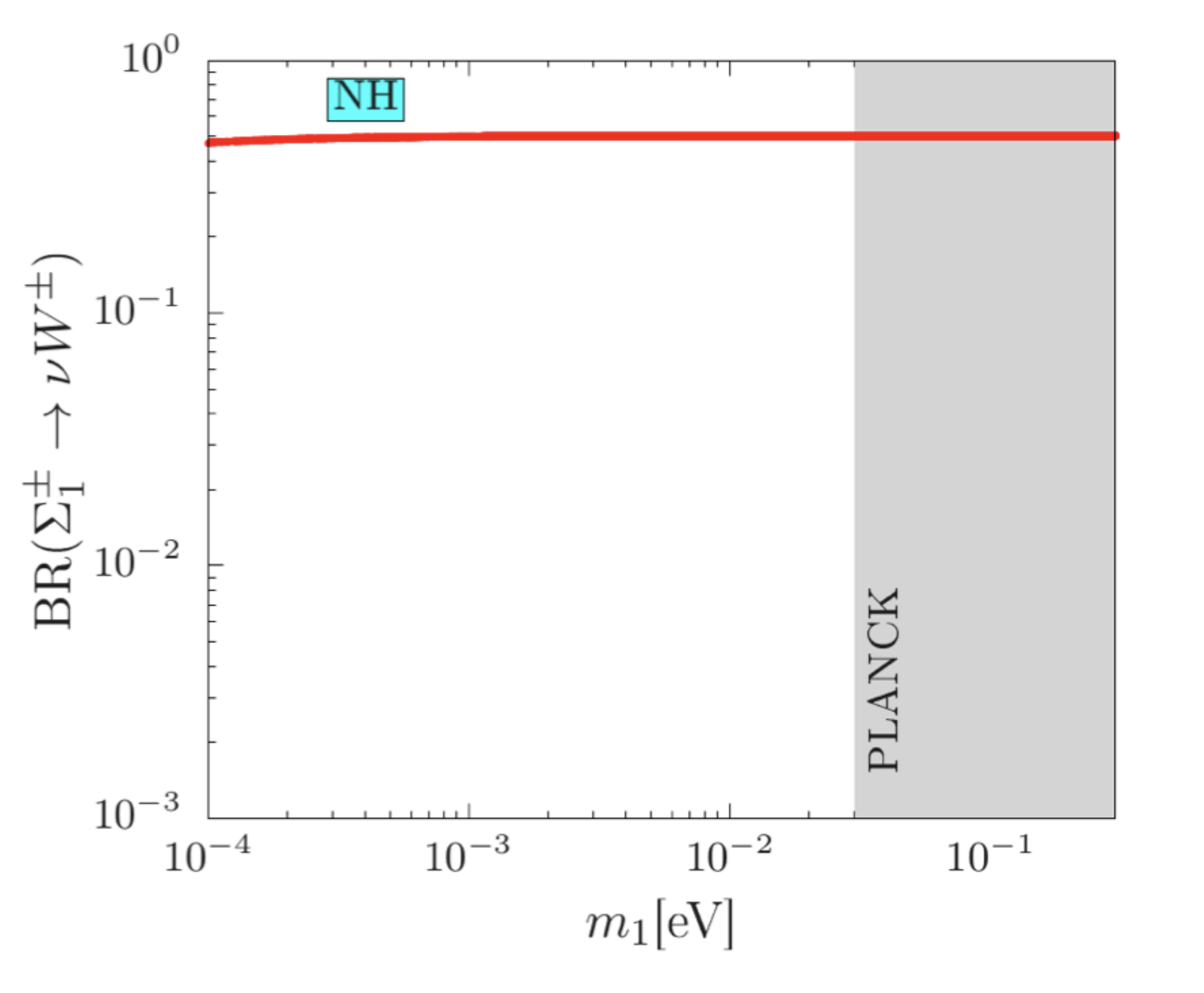}
\includegraphics[width=0.31\textwidth]{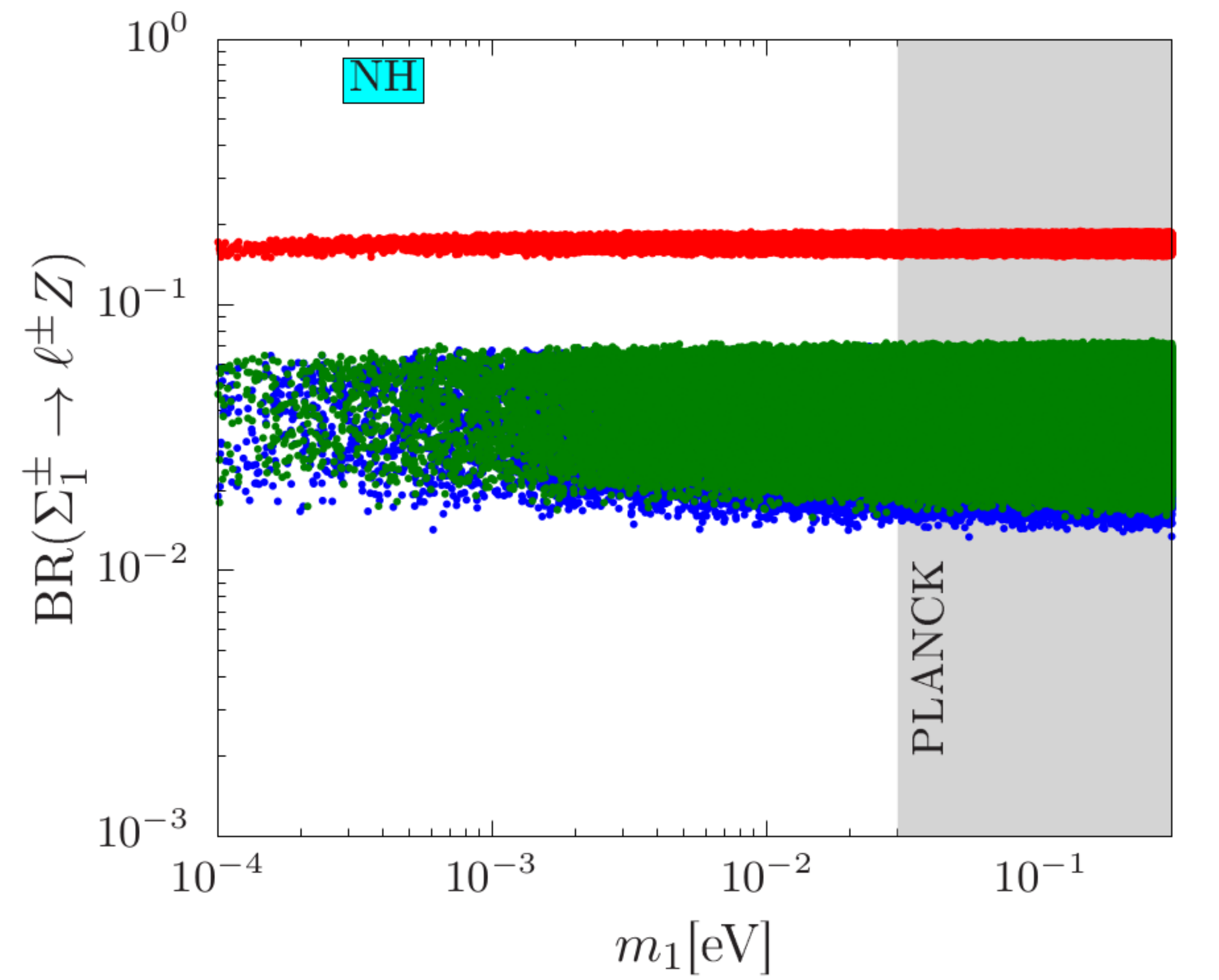}
\includegraphics[width=0.31\textwidth]{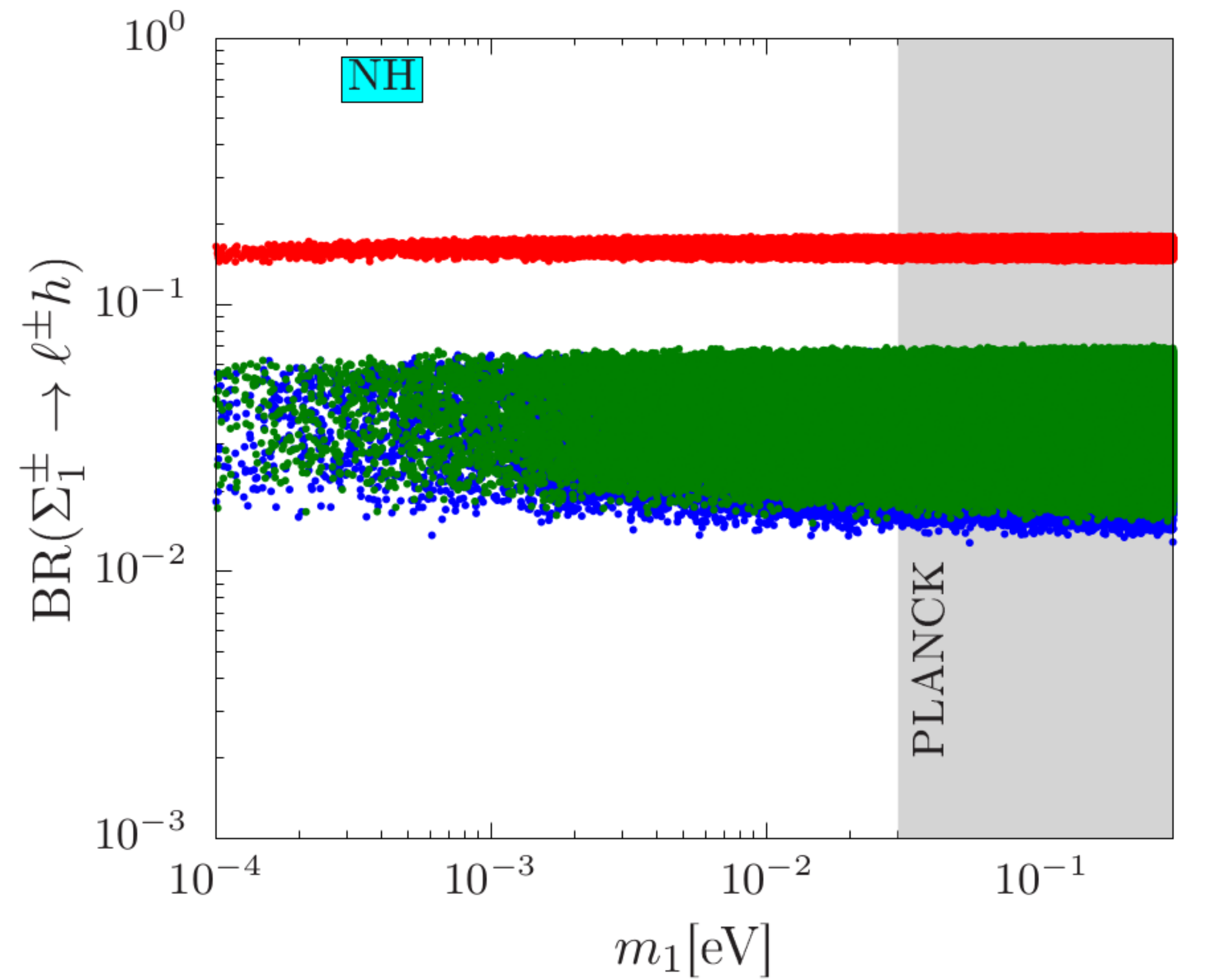}\\
\includegraphics[width=0.31\textwidth]{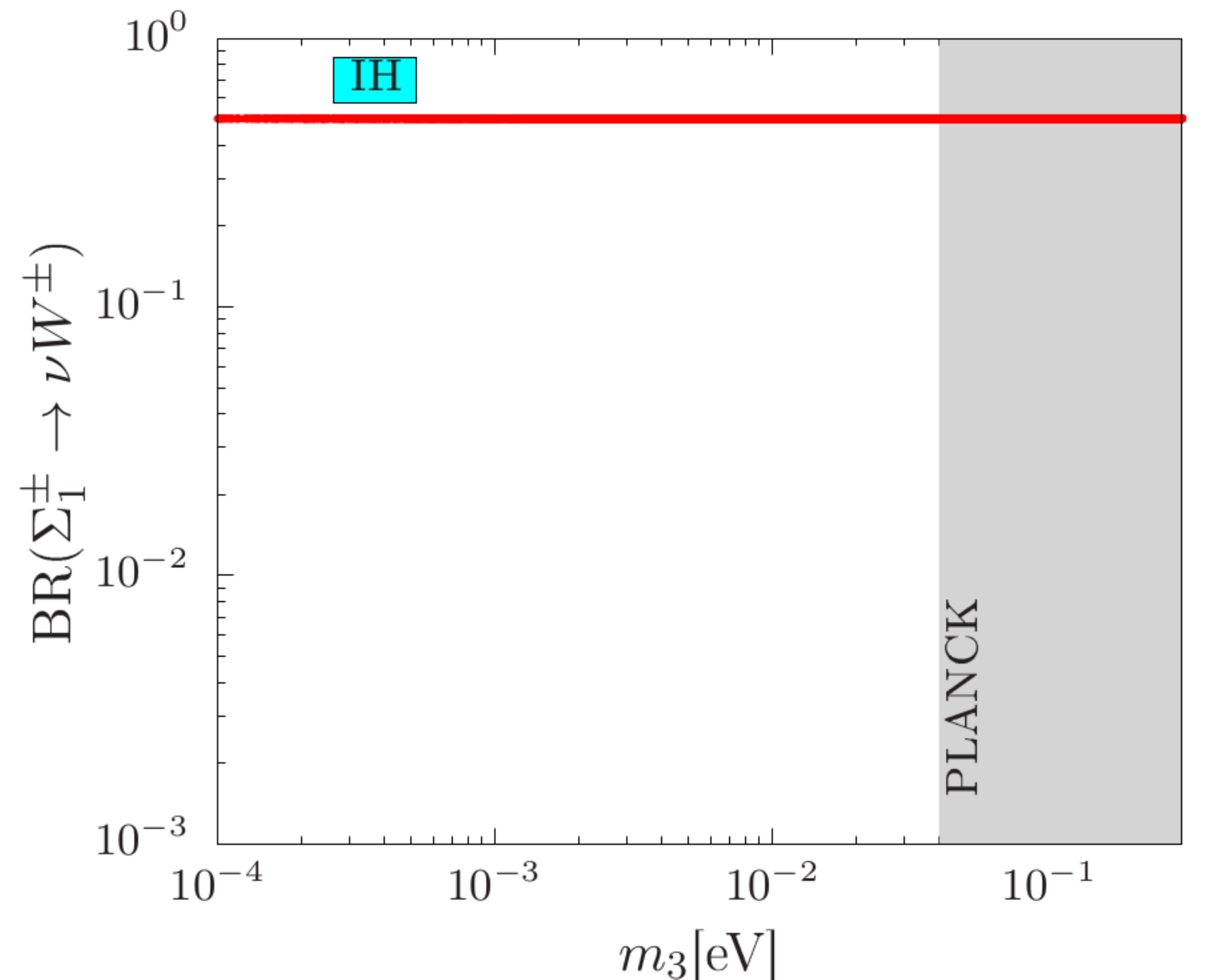}
\includegraphics[width=0.31\textwidth]{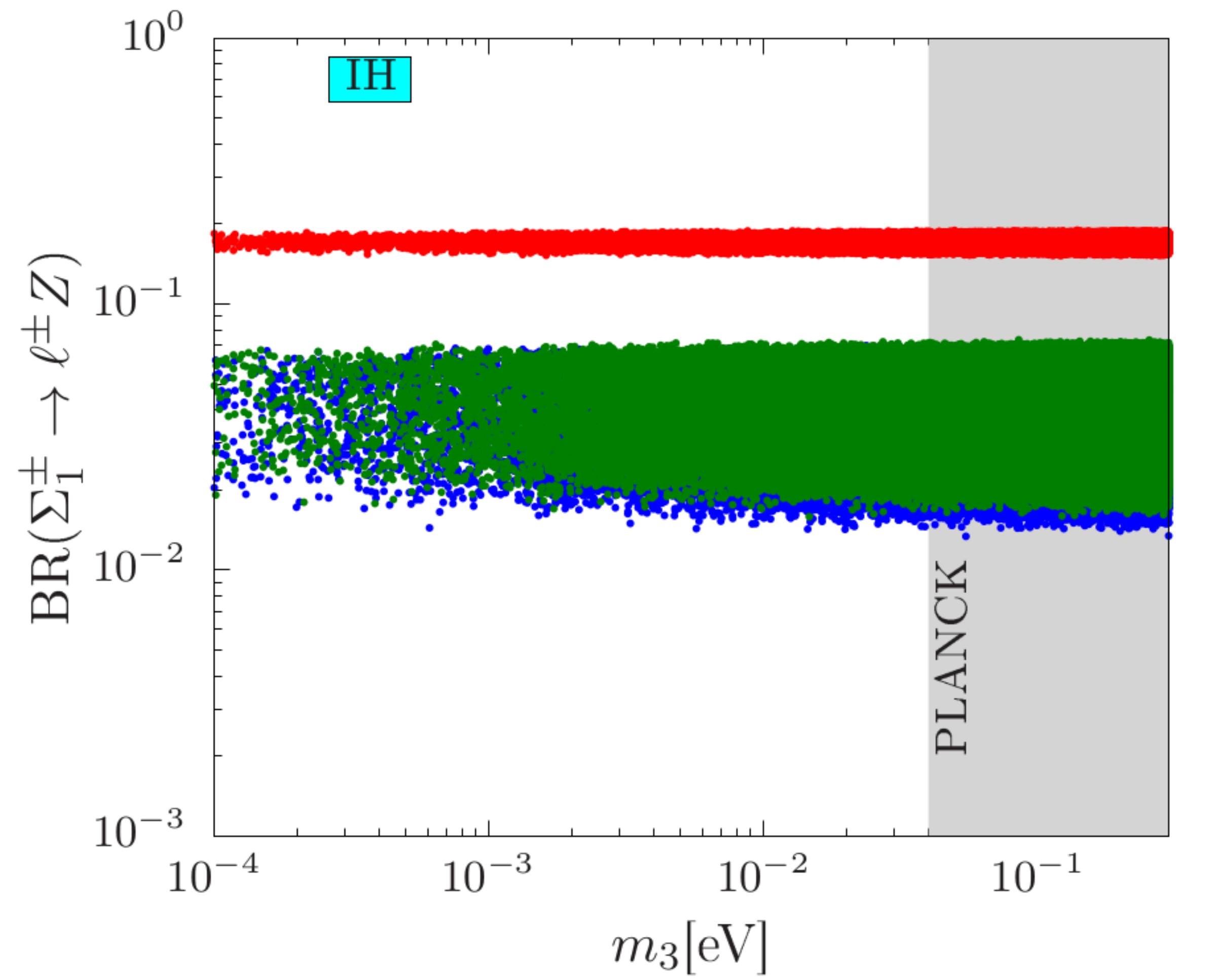}
\includegraphics[width=0.31\textwidth]{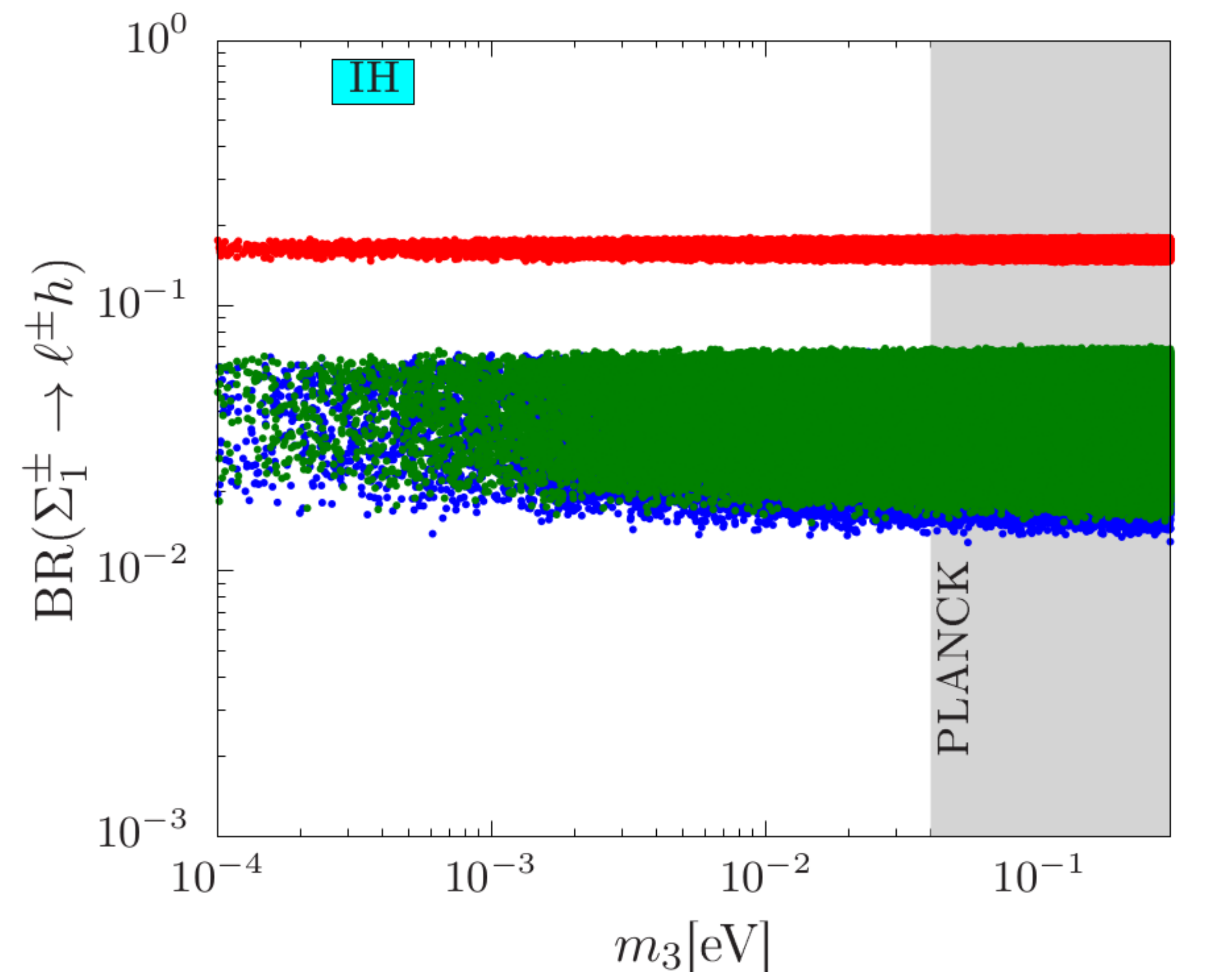}
\caption{Individual branching ratio of  $\Sigma^\pm_1$ into the leading $\nu W$ (left column) and subleading $\ell^\pm Z$ (middle column), $\ell^\pm h$ (right column) modes with respect to the lightest neutrino mass for the NH $(m_1)$ and IH $(m_3)$ cases in the top and bottom panels respectively for the orthogonal matrix considered to be a identity matrix. The decay modes contain electron (red), muon (blue) and tau (green). The shaded region in gray is excluded by the PLANCK data. We consider $M=1$ TeV.}
\label{Mix8}
\end{figure}

We study the individual branching ratios of the three different generations of the charged multiplet $\Sigma_{1}^\pm$ into $\nu W$, $\ell^\pm Z$ and $\ell^\pm h$ modes respectively where the orthogonal matrix has been considered as a identity matrix. The $\nu W$ mode is dominant over the $\ell^\pm Z$ and $\ell^\pm h$ modes. 
In Fig.~\ref{Mix8} we show the different decay modes of $\Sigma_i^\pm$ for the NH (IH) case in the top (bottom) panel. 
For the $\nu W$ mode we do not distinguish between the neutrinos as the neutrinos will be considered as the missing momenta and hence we summed over all flavor of neutrinos.
Therefore we have the single line in the first column for both of the NH and IH cases. In the $\ell^\pm Z$ and $\ell^\pm h$ modes we have almost the same nature 
in both of the neutrino mass hierarchies where the electron mode (red) dominates over the muon (blue) and tau (green) modes. 

\begin{figure}[]
\centering
\includegraphics[width=0.31\textwidth]{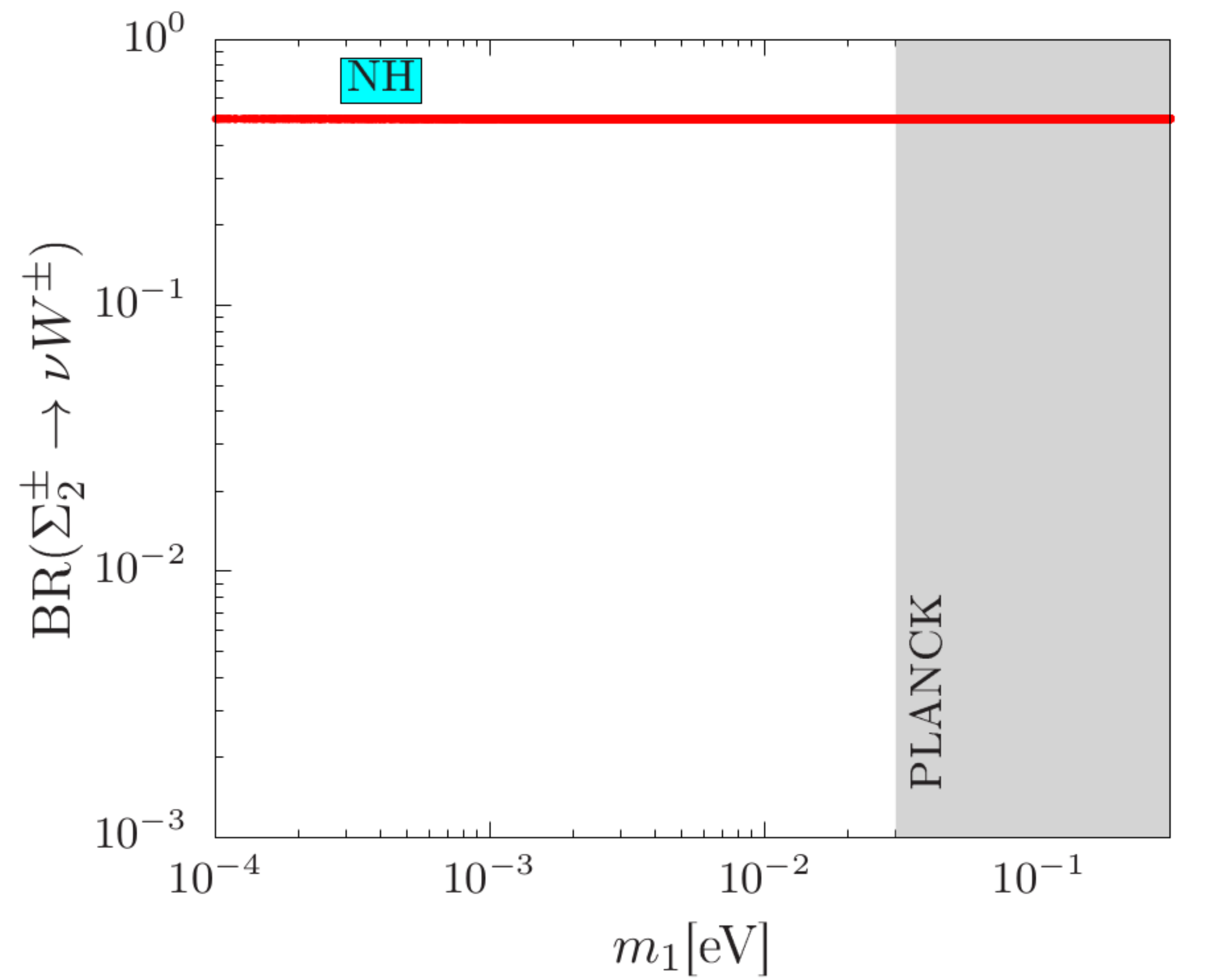}
\includegraphics[width=0.31\textwidth]{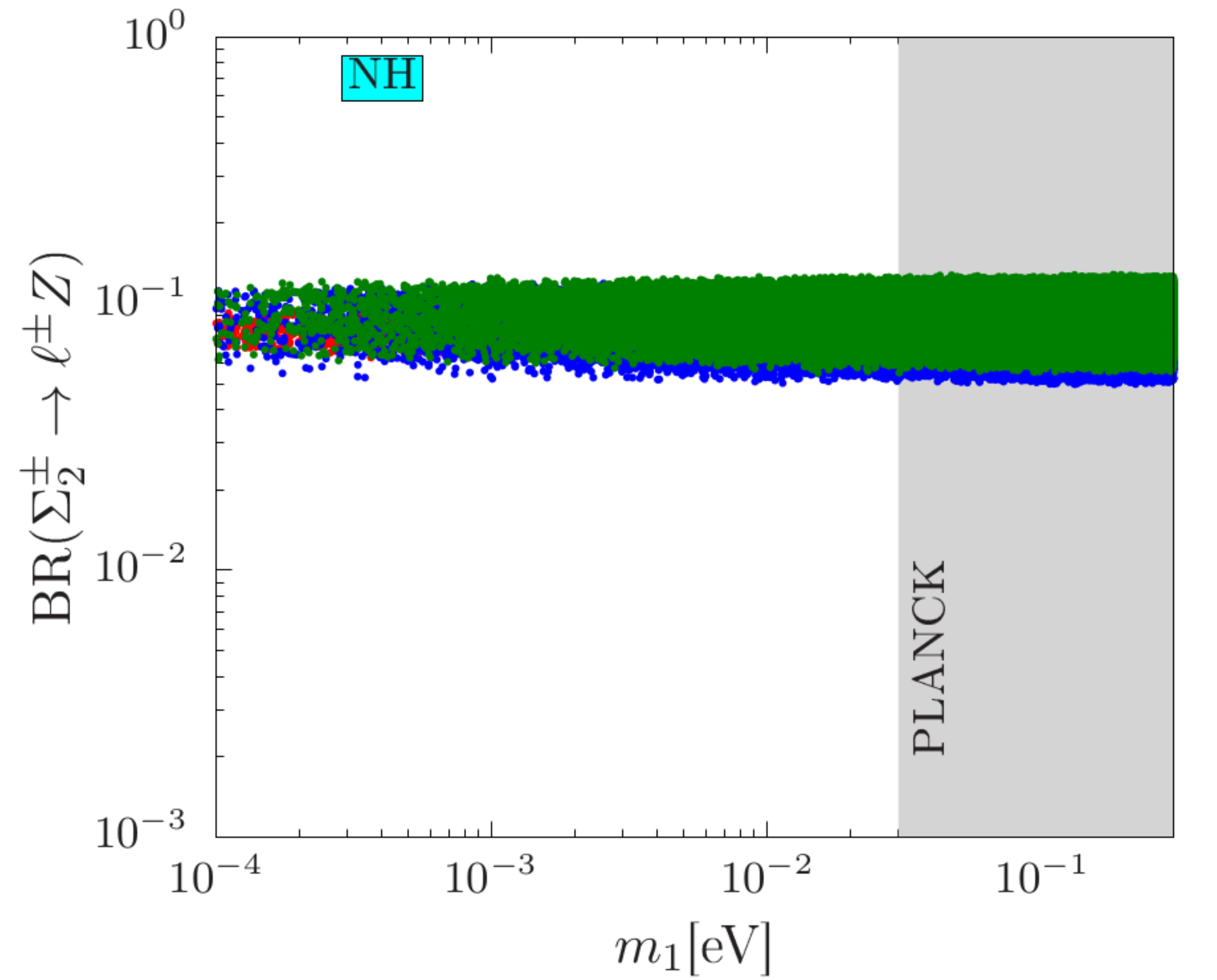}
\includegraphics[width=0.31\textwidth]{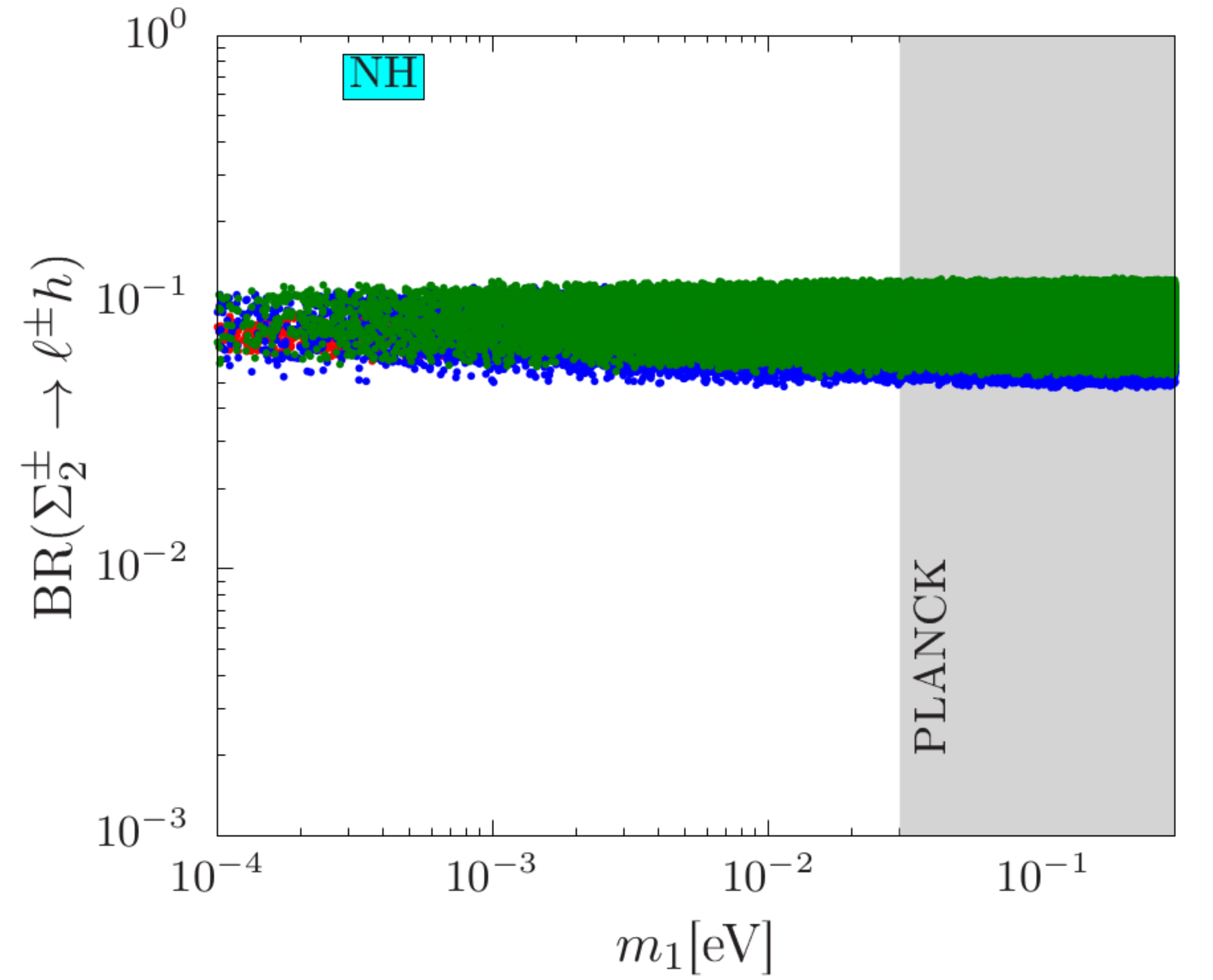}\\
\includegraphics[width=0.31\textwidth]{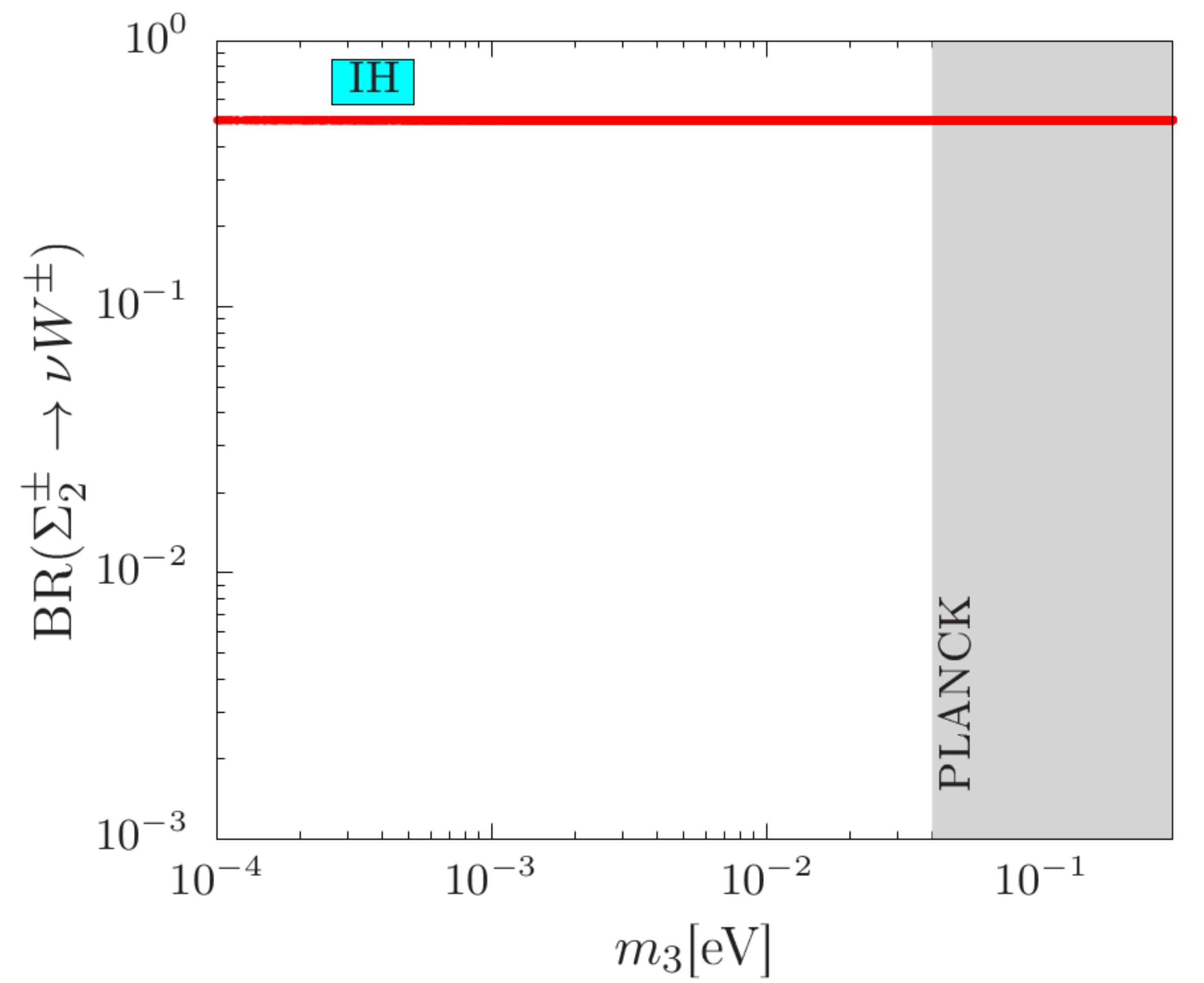}
\includegraphics[width=0.31\textwidth]{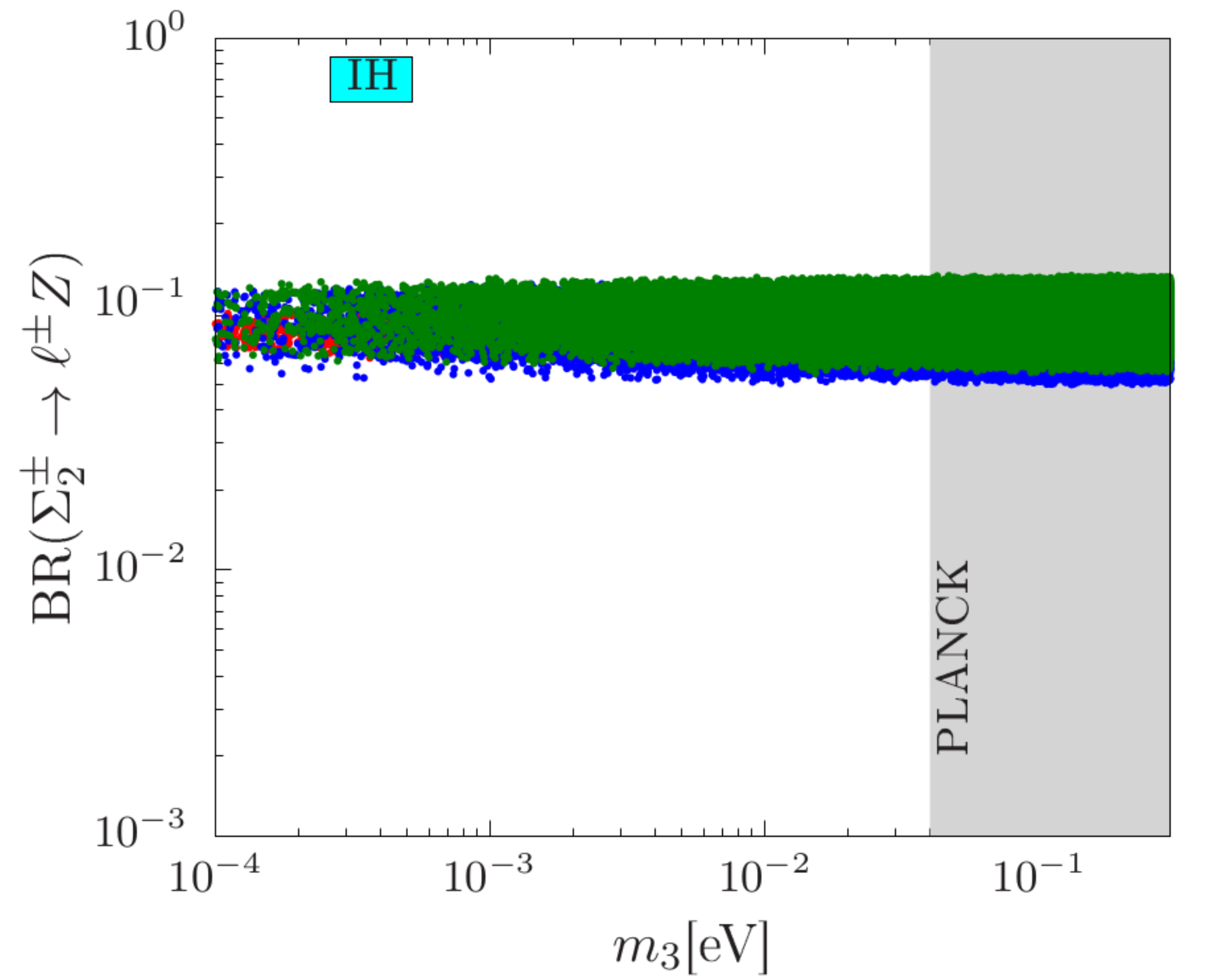}
\includegraphics[width=0.31\textwidth]{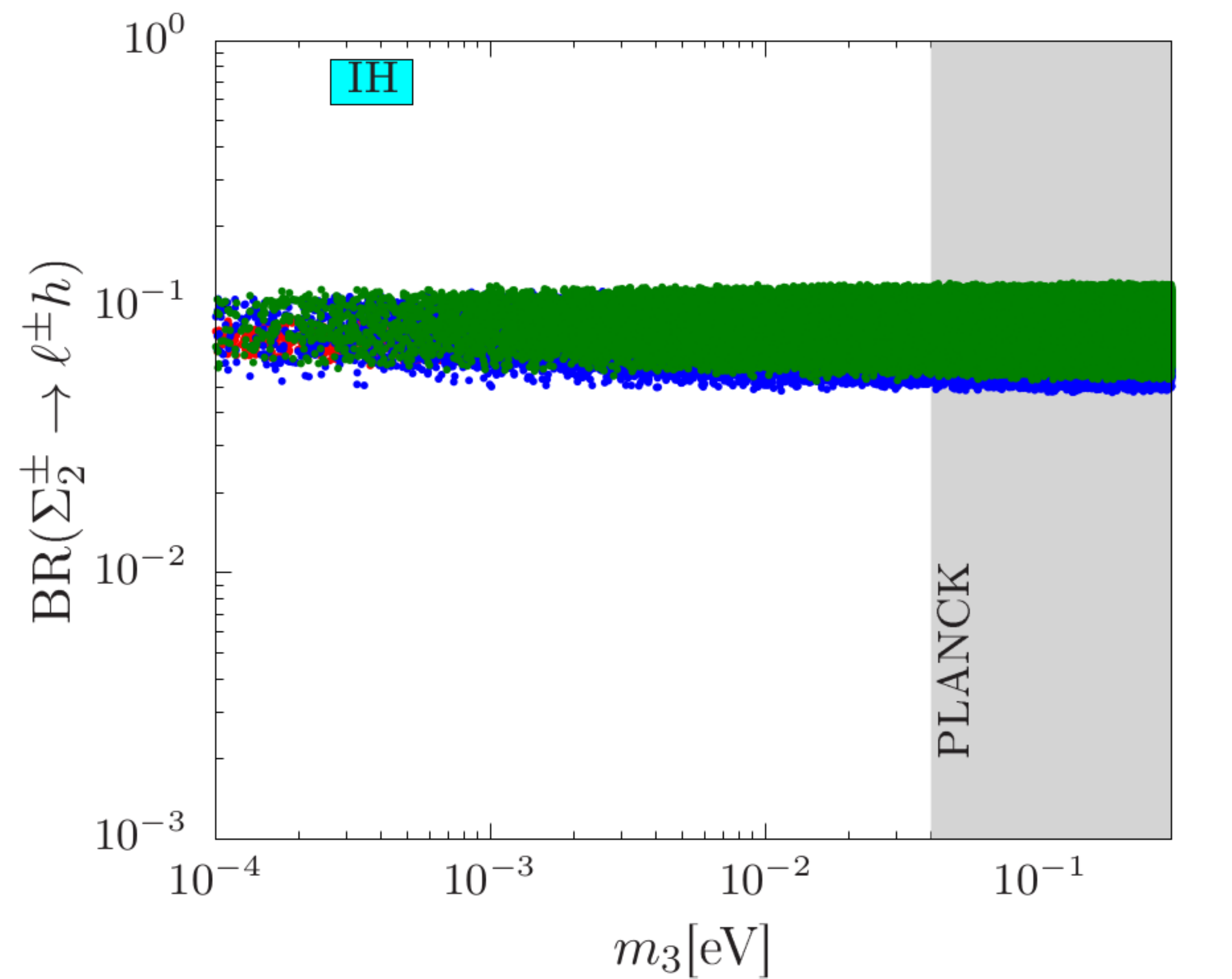}
\caption{Same as Fig.~\ref{Mix8} but now for $\Sigma_2^\pm$.}
\label{Mix9}
\end{figure}

The behavior for the $\nu W$ mode can be obtained for the second generation of the charged multiplet $\Sigma_2^\pm$ in the first column of the Fig.~\ref{Mix9} for the NH and IH case. 
We also study the $\ell^\pm Z$ and $\ell^\pm h$ modes. We notice that the parameter regions for the three flavors coincide with each other for both the NH and IH case, see top and bottom panel of the middle column in Fig.~\ref{Mix9}. 

\begin{figure}[]
\centering
\includegraphics[width=0.31\textwidth]{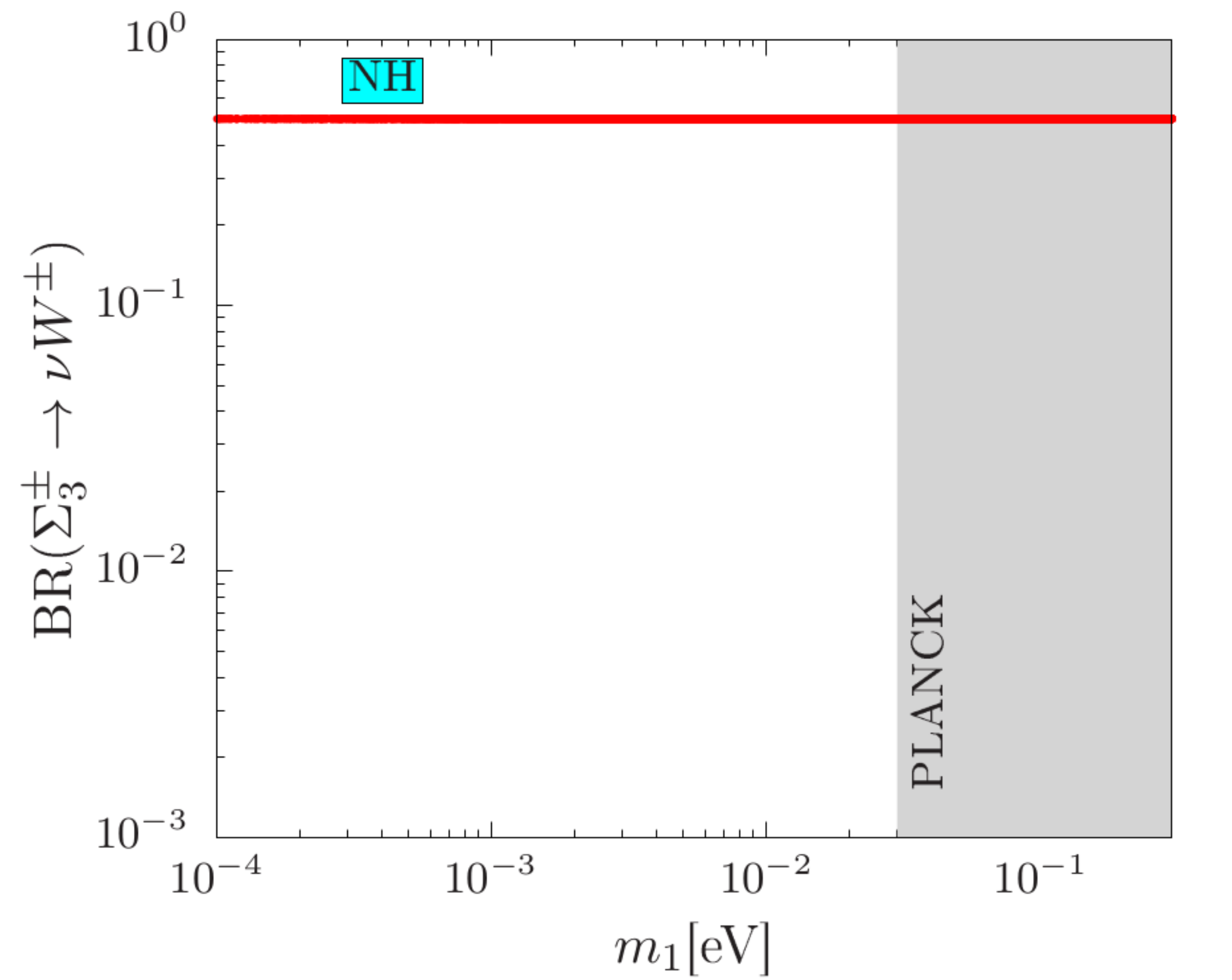}
\includegraphics[width=0.31\textwidth]{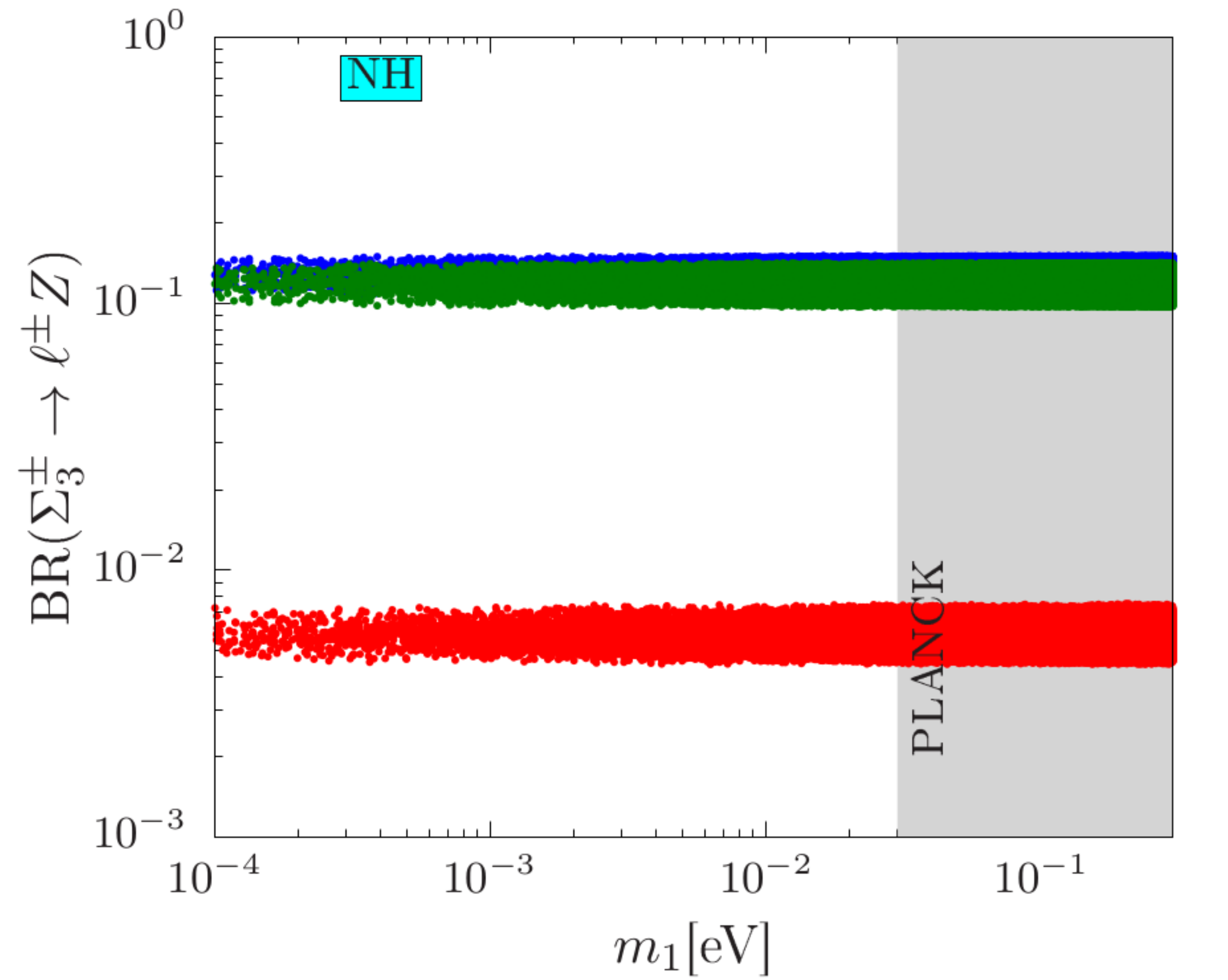}
\includegraphics[width=0.31\textwidth]{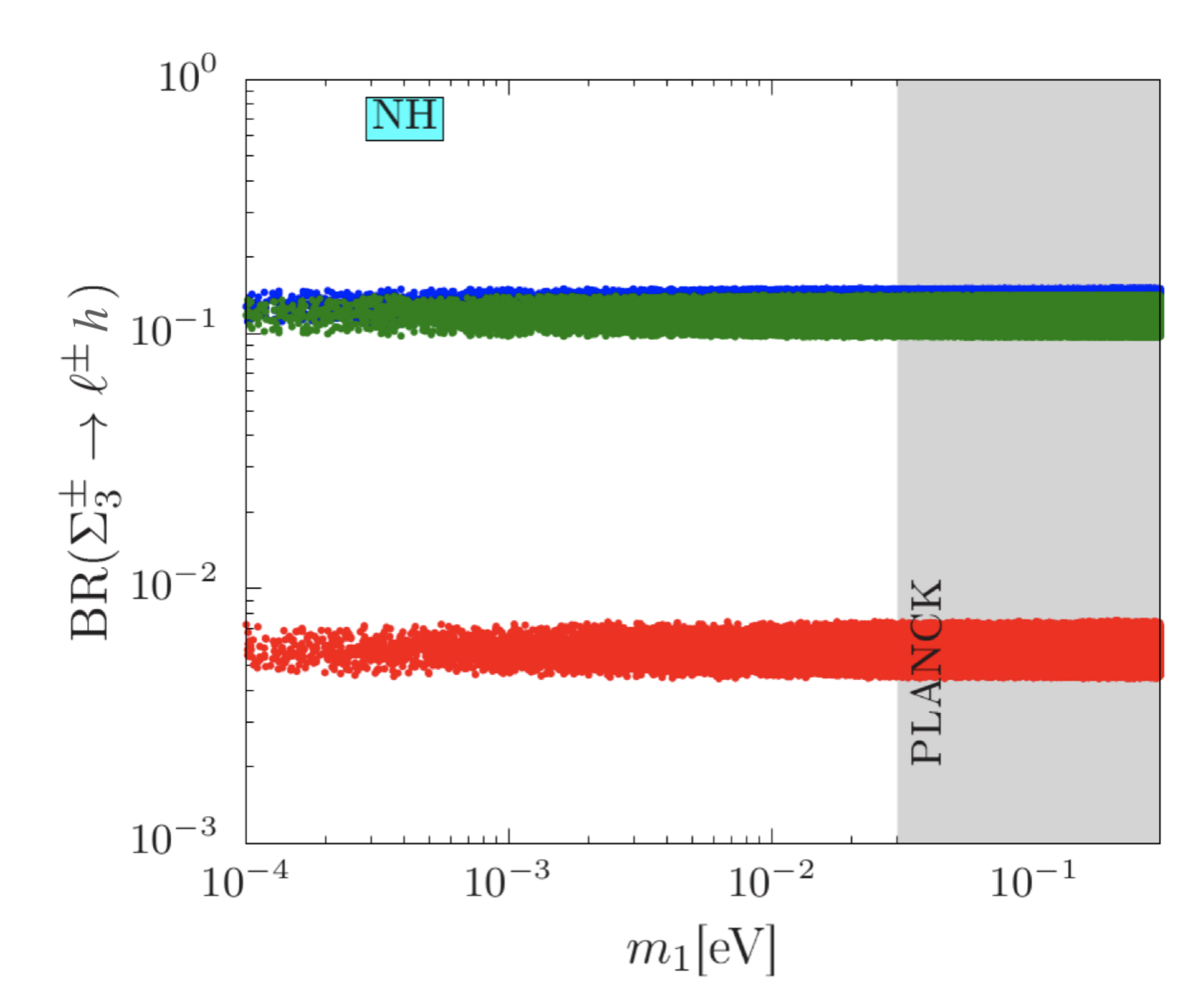}\\
\includegraphics[width=0.31\textwidth]{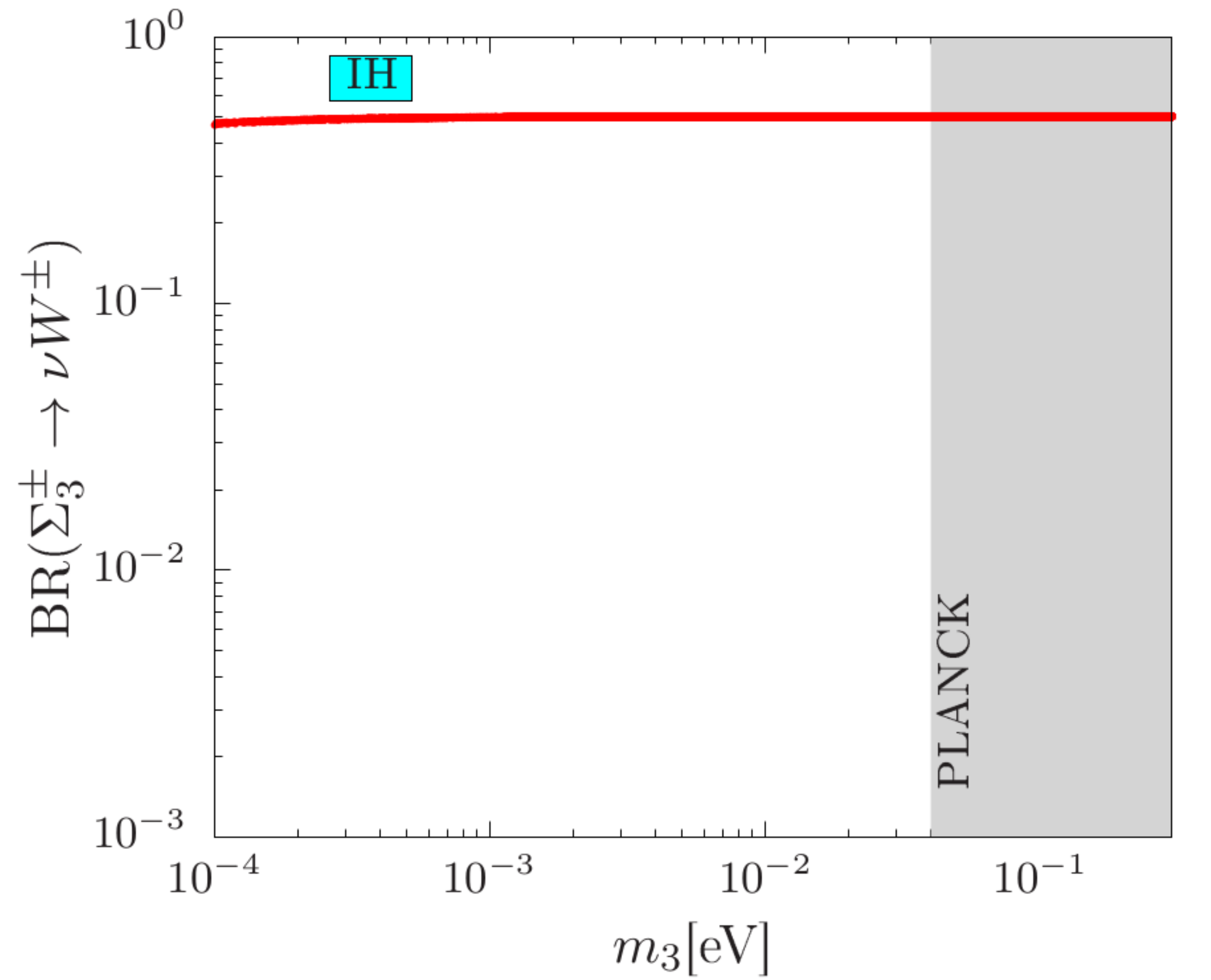}
\includegraphics[width=0.31\textwidth]{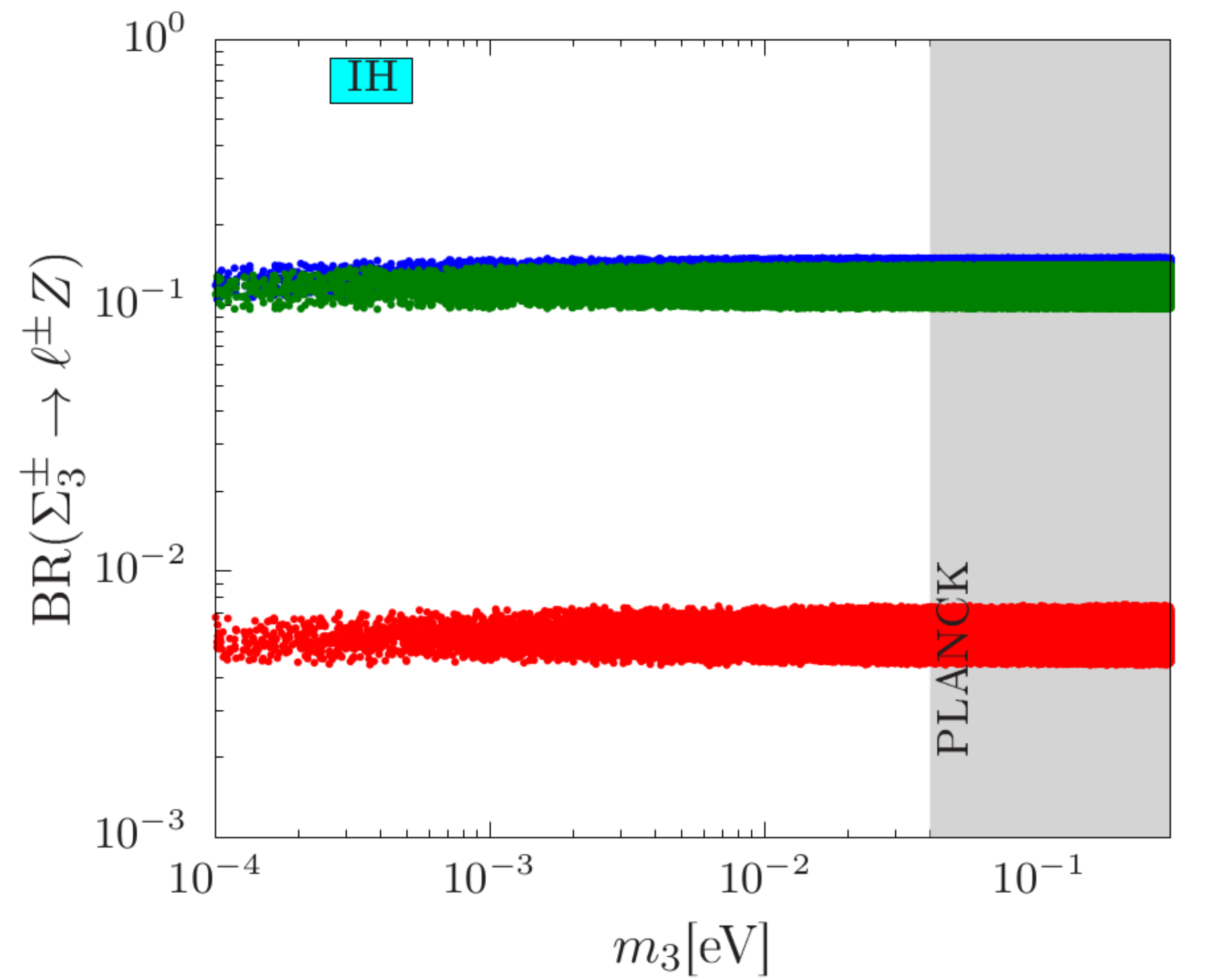}
\includegraphics[width=0.31\textwidth]{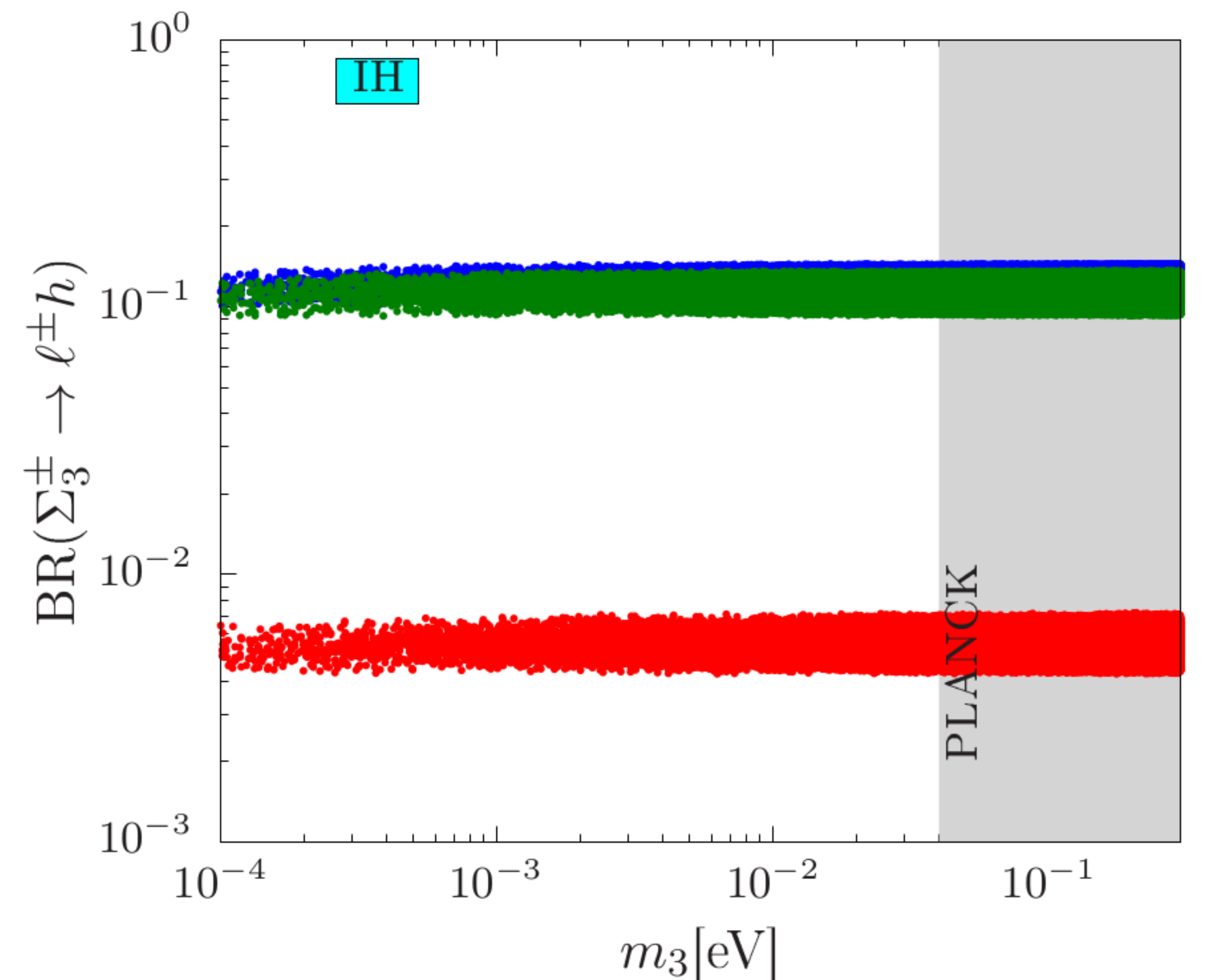}
\caption{Same as Fig.~\ref{Mix8} but now for $\Sigma_3^\pm$.}
\label{Mix10}
\end{figure}

The third generation charged triplet $\Sigma_3^\pm$ decaying into $\nu W$ show the same behavior as the other two generations, see top and bottom panel of first column in
Fig.~\ref{Mix10}. For the $\ell^\pm Z$ and $\ell^\pm h$ modes we see a different behavior unlike the other two generations. In case of $\Sigma_{3}^\pm$ the muon (blue) and tau (green) modes dominate over the electron (red) mode. The corresponding parameter spaces for the NH (IH) case is shown in the top (bottom) panel of the second column in Fig.~\ref{Mix10}.


\begin{figure}[]
\centering
\includegraphics[width=0.31\textwidth]{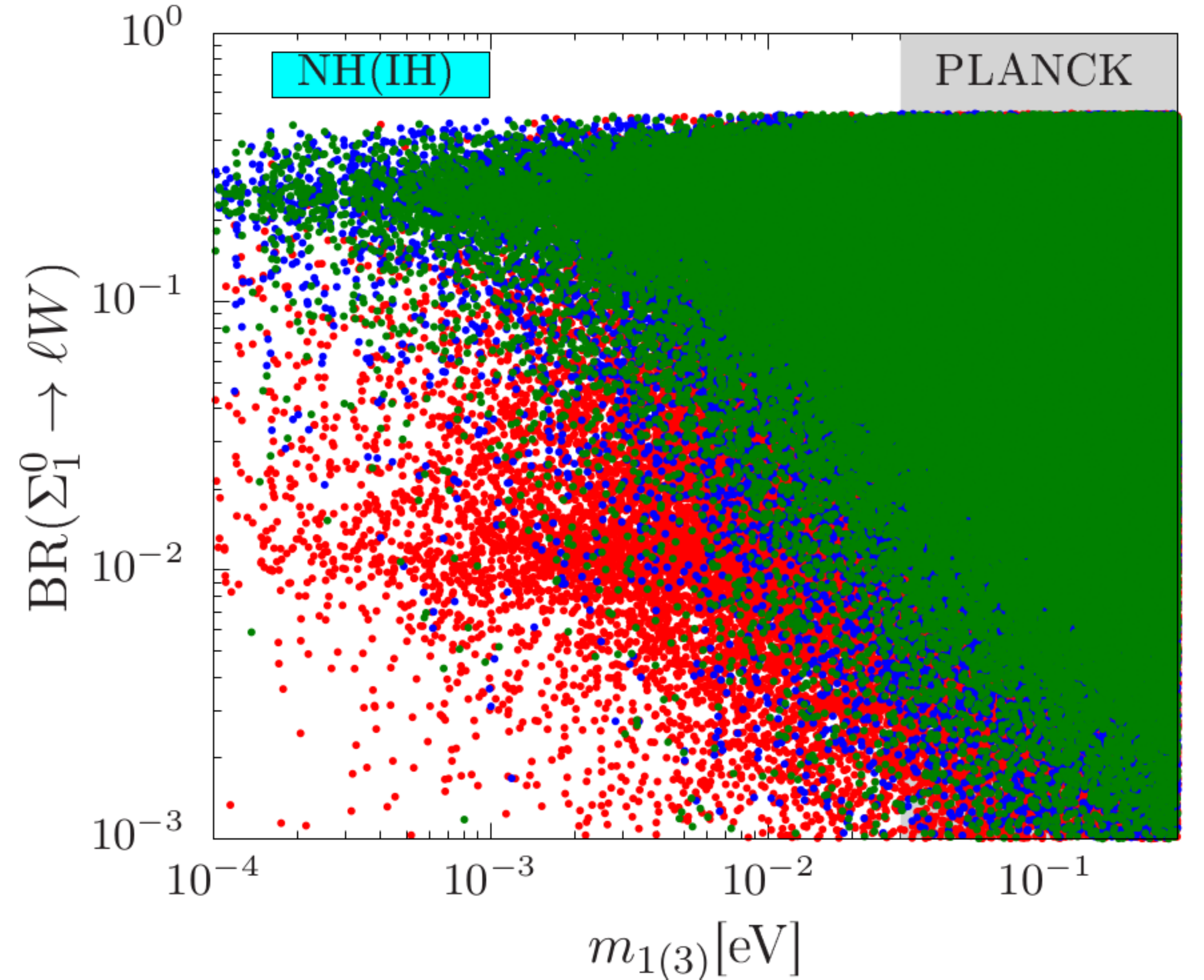}
\includegraphics[width=0.31\textwidth]{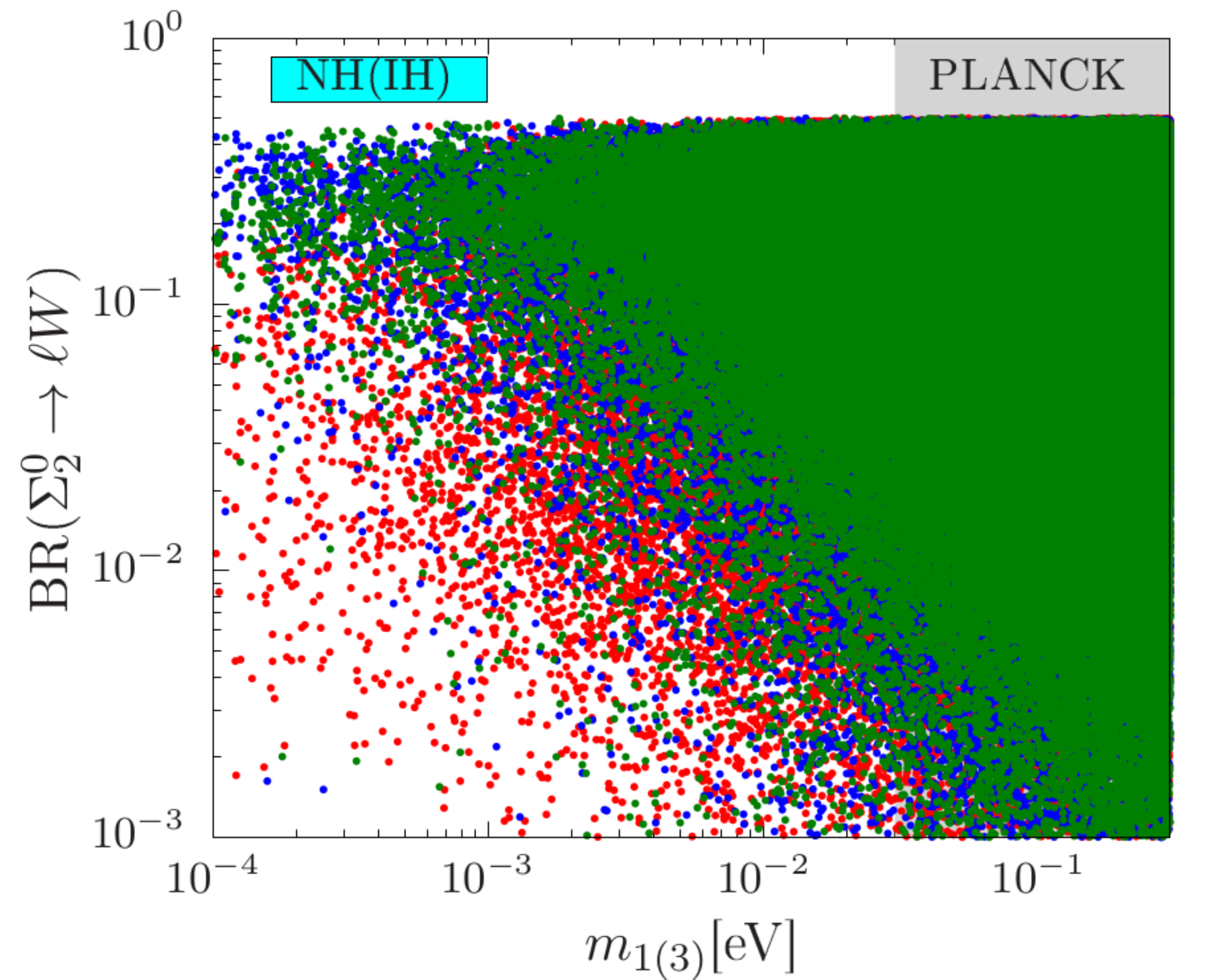}
\includegraphics[width=0.31\textwidth]{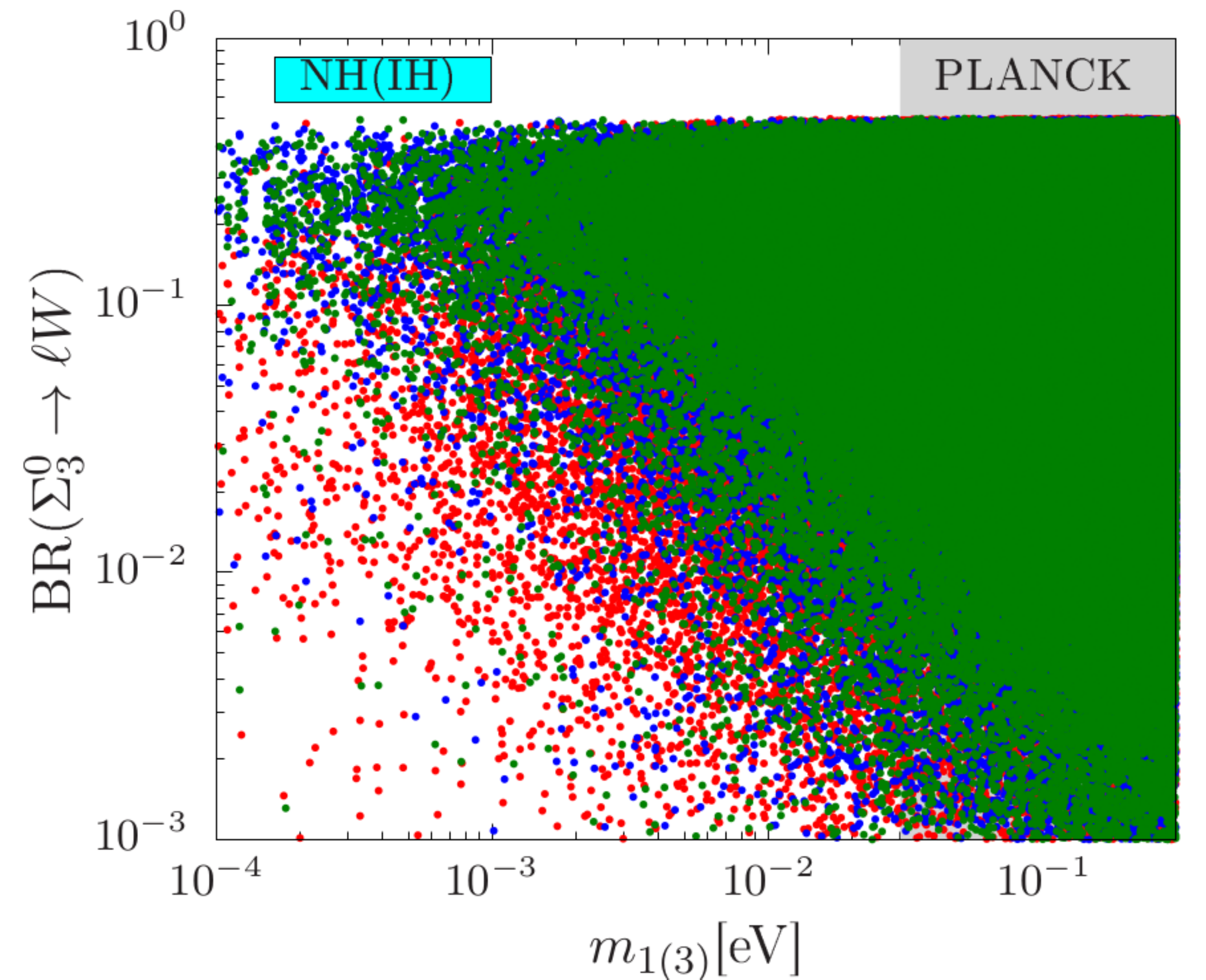}
\caption{Individual branching ratio of $\Sigma^0_i$ into the leading $\ell W$ mode as a function of the lightest neutrino mass $(m_1, m_3)$ for the two hierarchic cases (NH, IH) for the real orthogonal matrix. Red, blue and green color stand for electron, muon and tau modes, respectively. The shaded region in gray is excluded by the PLANCK data. We consider $M=1$ TeV.}
\label{Mix11}
\end{figure}

\begin{figure}[]
\centering
\includegraphics[width=0.31\textwidth]{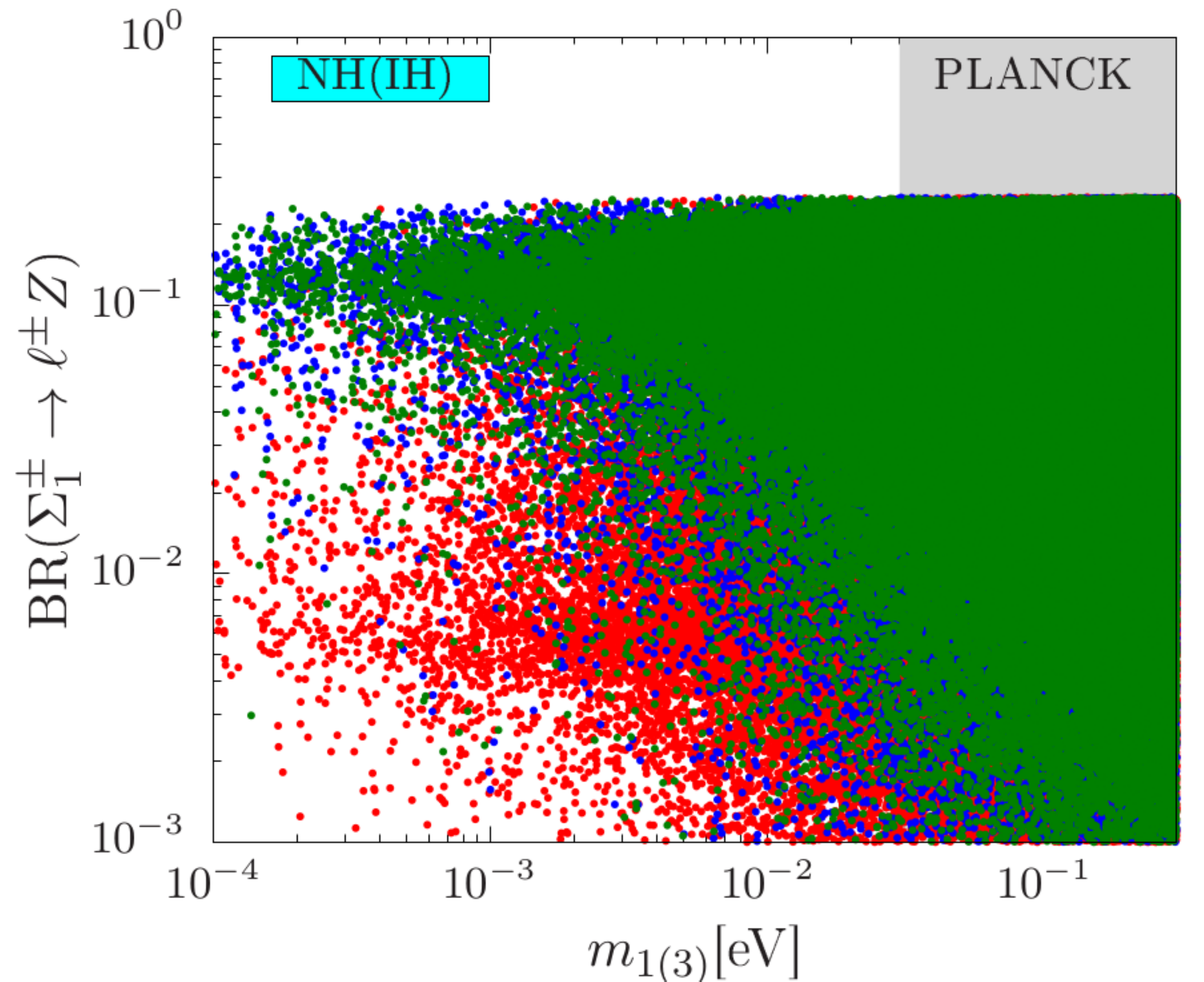}
\includegraphics[width=0.31\textwidth]{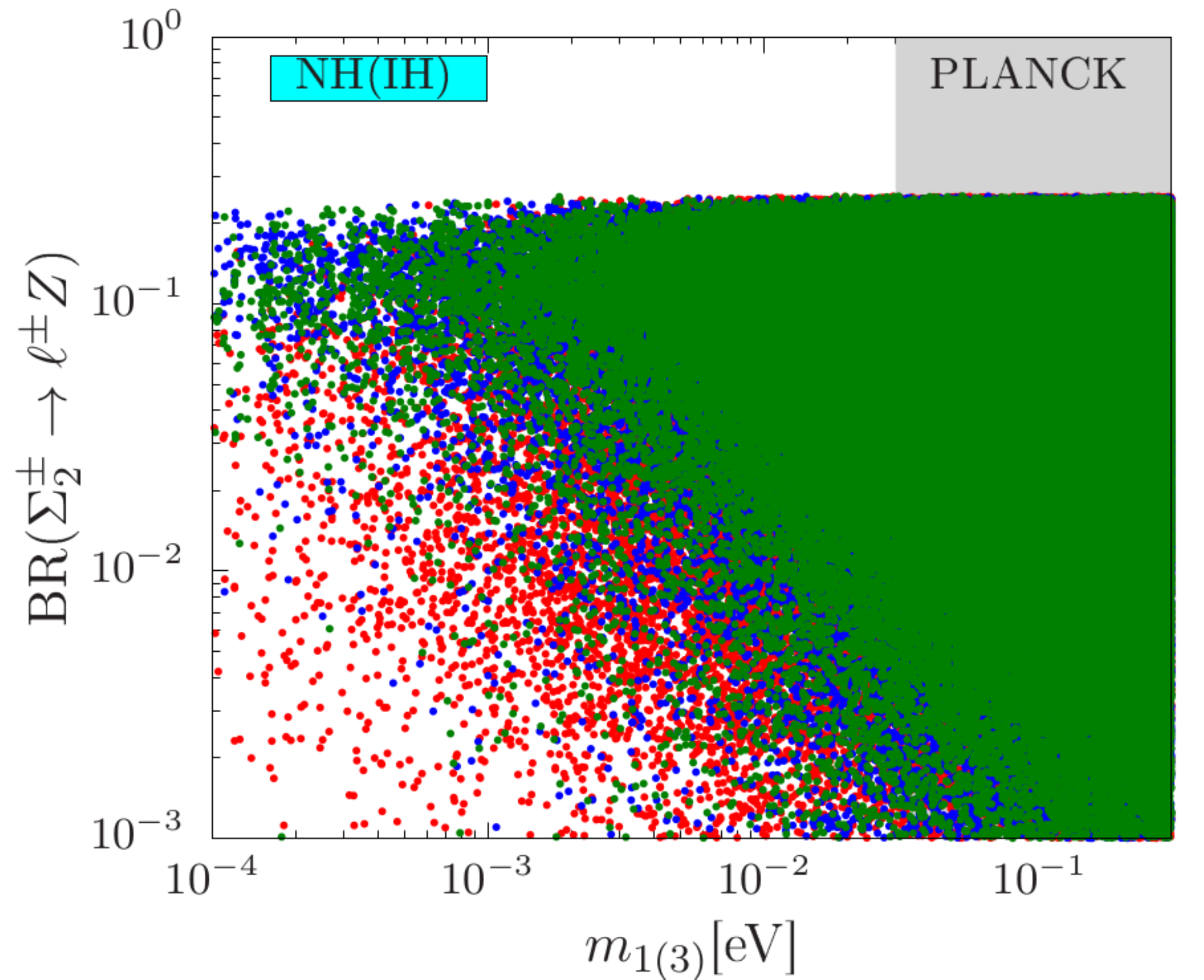}
\includegraphics[width=0.31\textwidth]{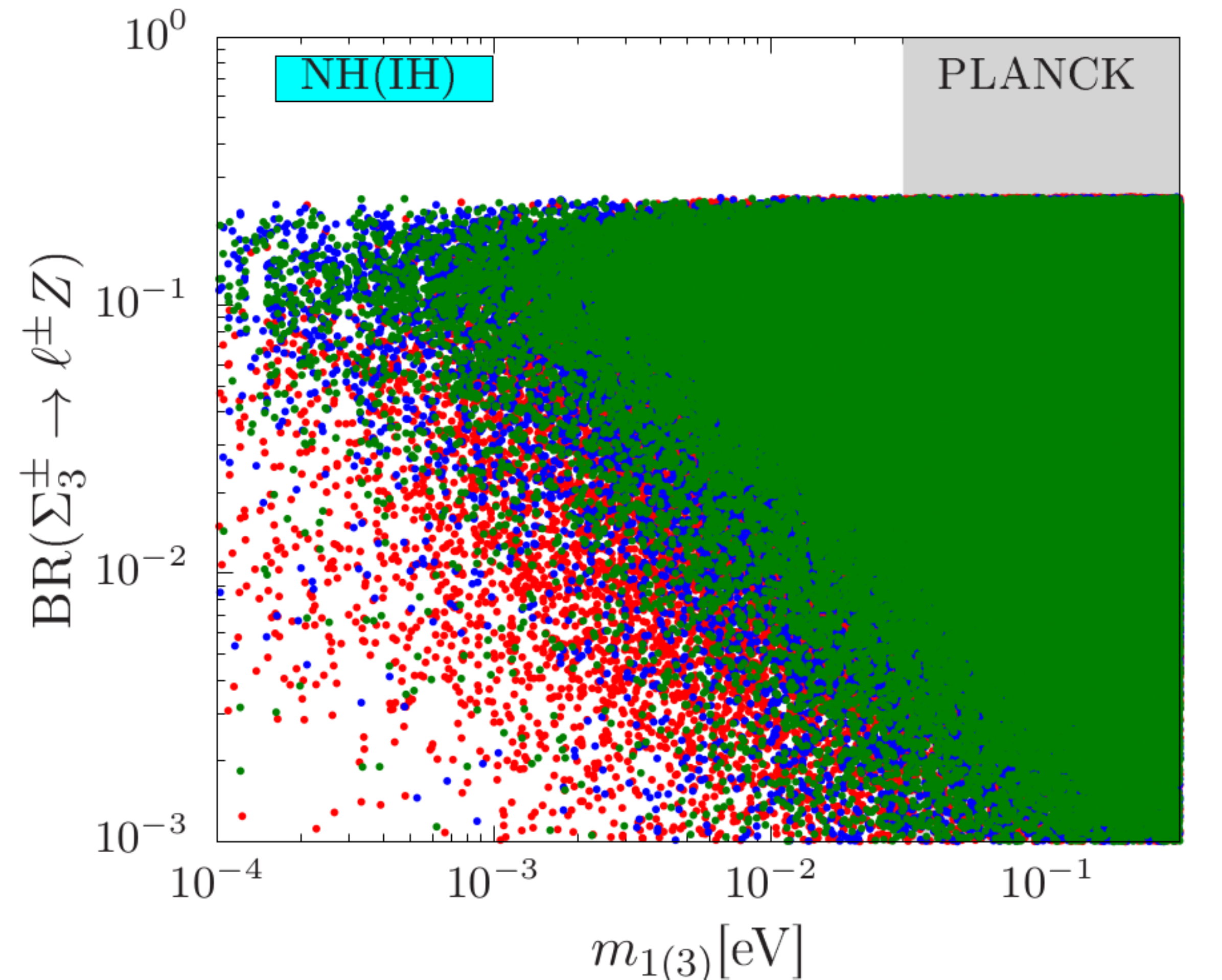}\\
\includegraphics[width=0.31\textwidth]{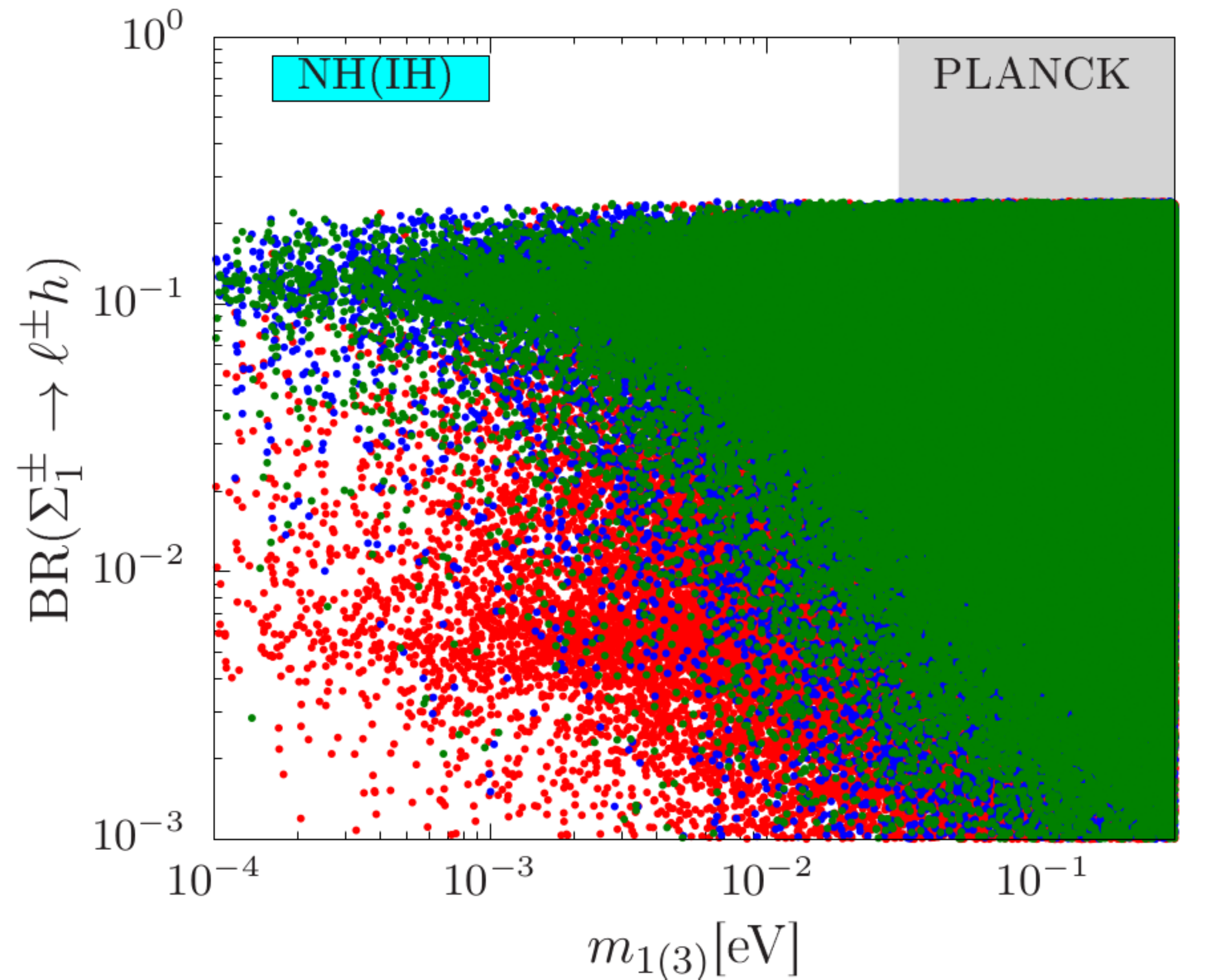}
\includegraphics[width=0.31\textwidth]{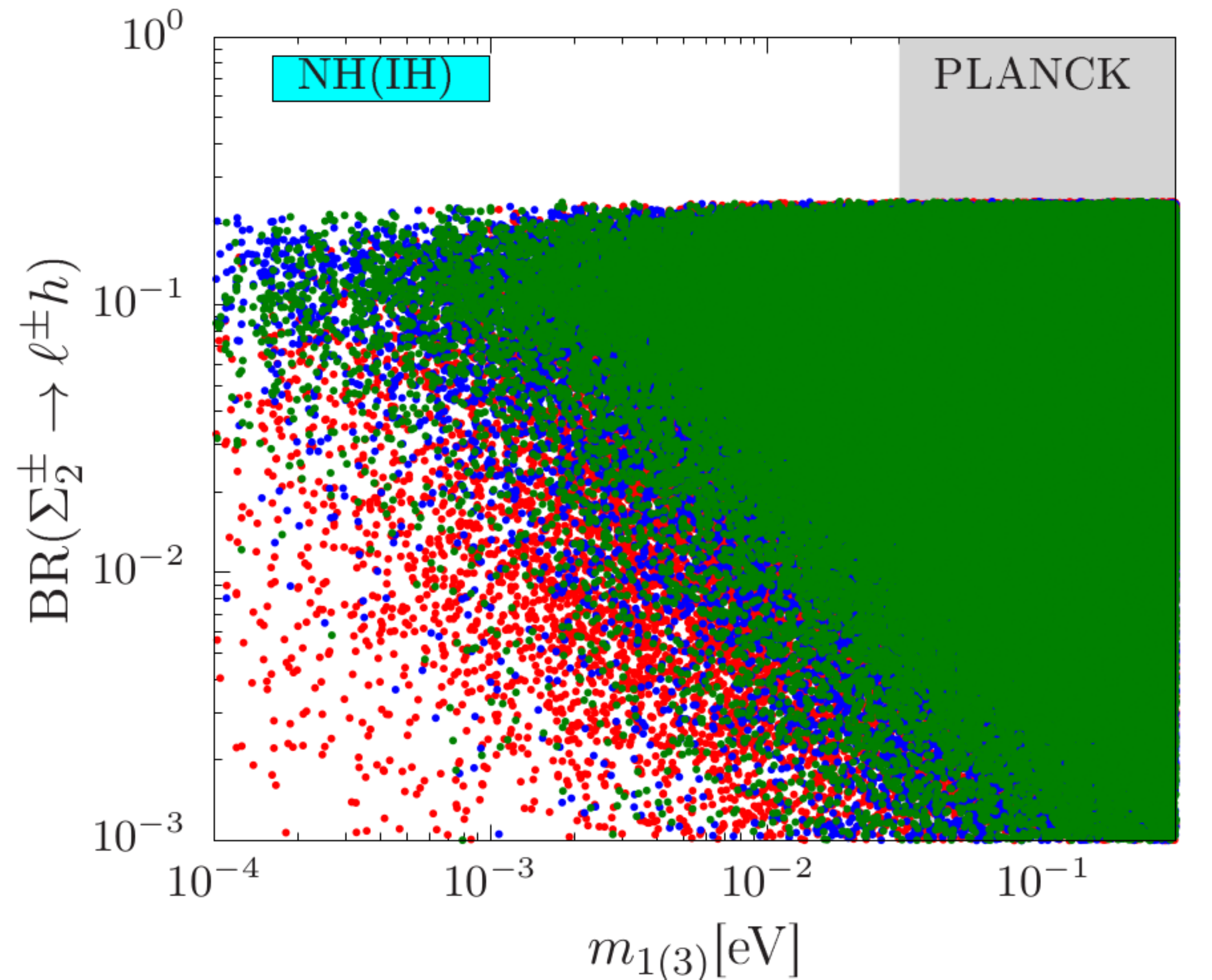}
\includegraphics[width=0.31\textwidth]{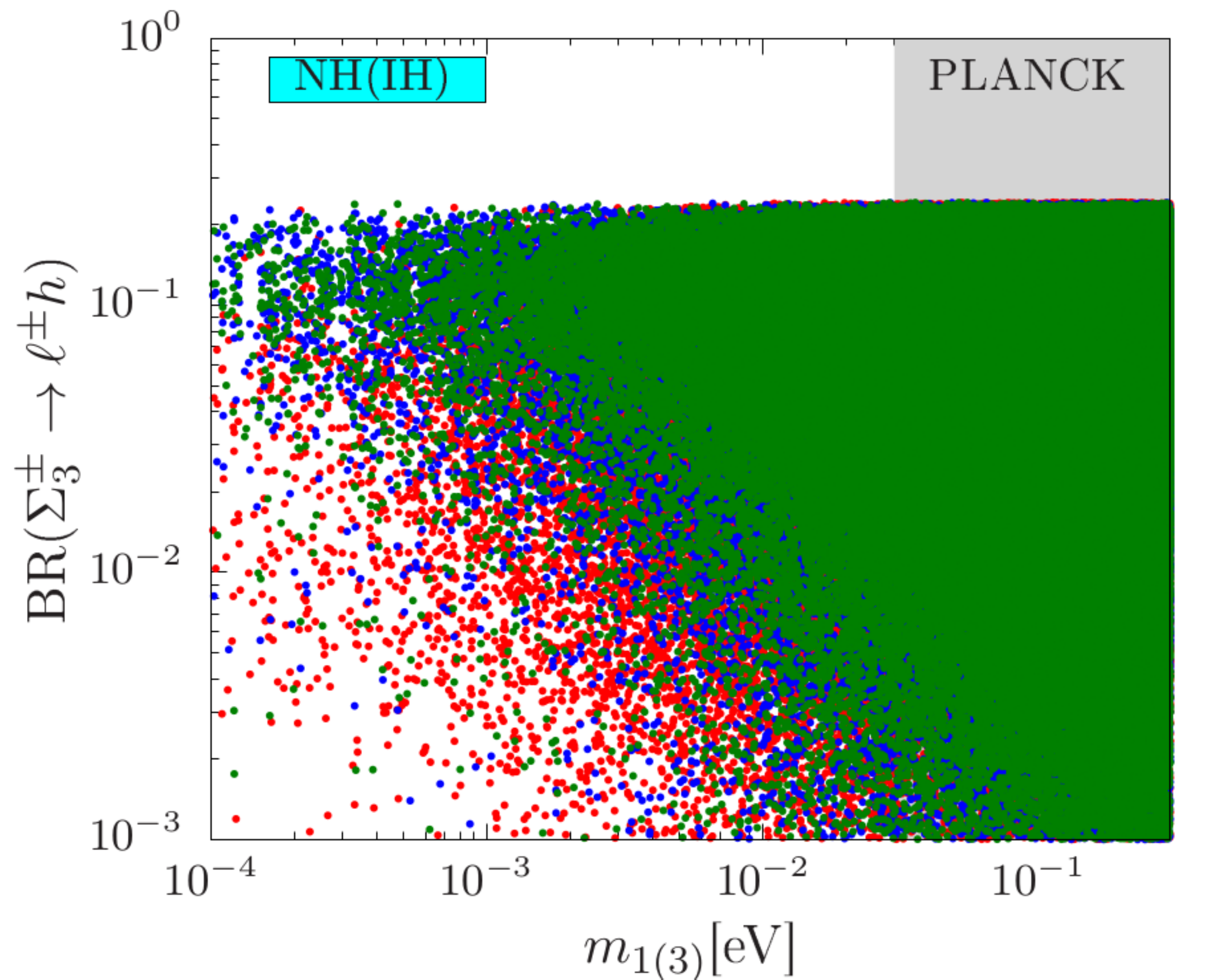}
\caption{Individual branching ratio of $\Sigma^\pm_i$ into the subleading and visible $\ell^\pm Z$ (top panel) and $\ell^\pm h$ (bottom panel)  modes with respect to the lightest neutrino mass $(m_1, m_3)$ for the two hierarchic cases (NH, IH) for the real orthogonal matrix.Red, blue and green color stand for electron, muon and tau modes, respectively. The shaded region in gray is excluded by the PLANCK data. We consider $M=1$ TeV.}
\label{Mix12}
\end{figure}

We have studied the case where $O$ is a general real orthogonal matrix. In this case the branching ratios of the three generations of $\Sigma_i^0$ and $\Sigma^\pm_i$ are shown in Figs.~\ref{Mix11} and \ref{Mix12}, respectively. For $\Sigma^0_i$ we show the leading visible mode in Fig.~\ref{Mix11}. We found that the NH and IH cases show same parameter spaces for the real orthogonal matrix. For the $\Sigma_i^\pm$ we demonstrate the subdominant $\ell^\pm Z$ and $\ell^\pm h$ cases because they are the visible final states with the charged leptons. 

\begin{figure}[]
\centering
\includegraphics[width=0.31\textwidth]{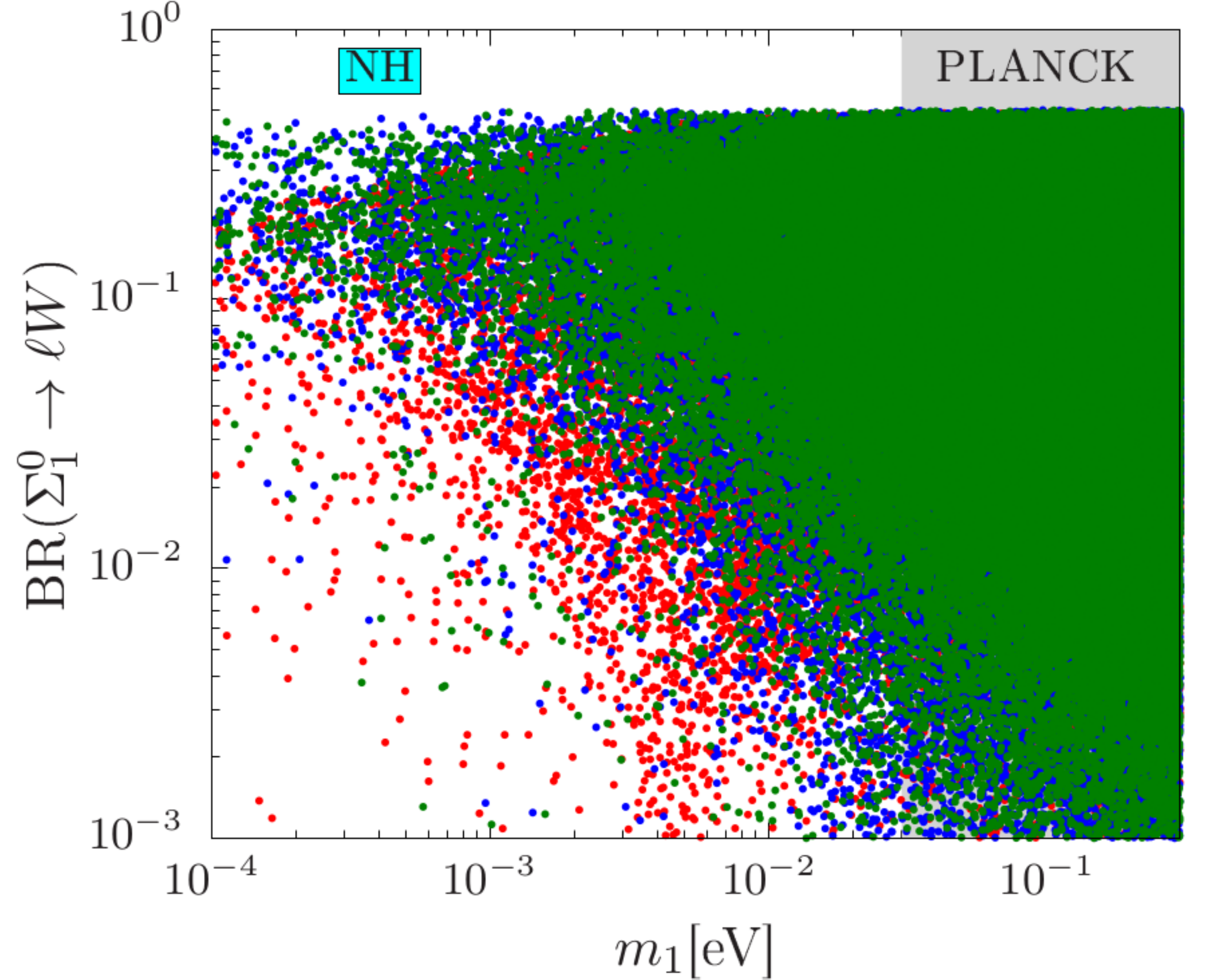}
\includegraphics[width=0.31\textwidth]{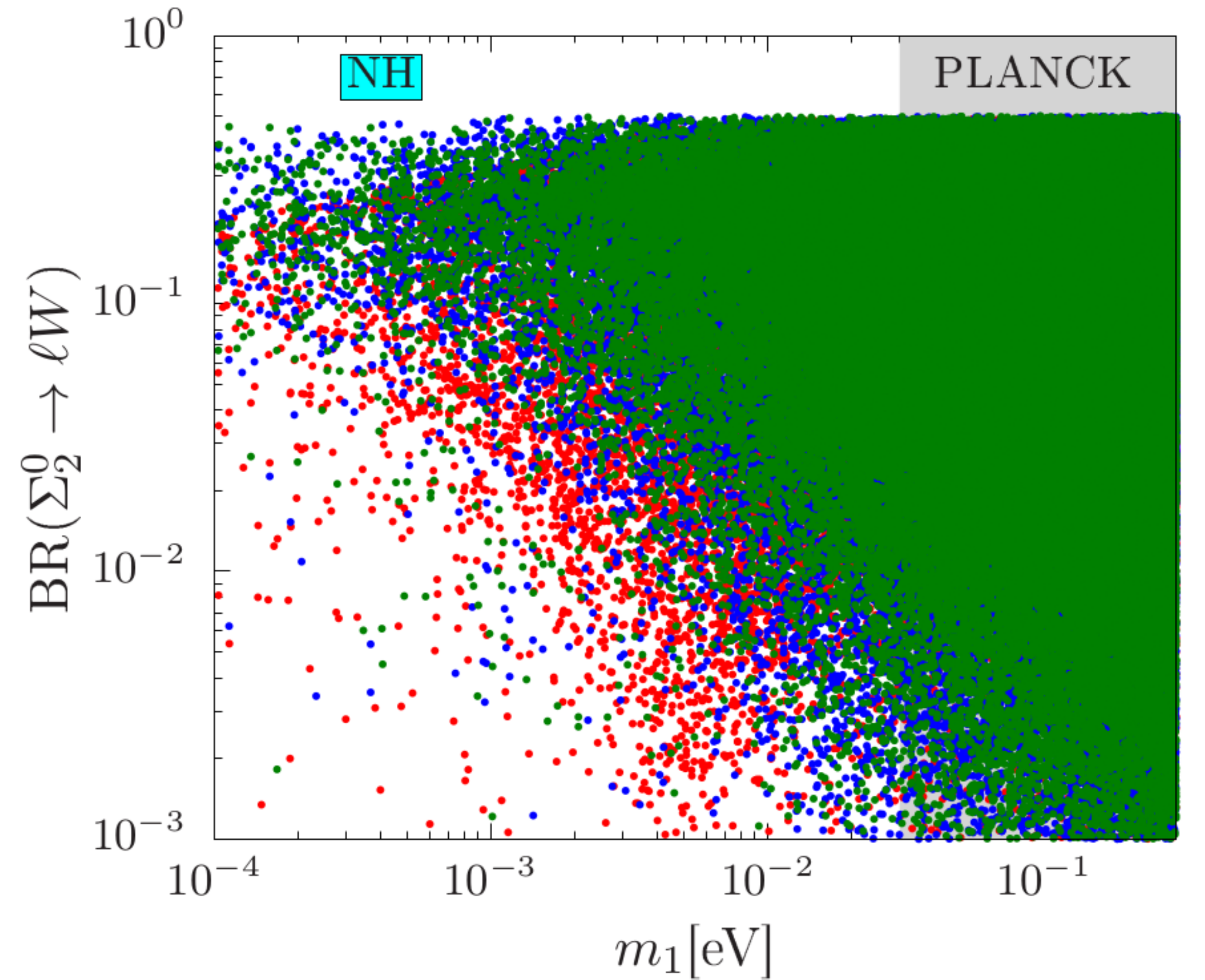}
\includegraphics[width=0.31\textwidth]{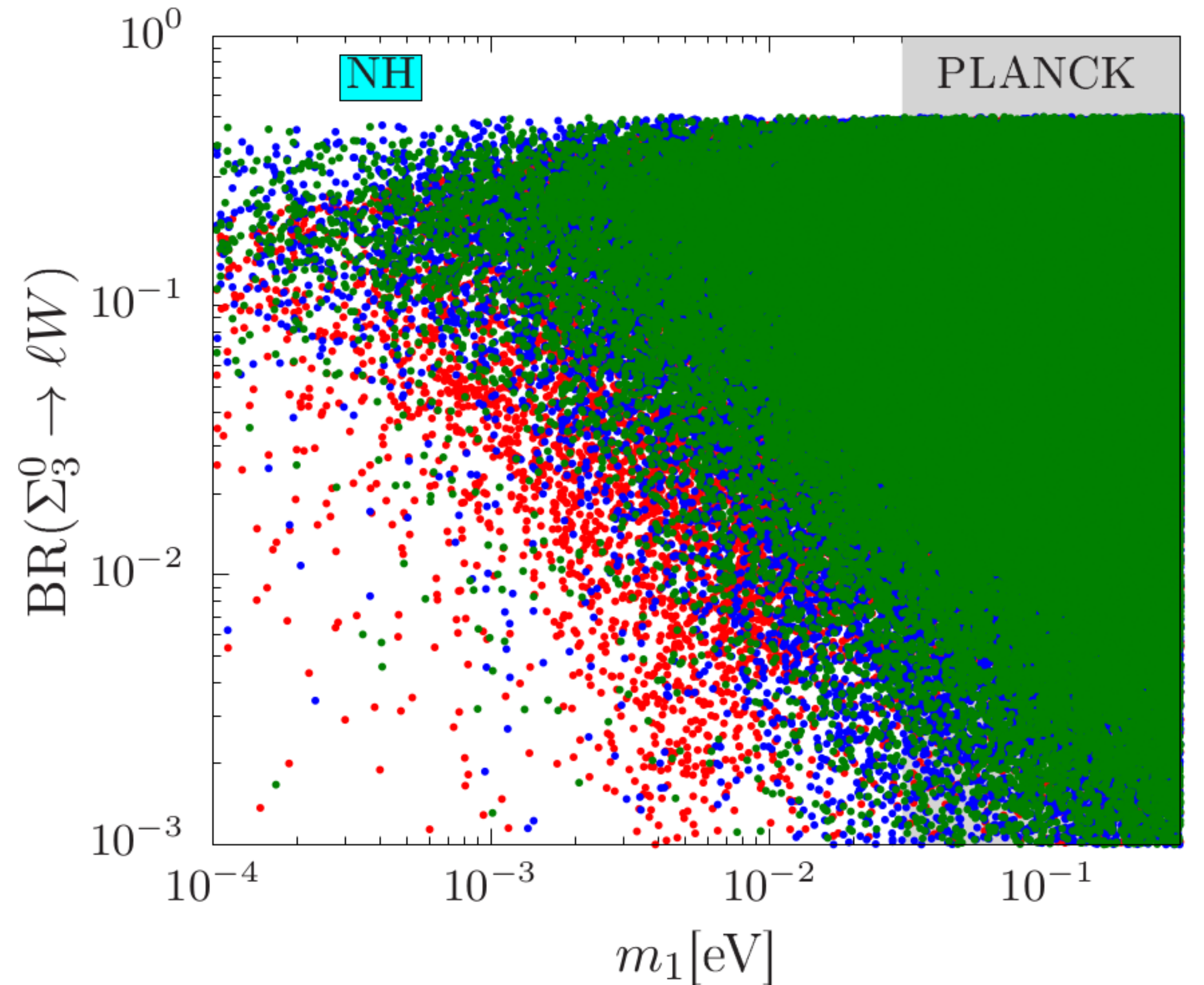}\\
\includegraphics[width=0.31\textwidth]{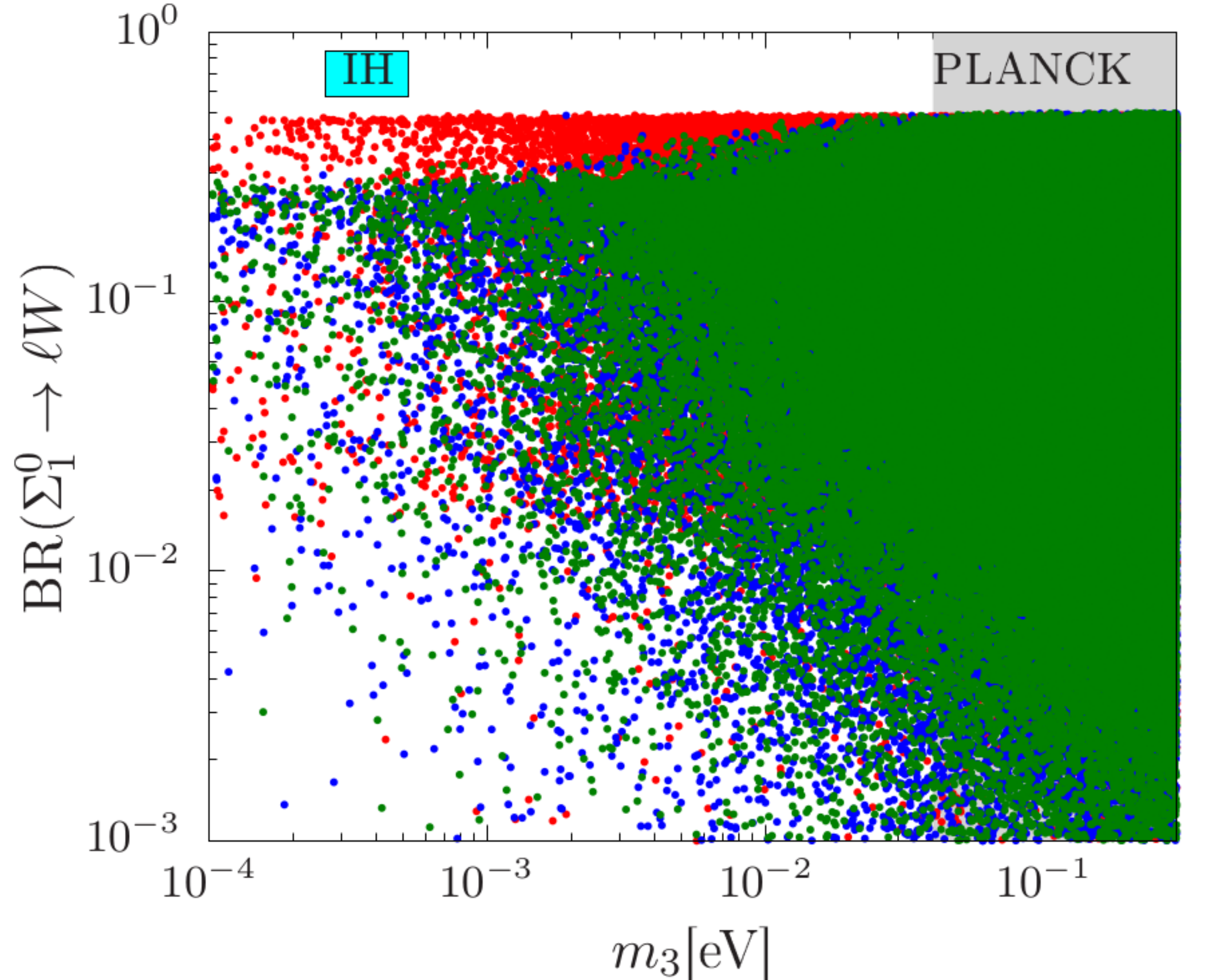}
\includegraphics[width=0.31\textwidth]{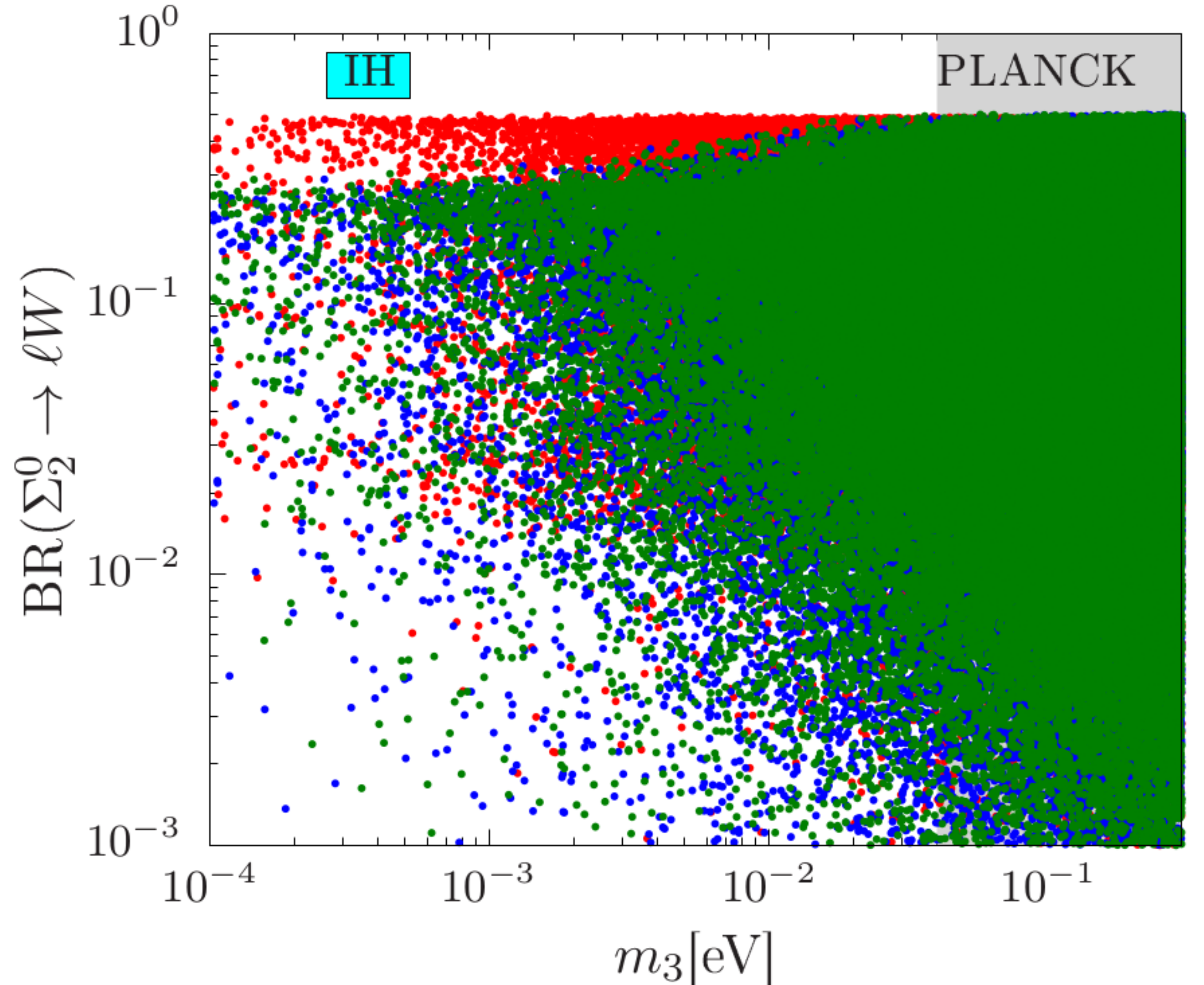}
\includegraphics[width=0.31\textwidth]{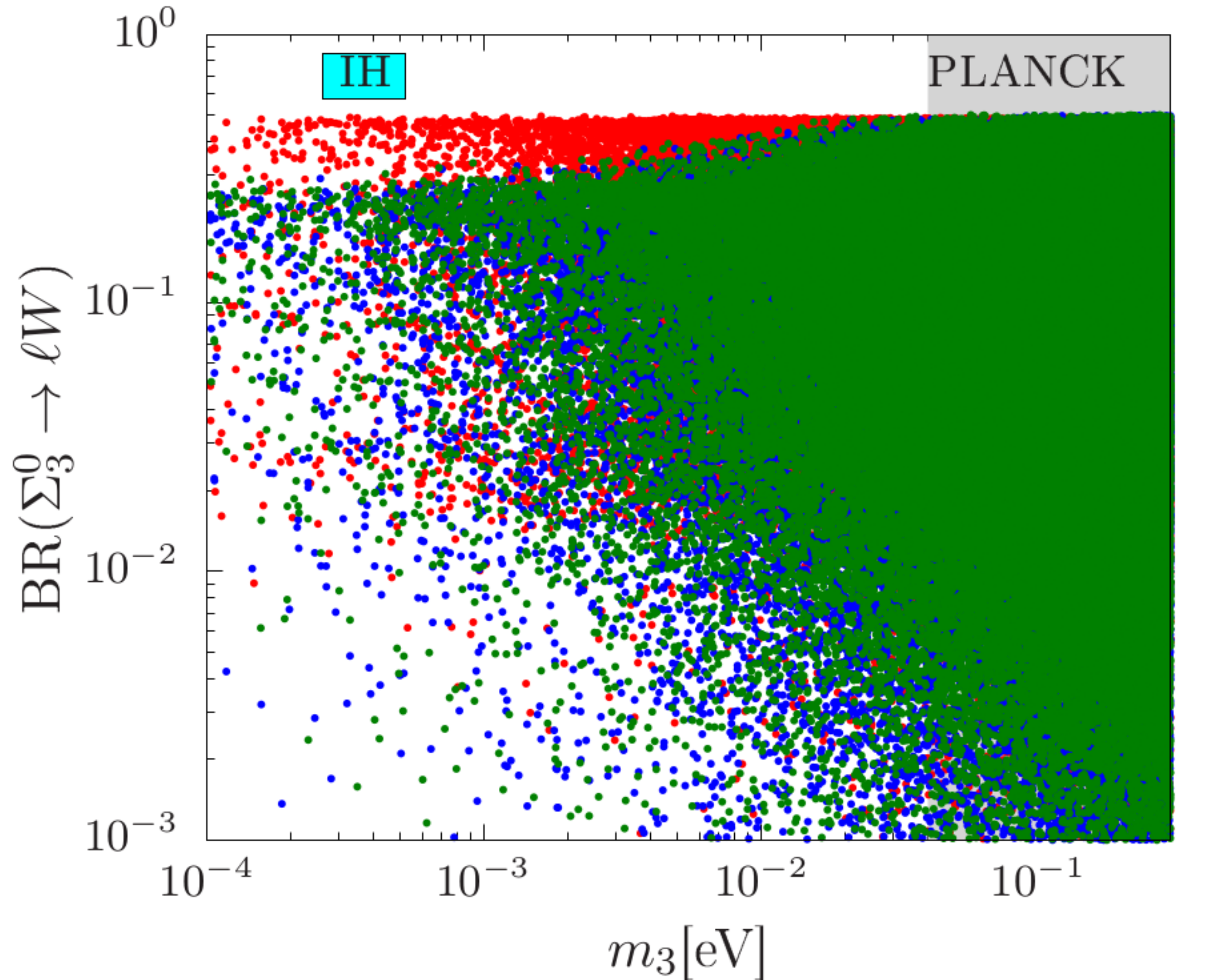}
\caption{Individual branching ratio of $\Sigma^0_i$ into the leading $\ell W$ mode as a function of the lightest neutrino mass $m_1$ $(m_3)$ using the NH (IH) case for the general complex orthogonal matrix in the upper (lower) panel. Red, blue and green color stand for electron, muon and tau modes, respectively. The shaded region in gray is excluded by the PLANCK data. We consider $M=1$ TeV.}
\label{Mix11c}
\end{figure}

We have also studied the case where $O$ is a general complex orthogonal matrix. We show the total branching ratio of the neutral multiplet into the leading mode $(\text{BR} (\Sigma^0_{\text{Tot}} \to \ell W))$ in the bottom-left~(bottom-right) panel of Fig.~\ref{Mix5} for the NH (IH) cases. The muon (blue) and tau (green) modes are dominant over the electron (red) mode of the lepton flavors in the NH case. The IH case is opposite to the NH case. We show the individual branching ratio of the three generations of $\Sigma_i^0$ for the NH (IH) case in the upper(lower) panel of the Fig.~\ref{Mix11c}. The decay of the three generations of the triplets into the electron dominate over the decay mode into the other two leptons in the IH case whereas the result is opposite in the NH case. Here we would like to comment that we do not show the individual or total branching ratio into the different modes for the $\Sigma^{\pm}_i$s because they will have exactly the same repertoire like the $\Sigma_i^0$s when $O$ is a complex orthogonal matrix.    
\section{Implications on the displaced decay of the triplet fermions}
\label{DV}
\begin{figure}[]
\centering
\includegraphics[width=0.49\textwidth]{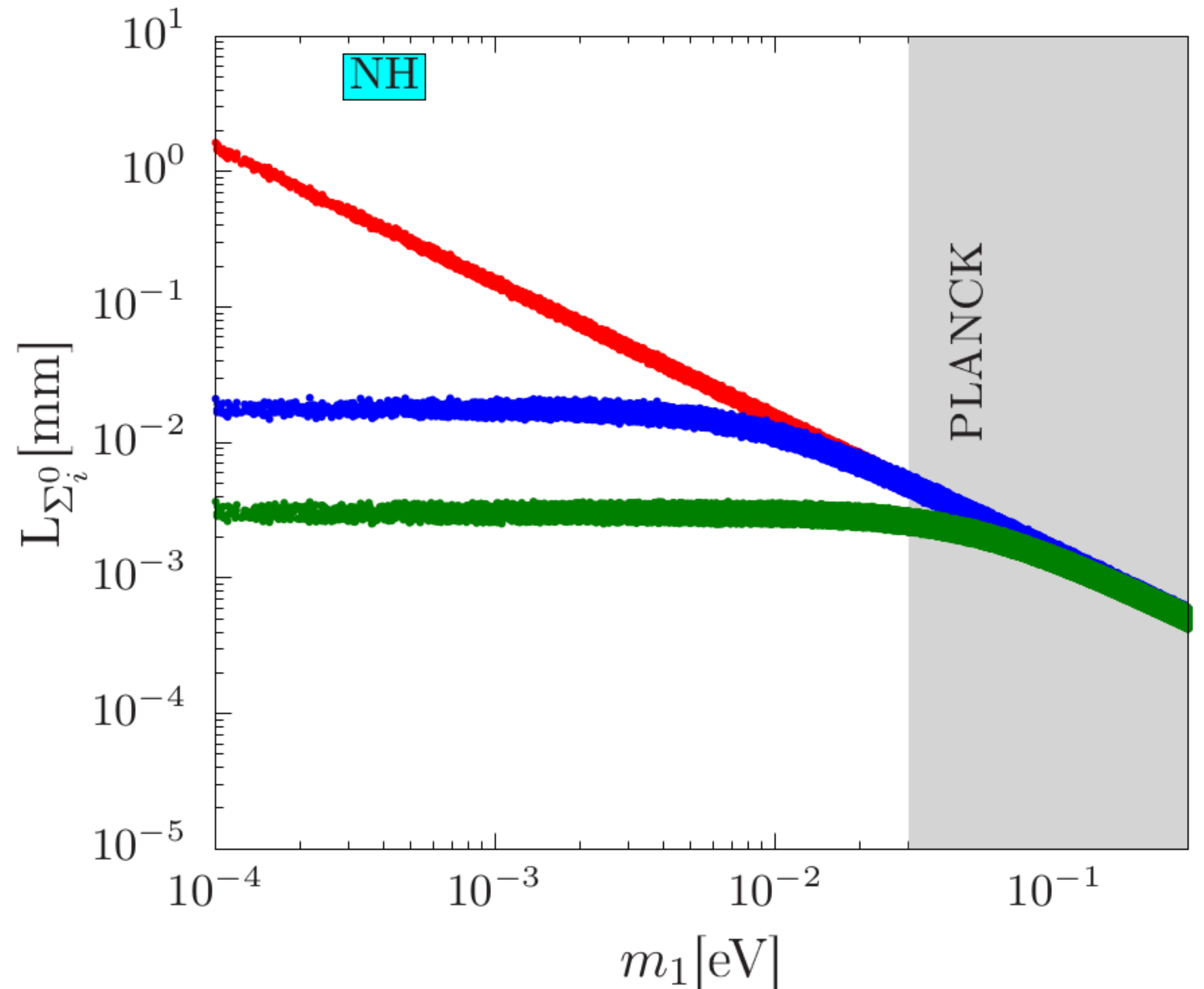}
\includegraphics[width=0.49\textwidth]{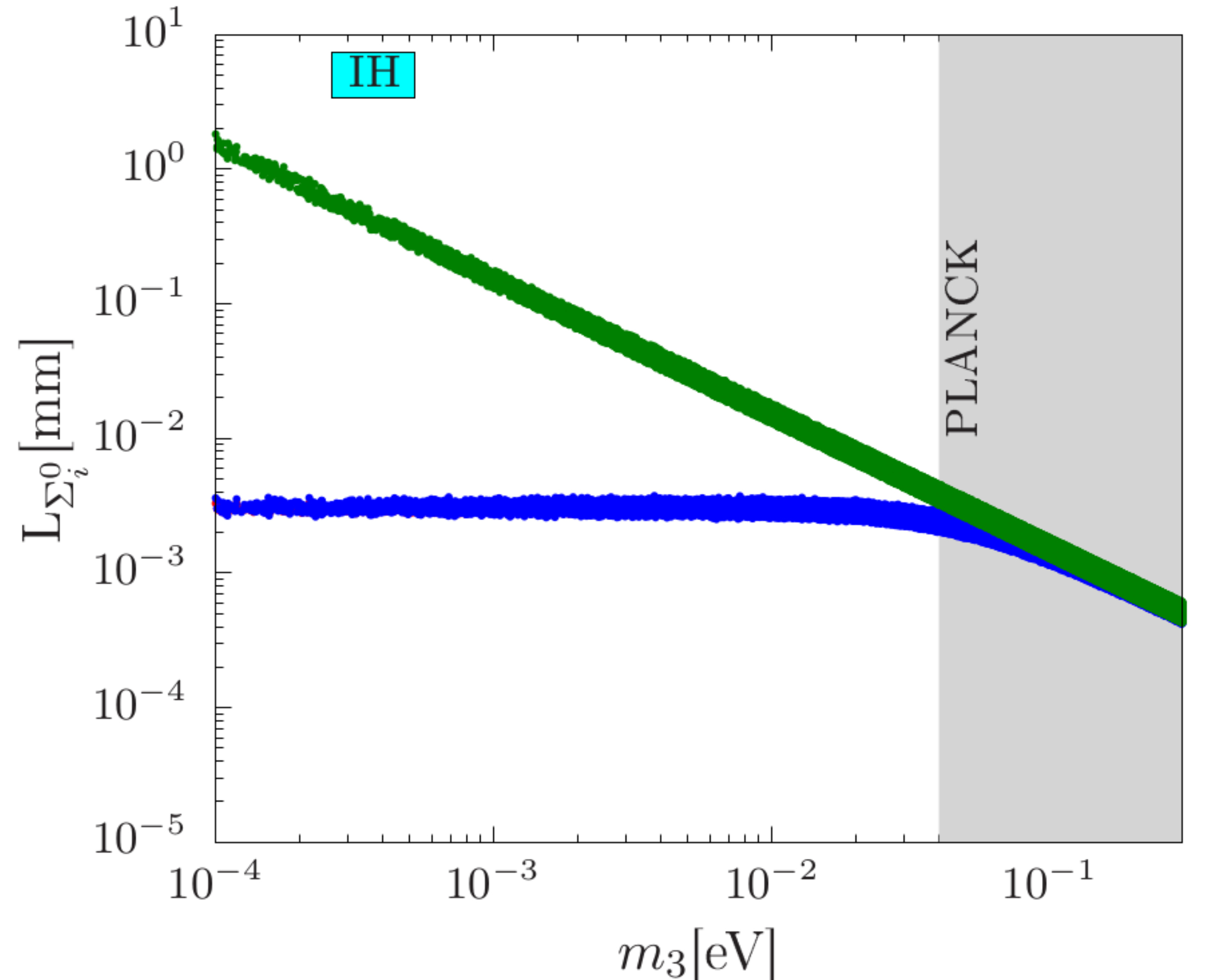}\\
\includegraphics[width=0.49\textwidth]{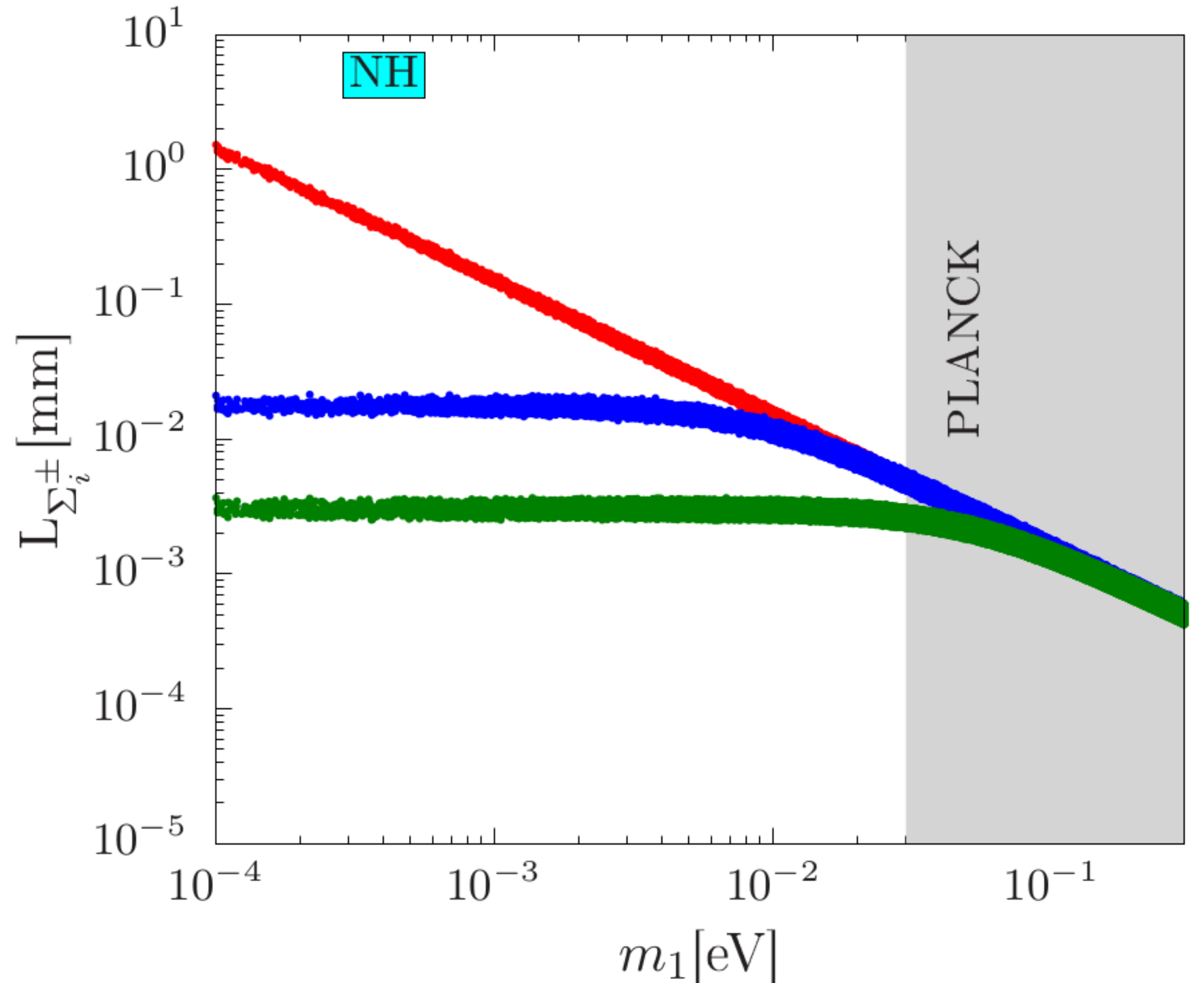}
\includegraphics[width=0.49\textwidth]{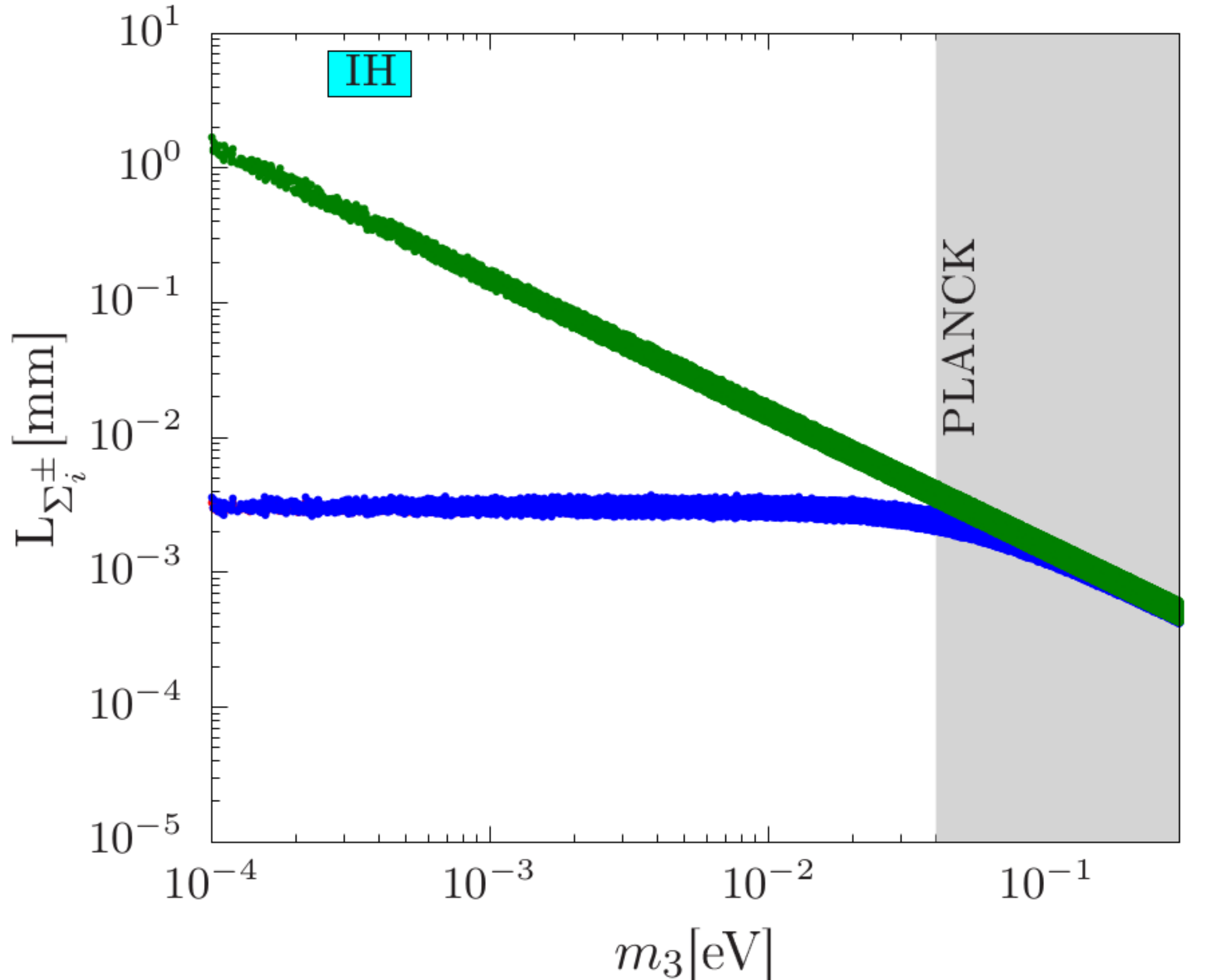}
\caption{Proper decay length of $\Sigma^0_i (\Sigma^{\pm}_i)$ for $O=\textbf{1}_{3\times3}$ with respect to the lightest neutrino mass in the upper (lower) panel.
We show the NH (IH) case in the left (right) panel using the neutrino oscillation data in Eq.~\ref{data}. The first generation triplet is represented by the red band, the second generation is represented by blue band and the third generation is represented by green band respectively. We consider $M=1$ TeV. The shaded region is excluded by the PLANCK data.}
\label{L1}
\end{figure}

We can write the proper decay lengths of the $\Sigma_i^0$ and $\Sigma_i^\pm$ in milimeter for the NH and IH cases as follows:
\bea
\text{L}^{\Sigma_i^0{^{\text{NH/IH}}}} = \frac{1.97\times10^{-13}}{\Gamma_{\Sigma_i^0}^{\text{NH/IH}} [\text{GeV}]} [\text{mm}] \,\, \, \text{and} \,\, \,
\text{L}^{\Sigma_i^\pm{^{\text{NH/IH}}}} = \frac{1.97\times10^{-13}}{\Gamma_{\Sigma_i^\pm}^{\text{NH/IH}} [\text{GeV}]} [\text{mm}] 
\label{decaylength}
\eea
where $i$ stands for the three generations of the triplets.
In this analysis we consider three types of the general orthogonal matrix $(O)$ as described in Sec.~\ref{CI}.
When $O$ is an identity matrix the proper decay lengths are shown in Fig.~\ref{L1} for the NH (IH) case
with respect to the lightest neutrino mass $m_1(m_3)$. The decay lengths of the $\Sigma_i^0 (\Sigma_i^\pm)$
are shown in the upper (lower) panel of Fig.~\ref{L1}. 
In the NH (upper, left panel) case we see that the proper decay length of $\Sigma_1^0$ becomes inversely proportional to $m_1$ which has been represented in red.
The proper decay lengths for the other two generations $\Sigma_2^0$ represented be blue and $\Sigma_3^0$ represented by green become constant 
when $m_1 < 10^{-2}$ eV. We estimate that for $m_1= 10^{-4}$ eV, $\text{L}_{\Sigma_1^0}^\text{max}\sim 1.5$ mm whereas that for 
$\Sigma_2^0$ $(\Sigma_3^0)$ is two (three) orders of magnitude less. This nature of $\text{L}_{\Sigma_1^0}$ can be realized from the Eq.~\ref{m1}.
The mixings between $\Sigma_1$ and the SM leptons, $|V_{\ell \Sigma_1}|^2$, are proportional to $m_1$.
Therefore when $m_1 \to 0$ the corresponding decay length of $\Sigma_1^0$ becomes very large. 
We have also tested this nature considering the lightest light neutrino mass $m_1$ $(m_3)$ for the NH (IH) case at $10^{-6}$ eV and $10^{-10}$ eV respectively.
The results are shown in the first row of Tab.~\ref{tab3}. The corresponding lengths are two and six orders magnitude larger than that for $m_1=10^{-4}$ eV.  
In the IH (upper, right panel)case we have the same scenario for the $\Sigma_3^0$ where as decay lengths of $\Sigma_1^0$ and $\Sigma_2^0$ coincide.
For $\Sigma_3^0$ we notice the form of $|V_{\ell \Sigma_3}|^2$, proportional to $m_3$, in Eq.~\ref{m1}.
Therefore the decay length of $\Sigma_3^0$ becomes very large when $m_3$ is very small and the corresponding benchmarks are given in the second row of Tab.~\ref{tab3}. 
Lower panel of Fig.~\ref{L1} shows that at least for lightest neutrino mass $m_1(m_3)> 10^{-4}$ eV, the decay length of $\Sigma^{\pm}_1$ $(\Sigma^\pm_3)$ in NH(IH) case has the same nature as the decay length of $\Sigma^{0}_1$ $(\Sigma^0_3)$ in NH(IH) case. For the lightest neutrino mass range $m_1(m_3)\leq 10^{-4}$ eV, the behavior of the decay length of $\Sigma^{\pm}_1$ $(\Sigma^\pm_3)$ in NH(IH) case is completely different from the decay length of $\Sigma^{0}_1$ $(\Sigma^0_3)$ in NH(IH) case, see the third and fourth rows of Tab.~\ref{tab3}. This implies that for $m_1(m_3)\leq 10^{-4}$ eV, $\text{L}_{\Sigma_1^\pm}(\text{L}_{\Sigma_3^\pm})$ is more or less constant. The reason for this is, in NH(IH) case as $m_1(m_3)\to 0$, mixing angle $|V_{\ell\Sigma_1}|^2(|V_{\ell\Sigma_3}|^2)\to 0$ and as a result the decay width for $\Sigma_1^\pm(\Sigma_3^\pm)$ will be dominated by the decay modes given in Eq.~\ref{decay3} which is controlled by the $\Delta M$ parameter. Hence this decay width or decay length is constant which can be noted from the benchmarks in the third or fourth row of Tab.~\ref{tab3}. We notice that in this case one can obtain large decay lengths which indicate possibilities of the displaced vertex scenarios when the decay lengths are $\mathcal{O} (100 \text{mm})$. Possible scenarios of the further long-livedness can also be observed when the decay lengths are $\mathcal{O}(10^{6} \text{mm})$.
\begin{table*}[!htbp]
	\begin{tabular}{|c|c|c|} 
		\hline
		Decay Length~[mm] & $m_{\text{lightest}}=10^{-6}$ eV & $m_{\text{lightest}}=10^{-10}$ eV \\ \hline
		L$_{\Sigma_1^0}$~(NH) & \big[134.13, 171.03\big]   & \big[$1.35\times 10^{6}$, $1.74\times 10^{6}$\big]   \\
		L$_{\Sigma_3^0}$~(IH) & \big[129.04, 183.71\big]   & \big[$1.28\times 10^{6}$, $1.83\times 10^{6}$\big]   \\
		L$_{\Sigma_1^\pm}$~(NH) & \big[20.29, 20.99\big]   & \big[23.9321, 23.9322\big]   \\
		L$_{\Sigma_3^\pm}$~(IH) & \big[20.18, 21.17\big]   & \big[23.9321, 23.9322\big]   \\
		\hline
	\end{tabular}
\caption{Benchmark for the proper decay lengths of $\Sigma_1^{0, \pm}$ $(\Sigma_3^{0,\pm})$ for the NH and IH cases fitting the neutrino oscillation data in Eq.~\ref{data} when $O=\textbf{1}_{3\times 3}$.The variation of the proper decay length represents a band due to the variation of $\pm3\sigma$ the oscillation data, $\delta_{\text{CP}}$ and $\rho_i$. We consider $M=1$ TeV.}
\label{tab3}
\end{table*}  

Similarly we consider the case when $O$ is a real orthogonal matrix. In this case the analytical form for mixings are given in Eq.~\ref{m4}-\ref{m6}.
We notice that now $|V_{\ell\Sigma_1}|^2$ depends on all the light neutrino mass eigenvalues like $m_1$, $m_2$ and $m_3$. Therefore in the NH case for $m_1 \to 0$, $|V_{\ell \Sigma_1}|^2$ attains a 
limiting value but does not vanish. Which will be reflected in the nature of the proper decay lengths of $\Sigma^0_i$ and $\Sigma^\pm_i$ respectively.
Similar behavior can be observed for the IH case when $m_3 \to 0$, $|V_{\ell \Sigma_3}|^2$ does not vanish due to its dependence on $m_1$ and $m_2$. 
The decay lengths of $\Sigma_i^0(\Sigma_i^\pm)$ are shown in the upper (lower) panel of Fig.~\ref{L2}. We find that for lightest light neutrino mass range $10^{-4}\,\text{eV}\leq m_{1(3)}\leq 0.1$ eV, maximum decay length can be around $1$ mm.
\begin{figure}[]
\centering
\includegraphics[width=0.49\textwidth]{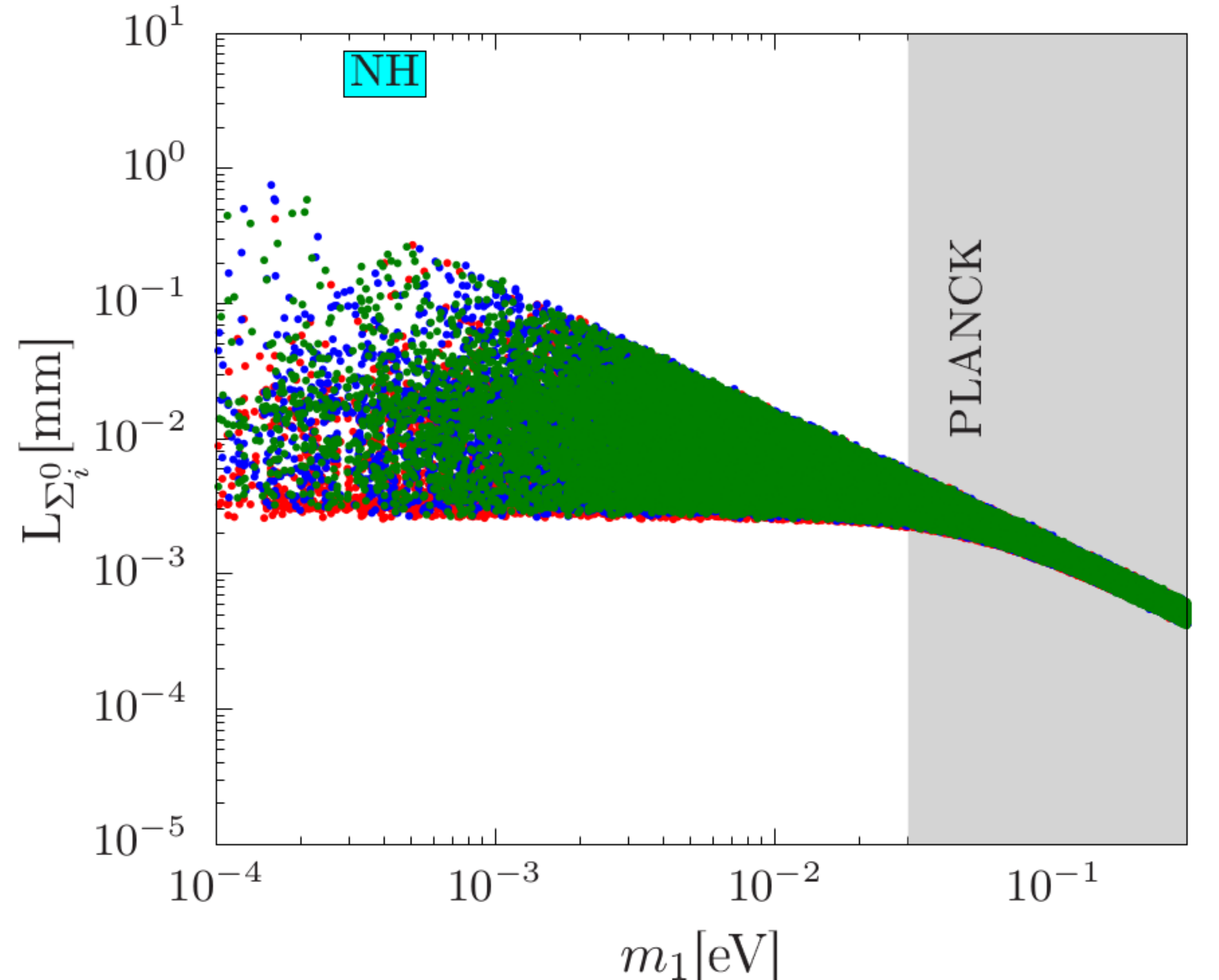}
\includegraphics[width=0.49\textwidth]{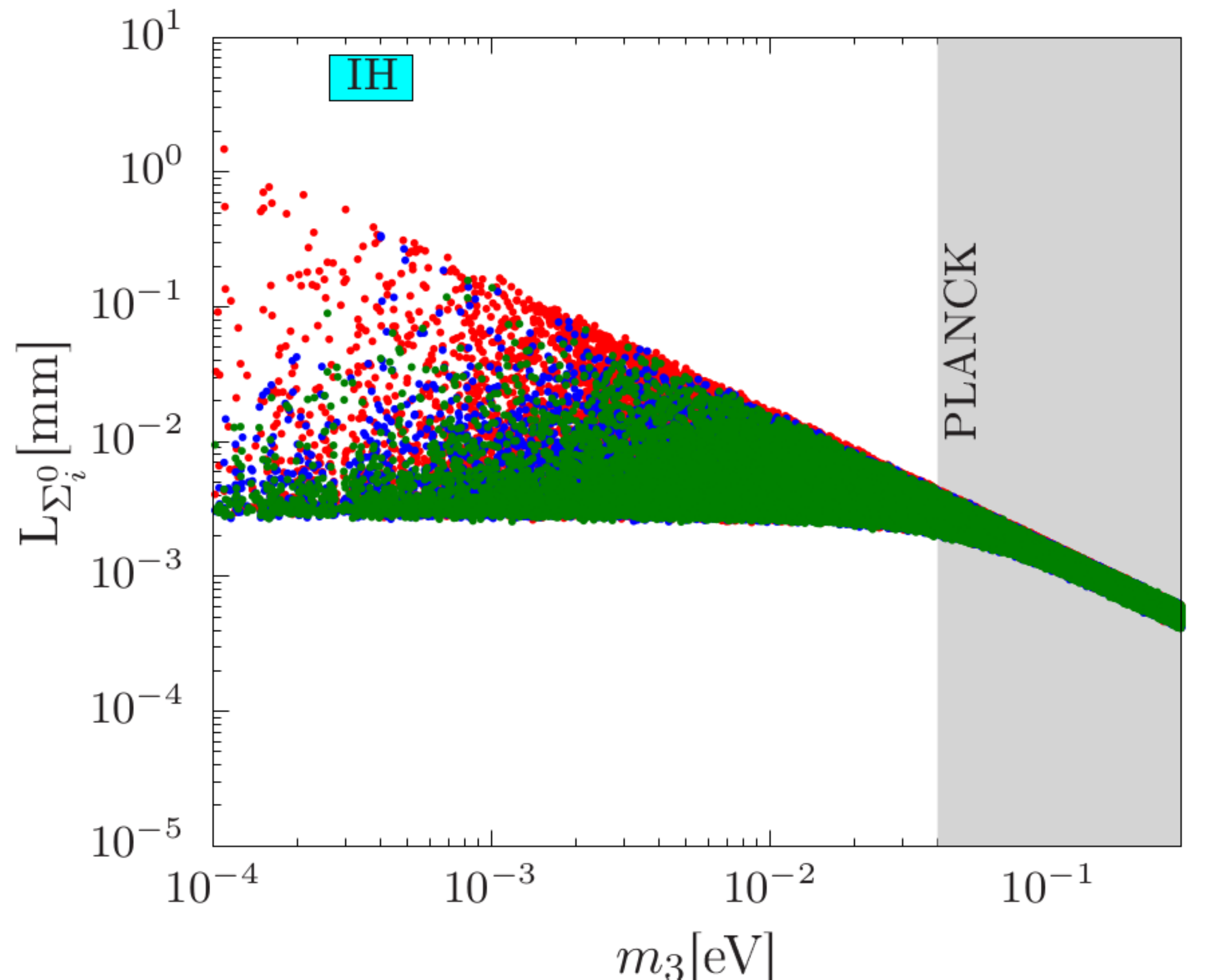}\\
\includegraphics[width=0.49\textwidth]{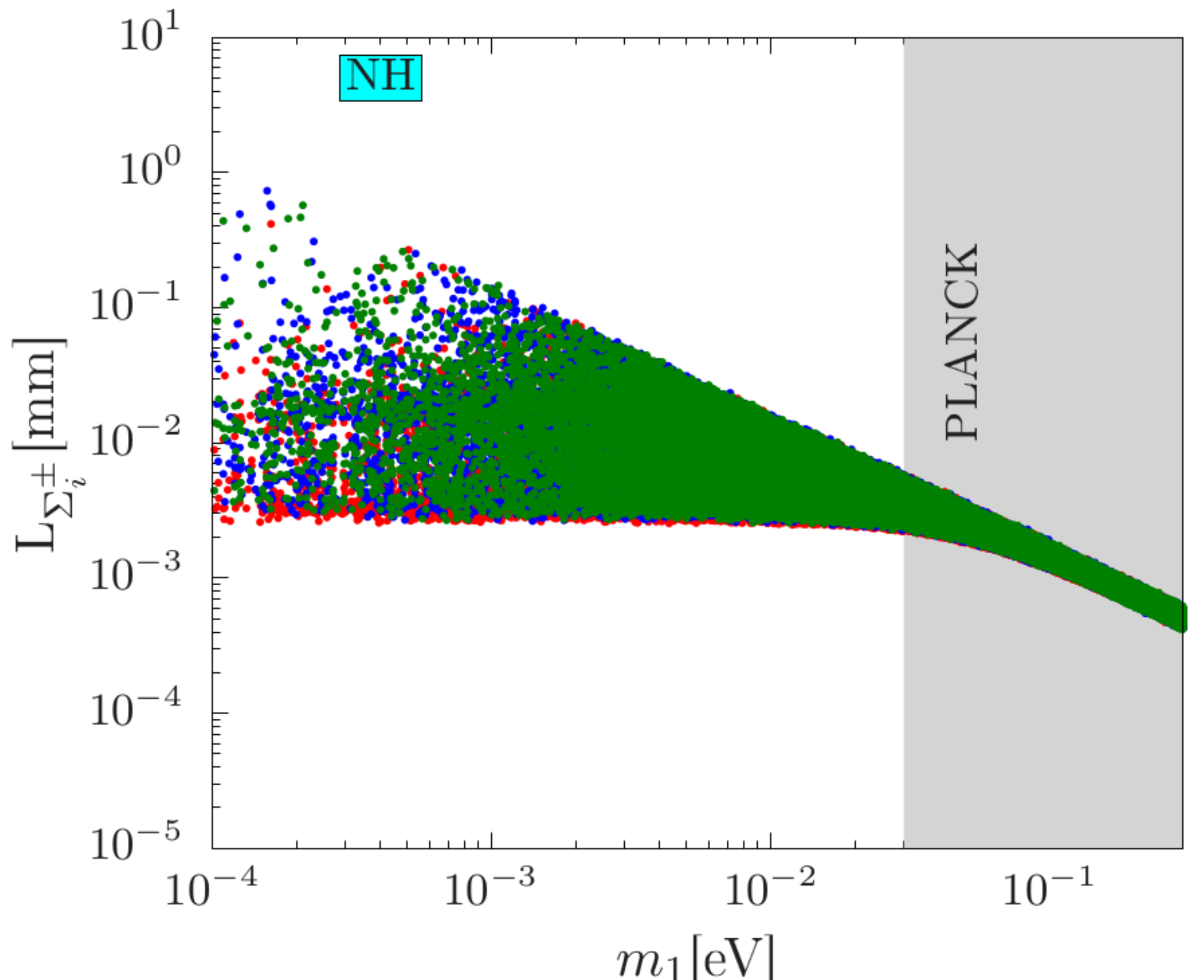}
\includegraphics[width=0.49\textwidth]{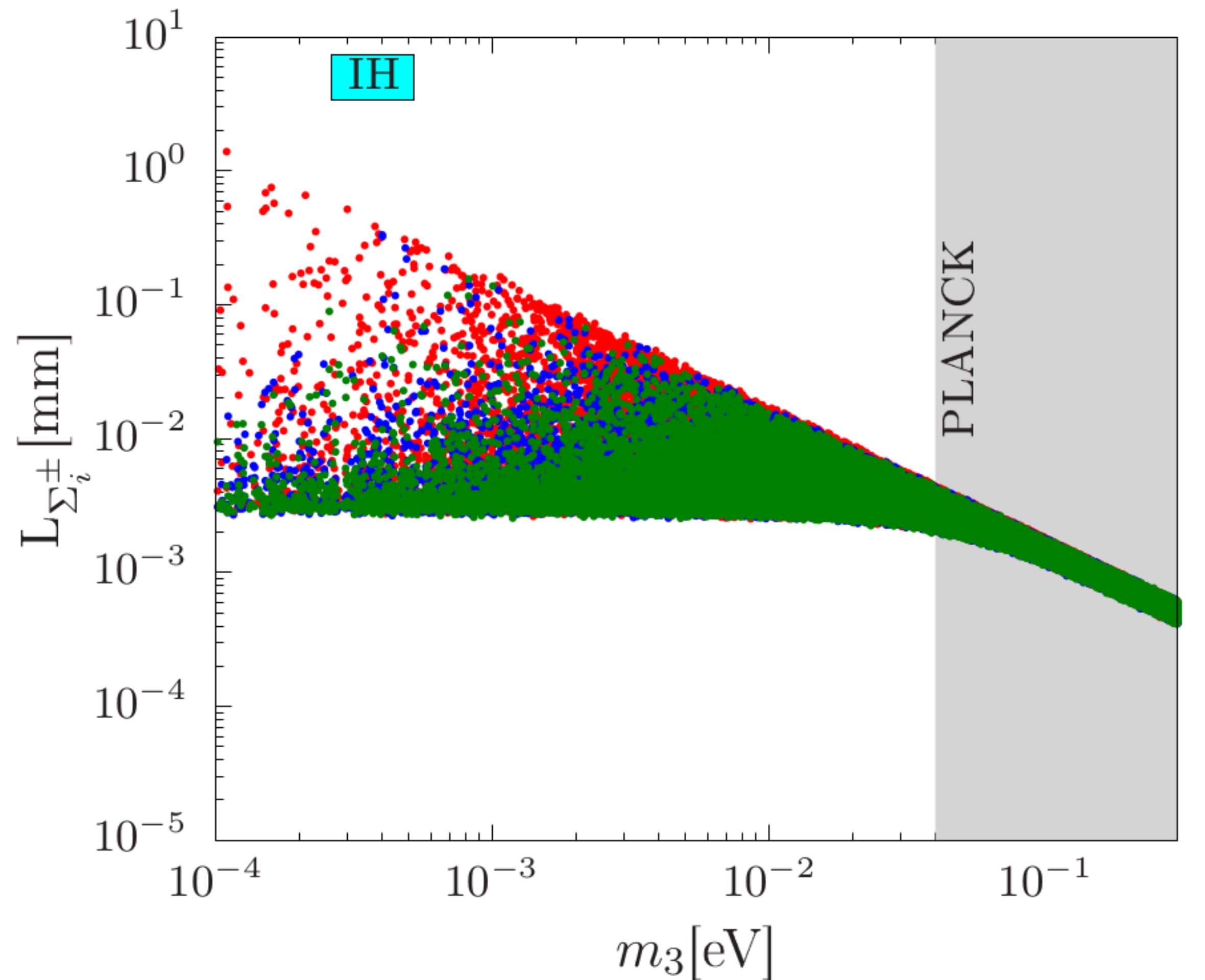}
\caption{Proper decay length of $\Sigma^0_i (\Sigma^{\pm}_i)$ when $O$ is a real and general orthogonal matrix with respect to the lightest neutrino mass in the upper (lower) panel. We show the NH (IH) case in the left (right) panel using the neutrino oscillation data in Eq.~\ref{data}. The first generation triplet is represented by the red band, the second generation is represented by blue band and the third generation is represented by green band respectively. We consider $M=1$ TeV. The shaded region is excluded by the PLANCK data.}
\label{L2}
\end{figure}
We have also considered some benchmark scenarios for very small lightest light neutrino mass, $m_1$ $(m_3)$ for the NH (IH) case. 
We fix $m_1$ $(m_3)$ at $10^{-6}$ eV and $10^{-10}$ eV respectively and find out the corresponding decay lengths in Tab.~\ref{tab4} fitting the neutrino oscillation data from Eq.~\ref{data}. In this case we have found that the minimum decay length can be as low as $\mathcal{O}(10^{-3}\,\text{mm})$ and the maximum decay length are of the same order as the case of identity orthogonal matrix $O$. The decay length can reach at a maximum value of $\mathcal{O}(10^6 \text{mm})$ showing the possibility of a long-lived scenario. When the decay length is $\mathcal{O}(10^{-3} \text{mm})$, the decay of the triplet can be prompt. In that case, a comparatively large mixing can be expected.  
\begin{table*}[!htbp]
	\begin{tabular}{|c|c|c|} 
		\hline
		Decay Length~[mm] & $m_{\text{lightest}}=10^{-6}$ eV & $m_{\text{lightest}}=10^{-10}$ eV \\ \hline
		L$_{\Sigma_1^0}$~(NH) & \big[0.0027, 171.1\big]   & \big[0.0028, $1.74\times 10^6$\big]   \\
		L$_{\Sigma_3^0}$~(IH) & \big[0.0026, 183.79\big]   & \big[0.0026, $1.84\times 10^6$\big]   \\
		L$_{\Sigma_1^\pm}$~(NH) & \big[0.0029, 20.84\big]   & \big[0.0025, $23.93$\big]   \\
		L$_{\Sigma_3^\pm}$~(IH) & \big[0.0027, 21.18\big]   & \big[0.0027, 23.93\big]   \\
		\hline
	\end{tabular}
\caption{Benchmark for the proper decay lengths of $\Sigma_1^{0, \pm}$ $(\Sigma_3^{0,\pm})$ for the NH and IH cases fitting the neutrino oscillation data in Eq.~\ref{data} when $O$ is a real and general orthogonal matrix. The variation of the proper decay length represents a band due to the variation of $\pm3\sigma$ the oscillation data, $\delta_{\text{CP}}, \rho_i$ and the parameters of the orthogonal matrix . We consider $M=1$ TeV.}
\label{tab4}
\end{table*}  

We have also studied the effect when $O$ is a complex orthogonal matrix. The real and imaginary parts of the elements of $O$ vary between $[-\pi, \pi]$.
Scanning over the $\delta_\text{CP}$ and $\rho_{1,2}$ within the interval $[-\pi, \pi]$ simultaneously we show the range of the proper decay length for some benchmark scenarios 
in Tab.~\ref{tab5} fitting the neutrino oscillation data from Eq.~\ref{data}. We adopt such a method for this case because there is no special pattern observed in this case after the scan. Therefore we fix the lightest light neutrino mass $m_1$ $(m_3)$ in the NH (IH) case at $10^{-6}$ eV and $10^{-10}$ eV respectively. Due to the presence of the complex orthogonal matrix there will be an improvement in the light-heavy mixings. As a result we can expect a prompt production of the triplets as expressed by the small decay lengths $\mathcal{O}(10^{-11} \text{mm})$ which represent a large mixing. On the other hand there will be some possibilities where small mixings can be observed and due to that large decay lengths $\mathcal{O}(100 \text{mm})$ can be obtained which ensure a possible displaced vertex scenario and if the decay lengths are $\mathcal{O}(10^{6} \text{mm})$ then a further long-lived case might be studied.    

\begin{table*}[!htbp]
	\begin{tabular}{|c|c|c|} 
		\hline
		Decay Length~[mm] & $m_{\text{lightest}}=10^{-6}$ eV & $m_{\text{lightest}}=10^{-10}$ eV \\ \hline
		L$_{\Sigma_1^0}$~(NH) & \big[$1.11\times 10^{-11}$, 171.2\big]   & \big[$1.08\times 10^{-10}$, $1.74\times 10^6$\big]   \\
		L$_{\Sigma_3^0}$~(IH) & \big[$1.74\times 10^{-11}$, 183.79\big]   & \big[$1.54\times 10^{-11}$, $1.84\times 10^6$\big]   \\
		L$_{\Sigma_1^\pm}$~(NH) & \big[$1.32\times 10^{-10}$, 20.84\big]   & \big[$1.47\times 10^{-10}$, $23.93$\big]   \\
		L$_{\Sigma_3^\pm}$~(IH) & \big[$1.41\times 10^{-11}$, 21.18\big]   & \big[$8.23\times 10^{-12}$, 23.93\big]   \\
		\hline
	\end{tabular}
\caption{Benchmark for the proper decay lengths of $\Sigma_1^{0, \pm}$ $(\Sigma_3^{0,\pm})$ for the NH and IH cases fitting the neutrino oscillation data in Eq.~\ref{data} when $O$ is a complex and general orthogonal matrix. The variation of the proper decay length represents a band due to the variation of $\pm3\sigma$ the oscillation data, $\delta_{\text{CP}}, \rho_i$ and the parameters of the orthogonal matrix. We consider $M=1$ TeV.}
\label{tab5}
\end{table*} 
\section{Implications on the collider searches}
\label{bds}
The triplets involved in the type-III seesaw mechanism are being studied at the LHC at different center of mass energies.
We also consider to give a discussion on the possible bounds and current limits on the light-heavy mixings which are important for the neutrino mass generation mechanism.
We consider these bounds to select an allowed triplet mass to study the discovery potential of the $\Sigma^\pm$ and $\Sigma^0$ 
at the hadron colliders at $\sqrt{s}=13$ TeV, $27$ TeV and $100$ TeV respectively. 
Finally we briefly discuss about the discovery potentials at the electron-positron $(e^- e^+)$ and electron-proton$(e^-p)$ colliders.
\begin{figure}[]
\centering
\includegraphics[width=1\textwidth]{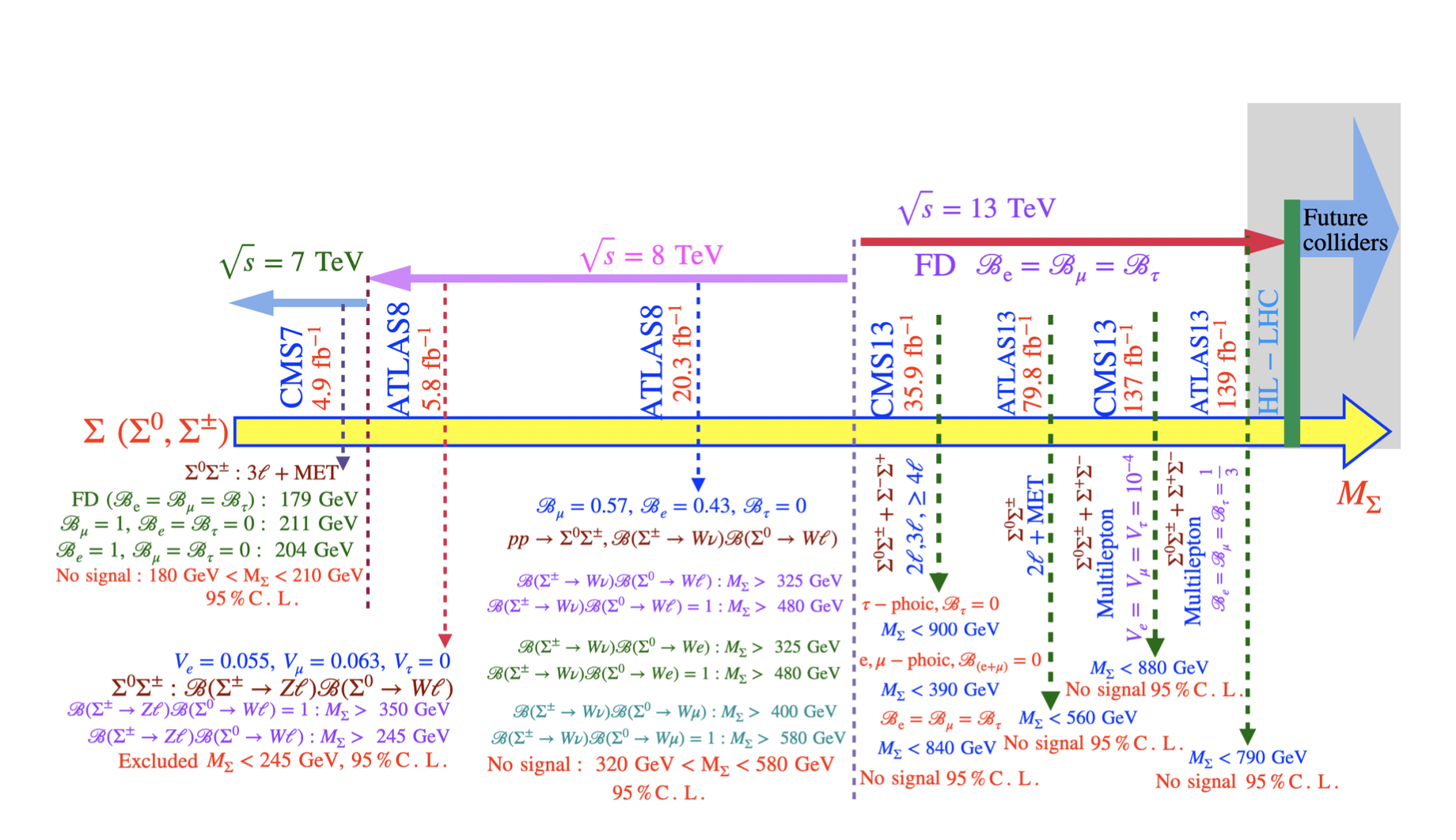}
\caption{The exclusion limits and the conditions on the triplet fermions from the LHC at $7$ TeV, $8$ TeV and $13$ TeV center of mass energies. 
The lower bounds CMS7\cite{CMS:2012ra}, ATLAS8 at $5.8$ fb$^{-1}$ luminosity \cite{ATLAS:2013hma},  ATLAS8 at $20.3$ fb$^{-1}$ luminosity \cite{Aad:2015cxa}, CMS13 at $35.9$ fb$^{-1}$ luminosity \cite{Sirunyan:2017qkz}, ATLAS13 at $79.8$ fb$^{-1}$ luminosity \cite{ATLAS:2018ghc}, CMS13 at $137$ fb$^{-1}$ luminosity \cite{Sirunyan:2019bgz} and ATLAS13 at $139$ fb$^{-1}$ luminosity \cite{Aad:2020fzq} considered the FD scenarios followed by the multilepton decay modes of the $\Sigma^\pm$ and $\Sigma^0$ respectively. HL-LHC and future colliders can set some stronger bounds in the near future.}
\label{L4}
\end{figure}
\subsection{LHC bounds}
A comprehensive roadmap has been pictorially represented in Fig.~\ref{L4} where we have shown modes of the triplet fermions, different exclusion bounds at different stages of LHC with some salient features of the signal and model parameters. The High-Luminosity LHC (HL-LHC) and the future colliders have no results so far and kept in the gray area. At the $7$ TeV LHC the CMS studied the $3$ lepton final state from the $\Sigma^0\Sigma^\pm$ process with $4.9$ fb$^{-1}$ luminosity \cite{CMS:2012ra} with Flavor Diagonal (FD) Yukawa coupling with equal branching ratios $(\mathcal{B})$ for the three flavors of the charged leptons, $\mathcal{B}_e=\mathcal{B}_\mu=\mathcal{B}_\tau$. CMS has also studied the cases with $100\%$ decay into the electron and muon flavors. Finally sets a lower bound the triplet mass at $M_\Sigma < 180$ GeV. At the $8$ TeV the ATLAS searched for the triplet fermion at $5.8$ fb$^{-1}$ luminosity and sets a lower bound $M_\Sigma < 245$ GeV at $95\%$ C.L \cite{ATLAS:2013hma} where they consider a two flavor case with electron and muon with corresponding mixings $V_e=0.055$ (with electron) and $V_\mu=0.063$ (with muon) and no mixing with the tau flavor. At the $8$ TeV with $20.3$ fb$^{-1}$ luminosity, ATLAS studied another scenario where the branching ratio into electron and muon are $57\%$ and $43\%$ respectively whereas no branching to tau flavor was considered \cite{Aad:2015cxa}. This analysis sets a lower limit of $320$ GeV on $M_\Sigma$ and finally finding no signature within $320$ GeV $< M_\Sigma < 580$ GeV at the $95\%$ C. L. 

At the $13$ TeV the LHC studied the flavor democratic scenario of the type-III scenario \cite{Sirunyan:2017qkz} where the branching ratio into the three lepton flavors are same and each branching ratio $(\mathcal{B}_\ell)$ is proportional to $\frac{|V_\ell|^2}{|V_e|^2+ |V_\mu|^2+ |V_\tau|^2}$ where $\ell$ represents the charged lepton flavor according to \cite{Sirunyan:2017qkz} from CMS. In this article the $\Sigma^\pm \Sigma^0$ and $\Sigma^+\Sigma^-$ productions have been studied for the multi-lepton final states. The bound on the triplet mass has been obtained at $840$ GeV at the $95\%$ CL for $35.9$ fb$^{-1}$. It has also been mentioned that limit on the $\tau$-phobic case where the branching ratio of the triplet fermion is set to be zero $(\mathcal{B}_\tau =0)$ sets a limit on the triplet mass $M_\Sigma= 900$ GeV at the $90 \%$ C. L. Studying the combined triplet production mode $\Sigma^\pm \Sigma^0$ and $\Sigma^+ \Sigma^-$ at the $13$ TeV LHC with $79.8$ fb$^{-1}$ luminosity ATLAS finds a bound on $M_\Sigma$ at $560$ GeV \cite{ATLAS:2018ghc} at the $95\%$ C. L. using the flavor democratic scenario. In this analysis a final state includes $e$ and $\mu$ flavors of two leptons with opposite and same sign combinations in association with the missing momentum and jets. In this case $\Sigma^\pm$ and $\Sigma^0$ dominantly decay into the modes containing $W^\pm$ and leptons $(\ell^\mp$ or $\nu)$. The missing momentum is mostly coming from the neutrinos and the jets are coming from the hadronic decay of the remaining $W^\pm$ decay obtained from the dominant decay of the triplet. Updates from both of CMS \cite{Sirunyan:2019bgz} and ATLAS \cite{Aad:2020fzq} at the $13$ TeV LHC using $137$ fb$^{-1}$ and $139$ fb$^{-1}$ luminosity find limits on $M_\Sigma$ at $880$ GeV and $790$ GeV at the $95\%$ C. L. respectively. In both of these cases a democratic flavor structure has been considered. The CMS considered a universal mixing of $10^{-4}$ and the ATLAS considered a scenario where the branching ratio of the triplet into each of the three flavors of the lepton is $\frac{1}{3}$.
 
\subsection{Limits on the mixing angles}
The global constraints on the triplet fermions can be found in \cite{Biggio:2019eeo}. 
The upper limits on the elements of symmetric $|V_{\ell \Sigma}|^2$ matrix have been calculated at the $2$-$\sigma$ level for the three generations of the triplets for the NH and IH cases.
The upper bounds on the matrix elements for the NH case are $|V_{11}|^2 < 6.2 \times 10^{-4}$, $|V_{22}|^2 < 2.8 \times 10^{-4}$, $|V_{33}|^2 < 1.3 \times 10^{-3}$, $|V_{12}|^2 < 6.0 \times 10^{-4}$, $|V_{31}|^2 < 5.0 \times 10^{-7}$, $|V_{32}|^2 < 2.8 \times 10^{-4}$. Similarly the upper bounds on the matrix elements for the IH case are $|V_{11}|^2 < 6.4 \times 10^{-4}$, $|V_{22}|^2 < 2.2 \times 10^{-4}$, $|V_{33}|^2 < 7.8 \times 10^{-4}$, $|V_{12}|^2 < 6.0 \times 10^{-7}$, $|V_{31}|^2 < 4.6 \times 10^{-4}$, $|V_{32}|^2 < 2.6 \times 10^{-4}$.

We consider a type-III seesaw scenario where the triplets are as heavy as $1$ TeV. In this scenario we first consider $O$ as a $3\times 3$ identity matrix to calculate the light-heavy mixing in terms of the lightest light neutrino mass. We calculate the bounds on the mixing for two hierarchic conditions of the neutrino mass, namely, NH and IH fitting the neutrino oscillation data. For identity or real orthogonal matrix $O$, observing the nature of the mixing summed over the triplet generation we notice that the mixing can reach up to a certain lower limit when varied with respect to the lightest light neutrino mass for the NH and IH cases under the PLANCK limit. The lower limit for the $\Sigma_i |V_{e \Sigma_i}|^2$ can go down to $3\times 10^{-15}$
whereas the upper limit can be one order of magnitude better under the PLANCK exclusion in the NH case. For the other two flavors in the NH case, the upper limit on the 
mixings can reach up to $5\times 10^{-14}$ whereas the lower limit is slightly better for them, namely $2\times 10^{-14}$ under the PLANCK exclusion. 
Alternatively if we look at the IH case, we notice that the limits on $\Sigma_i |V_{(\mu, \tau) \Sigma_i}|^2$ roughly remain the same, however, those on $\Sigma_i |V_{e \Sigma_i}|^2$  get improved.
The lower and upper limits on the mixing associated with the electron flavor also improves in the IH case up to $5\times 10^{-14}$ and $6 \times 10^{-14}$ respectively.
These limits can be observed from Fig.~\ref{Mix1}.

If we notice the individual mixing $|V_{\ell \Sigma_i}|^2$ in Fig.~\ref{Mix2} for identity orthogonal matrix $O$, we see that for the NH and IH cases the mixings $|V_{e\Sigma_1}|^2$ and $|V_{\tau \Sigma_3}|^2$ become zero as $m_1$ and $m_3$ go to zero. $|V_{\ell \Sigma_2}|^2$ has the same nature in both of the hierarchies. Below the PLANCK exclusion limit, $|V_{\ell\Sigma_2}|^2$ varies between $10^{-15}$ to $10^{-14}$ for the NH case and stays around $10^{-14}$ for the IH case. Following the same note, we notice that $|V_{e \Sigma_3}|^2$ stays around $10^{-15}$ below the PLANCK limit where as $|V_{\mu\Sigma_3}|^2$, $|V_{\tau\Sigma_3}|^2$ are one order of magnitude higher in the NH case. The scenario becomes opposite in the IH case for $|V_{\ell \Sigma_1}|^2$ where $|V_{e \Sigma_1}|^2$ stays around $2\times 10^{-14}$ and the other two mixings are one order lower than that below the PLANCK exclusion. The values could be observed from Fig.~\ref{Mix2}.

Second, we consider $O$ as a $3\times 3$ real orthogonal matrix. We find the same nature of $\Sigma_i |V_{\ell \Sigma_i}|^2$, as we found in the previous case and shown in Fig.~\ref{Mix1}. The individual mixing for this case is shown in Fig.~\ref{Mix3}. It is important to note that the depending upon the hierarchies the mixings are dependent upon the three light neutrino mass eigenvalues. In the NH case the mixing for the $\mu$ and $\tau$ can reach up to $2\times 10^{-14}$ and that for $e$ flavor can reach up to $3.5\times 10^{-15}$. This nature becomes opposite in the IH case where the maximum value of the mixing involving $e$ flavor can go up to $4\times 10^{-14}$ where the rest of the two remain some factor below around $10^{-14}$.

Third, we consider $O$ as a $3\times 3$ general orthogonal matrix where the entries of the matrix can be complex quantities. Using the neutrino oscillation data considering the PLANCK exclusion limit we have found that in this case there is no special correlation in the parameter space of the mixing and the lightest light neutrino mass. This happens due to the dependence of the mixing angles on the light neutrino mass eigenvalues and the complex entries of the general orthogonal matrix $O$. We notice that the highest mixing can reach $\mathcal{O}(10^{-5})$ depending up on the generations of the triplet and SM lepton which is very high compared to the other two choices of the orthogonal matrices. The lower limit in the mixing in the different cases reach around $10^{-18}$. We mention that the limits on the mixing from the EWPD have been given in \cite{delAguila:2008cj,delAguila:2008pw}. We quote limits as $|V_{e\Sigma}|^2= 3.61\times 10^{-4}$, $|V_{\mu \Sigma}|^2= 2.89\times 10^{-4}$ and $|V_{\mu \Sigma}|^2= 7.29 \times 10^{-4}$ respectively at $90\%$ CL.

\subsection{Multilepton channels}
\begin{figure}[]
\centering
\includegraphics[width=0.9\textwidth]{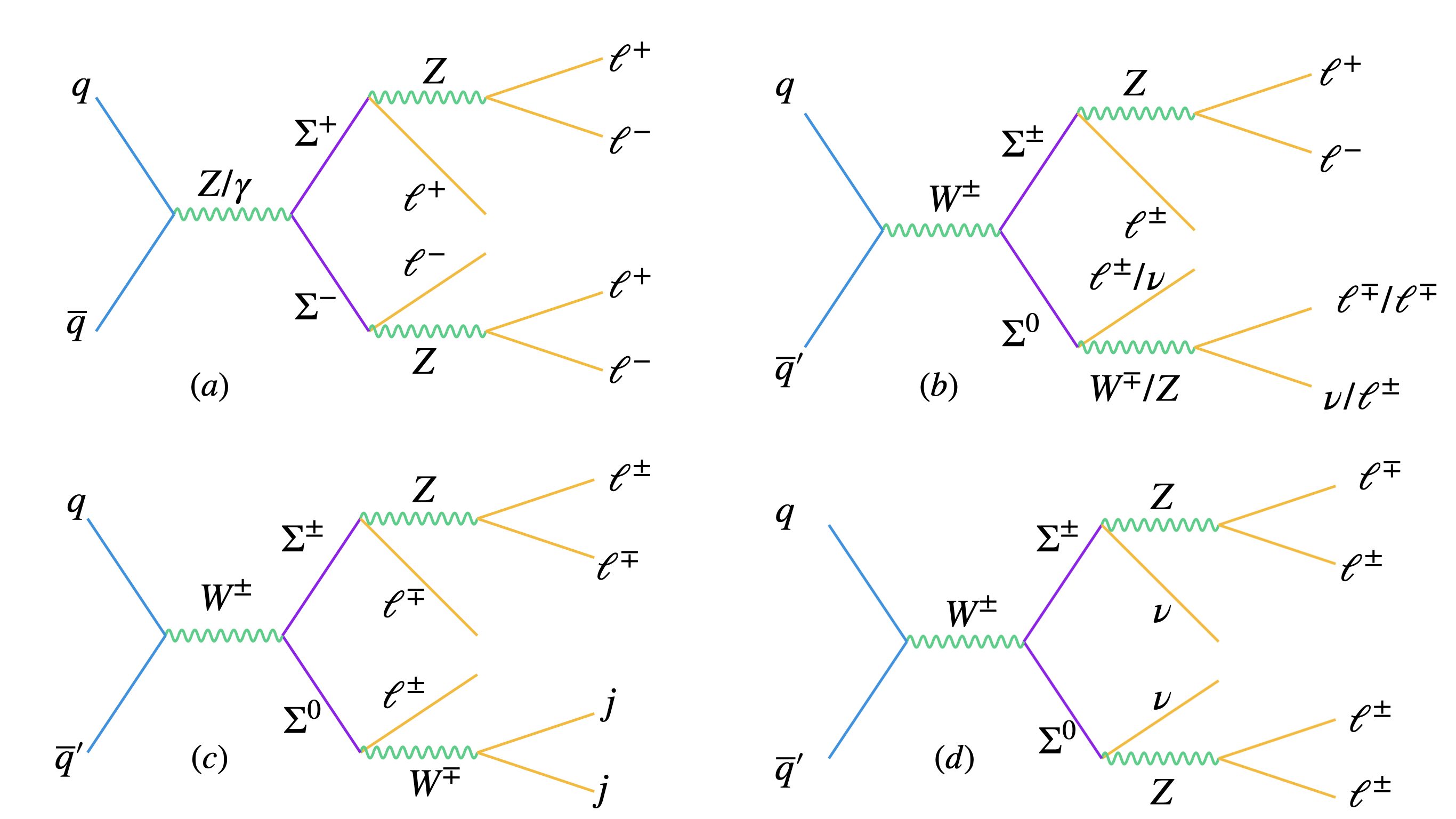}
\caption{Feynman diagrams for the multilepton channels at the hadron colliders: (a) $6$ lepton final state, $(b)$ $5$ lepton final state with missing momentum and $(c, d)$ $4$ lepton final state. Another $4$ lepton final state can be obtained from (a) when one $Z$ from $\Sigma^\pm$ decays into neutrinos.} 
\label{FD2}
\end{figure}

In this article we have shown the upper bound on the mixings as well as the lower bounds on the mixings generalizing the Yukawa coupling of the triplet fermion and applying the constraints from the direct searches. The LHC has given very strong bound on the triplet mass. At such strong bounds it becomes very crucial to study the triplets, however, interesting techniques including lepton-jet, displaced vertex, track search could be very useful for the heavier mass. These techniques can be useful in the future colliders using proton-proton, electron-positron and electron-proton in the near future. Here we focus on few multilepton final states at $pp$ collider coming from the $\Sigma^\pm \Sigma^0$ and $\Sigma^\pm \Sigma^\mp$ production channels. The decay of $\Sigma^\pm$, $\Sigma^0$ can produce $6$, $5$ and $4$ lepton final states. The Feynman diagrams of these channels are shown in Fig.~\ref{FD2}.

\begin{figure}[]
\centering
\includegraphics[width=1\textwidth]{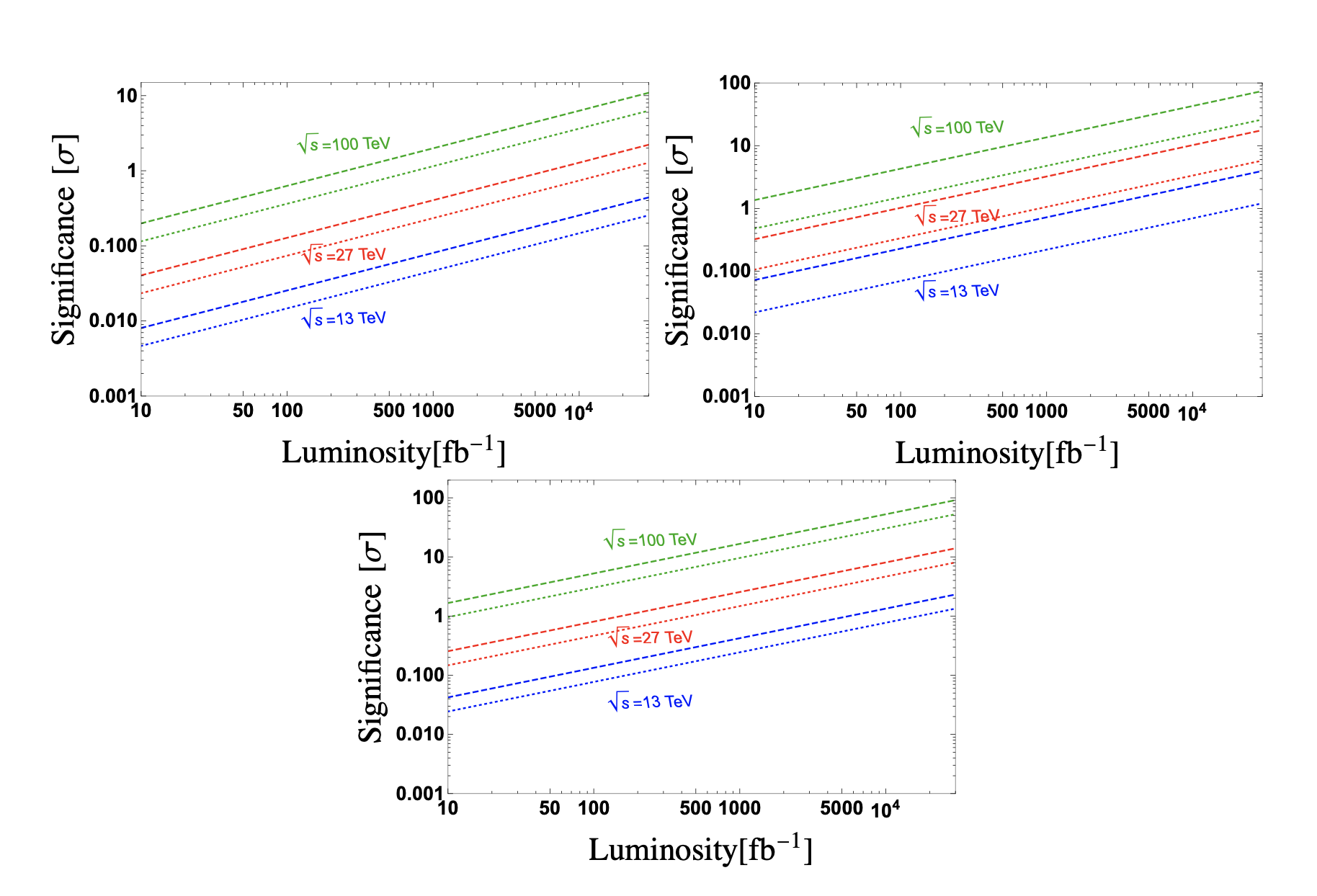}
\caption{The significance as a function of the luminosity. The luminosity goal for the $13$ TeV LHC is $3$ ab$^{-1}$ whereas that at the $27$ TeV and $100$ TeV are $15$ ab$^{-1}$ and $30$ ab$^{-1}$ simultaneously. The decay of the $W$ and $Z$ bosons from the triplets into a single generation lepton is represented by the dotted lines whereas that for the three generation case have been shown by the dashed lines. The $13$ TeV case is shown by the blue, the prospective $27$ TeV and $100$ TeV cases are shown by the red and green lines respectively. The $6$ lepton case is shown in the top left panel, the $5$ lepton case has been shown in the top right panel and the combined $4$ lepton case has been shown in the bottom panel respectively.}
\label{Sig1}
\end{figure}

For the $6$ charged lepton final state in Fig.~\ref{FD2}(a) the irreducible backgrounds can be obtained from $ZZZ$ and $ZZWW$ with missing momentum and jets at the proton proton collider using MadGraph \cite{Alwall:2014hca} followed by the hadronization \cite{Sjostrand:2014zea} and detector simulation \cite{deFavereau:2013fsa}. The production of this channel is independent of the mixing. The decay of the $Z$ and $W$ bosons into the charged leptons, namely, $Z \to \ell^+ \ell^-$ and $W \to \ell \nu$ will produce $6$ charged lepton final states.  The leading order production cross section of $ZZZ$ into $6$ charged leptons with $\ell=\mu$ is  $3.52\times10^{-4}$ fb at the $13$ TeV, $1.03 \times 10^{-3}$ fb at $27$ TeV and $5.3\times 10^{-3}$ fb at $100$ TeV respectively. The same for the $ZZWW$ final state also produces $6$ lepton final state with missing momentum and the cross section at $13$ TeV  is $5.02\times10^{-6}$ fb, at $27$ TeV is $2.06\times 10^{-5}$ fb and at $100$ TeV is $1.52 \times 10^{-4}$ fb respectively. The $6\ell$ production cross section from $\Sigma^\pm$ with $M_\Sigma=1$ TeV is $2.9\times 10^{-5}$ fb which is small to be observed at the HL-LHC era where as that at the $27$ TeV is $2.3\times 10^{-4}$ fb. Finally at the $100$ TeV, the $6\ell$ production cross section is $3.4\times 10^{-3}$ fb. To estimate the signal and background events we appled a transverse momentum cut of the lepton with $p_T^\ell > 50$ GeV, pseudo-rapidity with $|\eta_\ell| < 2.5$, separation between the leptons in the $\eta-\phi$ plane with $\Delta R_{\ell\ell} > 0.4$, transverse momentum cut on the jets with $p_T^j > 30$ GeV, pseudo-rapidity cut on the jets with $|\eta_j| < 2.5$ and the jet-lepton separation cut in the $\eta-\phi$ plane with $\Delta R_{\ell j} > 0.4$. Additionally we consider that the missing momenta should be less than $30$ GeV. We calculate the significance for the $6\ell$ with $\ell=\mu$ in Fig.~\ref{Sig1} using $\frac{\rm{Signal}}{\sqrt{\rm{Signal+ Background}}}$, where the signal and background events have been calculated with luminosities in fb$^{-1}$ as a free parameter. The $13$ TeV  and $27$ TeV cases can not give significant results even at a very high luminosity whereas the projected significance can reach up to $5~\sigma$ at $16.47$ ab$^{-1}$ at the $100$ TeV. If we combine three generations of the leptons from the $Z$ decay, the situation becomes better than the single flavor case for the $100$ TeV collider where $5 \sigma$ significance can be reached at $5.6$ ab$^{-1}$ luminosity. The projected significance for the single flavor case has been shown by the dotted lines and that for the three flavor case can be shown by the dashed lines in the top left panel of the Fig.~\ref{Sig1}.

The production of the $5$ lepton final state can be obtained from the $\Sigma^+ \Sigma^0$ and $\Sigma^- \Sigma^0$ channels in Fig.~\ref{FD2} (b). 
We combine all the channels to calculate the prospective significance for $\ell=\mu$. The irreducible backgrounds are coming from $ZZW$ and $WWWZ$ channels.
The cross sections at the $13$ TeV are $3.24 \times 10^{-3}$ fb and $2.43 \times 10^{-5}$ fb respectively. The cross sections of these backgrounds at the $27$ TeV are 
$9.61\times 10^{-3}$ fb and $9.73\times 10^{-5}$ fb respectively. The same at the $100$ TeV will be $5.00\times 10^{-2}$ fb and $6.3\times10^{-4}$ fb respectively.
Finally at the $100$ TeV, the $6\ell$ production cross section is $3.4\times 10^{-3}$ fb. To estimate the signal and the background events we applied same cuts as we applied in the $6$ lepton case. Considering a single generation of the leptons from the $Z$ and $W$ bosons can be observed at the $100$ TeV collider with $5 \sigma$ significance with $930$ fb$^{-1}$. The three generation cases from the $Z$ and $W$ bosons can be observed at the $27$ TeV collider with $2.073$ ab$^{-1}$ and at the $100$ TeV collider with $111$ fb$^{-1}$ with $5 \sigma$ significance. In this analysis we did not consider the misidentification of the leptons for simplicity. Generally the misidentification of the leptons at the LHC is below $1\%$ and not known for the prospective $27$ TeV and $100$ TeV colliders. The projected significance for the single flavor case has been shown by the dotted lines and that for the three flavor case can be shown by the dashed lines in the top right panel of the Fig.~\ref{Sig1}.

Finally we study the $4$ lepton final state from Fig.~\ref{FD2} (c) and (d). In addition to that Fig.~\ref{FD2} (a) will produce $4$ lepton final state when one of the $Z$ bosons from the $\Sigma^\pm$ decays into light neutrinos. We study the SM irreducible backgrounds including $W^+W^-W^+W^-$, $ZZ$,$ZZZ W^\pm$, $ZZZ$ and $ZZZZ$ to generic SM backgrounds of $4$ leptons plus missing momentum accompanied by jets. The $4$ lepton background from $ZZ$ is the leading one among these channels which contributes $0.142$ fb at the $13$ TeV, $0.32$ fb at the $27$ TeV and $1.18$ fb at the $100$ TeV colliders respectively. The $4$ lepton with missing momentum background is originated from $ZZZ$ and $W^+W^-W^+W^-$. This combined background provides a cross section of $2.03\times 10^{-3}$ fb at $13$ TeV, $6.00\times10^{-3}$ fb at $27$ TeV and $0.024$ fb at $100$ TeV. Additionally $4$ leptons and missing momentum background can be accompanied by the jets coming from $ZZZZ$ and $ZZZW$ providing $15.8\times10^{-6}$ fb at the $13$ TeV, $6.01\times 10^{-5}$ fb at the $27$ TeV and $4.16\times10^{-4}$ fb at the $100$ TeV collider. The combined signal cross sections for the $4$ lepton final states with missing momentum and jets $0.003$ fb at $13$ TeV, $0.028$ fb at $27$ TeV and $0.34$ fb at $100$ TeV colliders respectively. To estimate the signal and the background events we applied the same cuts as we applied in the $6$ lepton case. Hence the significance can reach at the $5\sigma$ level with $10.47$ ab$^{-1}$ at the $27$ TeV for one generation lepton and $3.32$ ab$^{-1}$ with three generations of the leptons from the final $W$ and $Z$ leptonic decay. Similarly the $5\sigma$ significance for this channel can be achieved at $247$ fb$^{-1}$ with single flavor and $74.5$ fb$^{-1}$ with three flavor leptons at the $100$ TeV collider. The projected significance for the single flavor case has been shown by the dotted lines and that for the three flavor case can be shown by the dashed lines in the bottom panel of the Fig.~\ref{Sig1}.
\subsection{Summary at the $e^-e^+$ and $e^-p$ colliders}
The triplet fermions can be produced at the $e^- e^+$ colliders in the form of the pair production $(\Sigma^+ \Sigma^-)$ and single production $(\Sigma^\pm$ and $\Sigma^0)$ in association with the SM particles. Hence, the above multilepton final states can be studied at the linear collider. Apart from the multilepton signatures, boosted objects from the $\Sigma^\pm$ and $\Sigma^0$ can also be studied as in linear collider the cross section remains almost same until the energy threshold. The boosted objects from the type-III seesaw has been studied in \cite{Das:2020gnt} with single charged lepton plus fat jet, dilepton plus fat b-jet, dilepton plus fat-jet (non-b) and two fat-jet plus missing energy for $M_\Sigma=1$ TeV. Different multilepton channels have been studied in \cite{Goswami:2017jqs} for $M_\Sigma < 1$ TeV, however, the strong LHC bounds rule out the triplets below $880$ GeV \cite{Sirunyan:2019bgz} and $790$ GeV \cite{Aad:2020fzq} for the flavor diagonal scenario. We have also studied the triplet production at the electron proton collider with the boosted objects \cite{Das:2020gnt} for the first time. 

We have found that the mixing can be probed at the $\sqrt{s}=1$ TeV and $\sqrt{s}=3$ TeV $e^-e^+$ and $\sqrt{s}=3.46$ TeV $e^-p$ colliders well below the eletroweak precision bounds \cite{delAguila:2008cj,delAguila:2008pw} at the $5\sigma$ significance level. The boosted objects can also be studied at the hadron colliders in the near future. Apart from the electron channels, the muon and tau channels will also be interesting to study at the lepton colliders at the $e^-e^+$ colliders and $e^-p$ colliders respectively. 
\section{Conclusions}
\label{conclusion}
We study the type-III seesaw model in this article where we mainly observe the allowed parameter regions for the light-heavy mixings as a function of the lightest neutrino mass.
Depending upon the choice of the general Dirac Yukawa coupling of the triplet fermion with the SM lepton doublet and the Higgs doublet the allowed parameter space of the mixing changes under a variety of of constraints. We also calculate the branching ratios of the neutral and charged multiplets of the triplet fermion into leading and sub-leading modes to investigate the correlation with the lightest light neutrino mass eigenvalue for two different light neutrino mass hierarchies. In a continuation we have also shown the parameter space of the proper decay length of the triplet generalizing the Dirac Yukawa coupling using three different choices. This leads to an interesting property of the displaced vertex search for the triplet fermions due to their sizable proper decay length, however, we predict that for some parameter choices prompt decay of the triplet fermion can also be possible. We evaluate the mixings, branching ratios of the triplets in the different modes and hence we predict that such parameter spaces can be probed studying the decay modes (prompt or displaced) of the triplets at the different high energy colliders in the near future.
\begin{acknowledgments}
The work of A. D. is supported by the JSPS, Grant-in-Aid for Scientific Research, No. 18F18321. The work of S. M. is supported by the Spanish grants FPA2017-85216-P (AEI/FEDER, UE), PROMETEO/2018/165 (Generalitat Valenciana) and the Spanish Red Consolider MultiDark FPA2017-90566-REDC. 
\end{acknowledgments}
\bibliography{bibliography}
\bibliographystyle{utphys}
\end{document}